\documentclass[preprint,5p,times,twocolumn]{elsarticle}

\usepackage[colorlinks=true, linkcolor=blue, citecolor=blue, urlcolor=blue]{hyperref}
\usepackage{makecell}
\usepackage{multirow}
\usepackage{enumitem}
\usepackage{booktabs}
\usepackage{tabularx,booktabs}
\usepackage{multicol}

\usepackage{longtable}

\usepackage{tikz}

\usepackage{subcaption}
\usepackage{pdflscape}
\usepackage{graphicx}

\usepackage{bm}

\usepackage{comment}

\usepackage{fontawesome}

\usepackage{xcolor}
\usepackage{color}
\definecolor{gray75}{gray}{.25}
\definecolor{gray70}{gray}{.3}
\definecolor{gray60}{gray}{.4}
\definecolor{gray50}{gray}{.5}
\definecolor{gray40}{gray}{.6}
\definecolor{gray30}{gray}{.7}
\definecolor{gray20}{gray}{.8}
\definecolor{gray15}{gray}{.85}
\definecolor{gray10}{gray}{.9}
\definecolor{gray05}{gray}{.95}

\definecolor{graytable}{gray}{.95}

\definecolor{headertable}{rgb}{0.0, 0.0, 0.0}

\definecolor{steBoxLine}{rgb}{0.0, 0.0, 0.0}

\usepackage[framemethod=tikz]{mdframed}
\newmdenv[
  linecolor=steBoxLine,
  backgroundcolor=gray!15,
  linewidth=1pt,
  roundcorner=5pt,
  skipabove=10pt,
  skipbelow=10pt,
  innertopmargin=6pt,
  innerbottommargin=6pt,
  innerleftmargin=10pt,
  innerrightmargin=10pt
]{takehomebox}

\usepackage[colorinlistoftodos]{todonotes}

\usepackage{amsthm}

\usepackage[]{mdframed}

\mdfdefinestyle{stebox4}{%
linecolor=steBoxLine,linewidth=1pt,%
leftmargin=0cm,rightmargin=0cm,%
roundcorner=5pt,
topline=false,bottomline=true,rightline=false,leftline=false,%
backgroundcolor=gray!15,
frametitlebackgroundcolor=steBoxLine
}

\mdfdefinestyle{steboxSummaryWhite}{%
linecolor=steBoxLine,linewidth=1pt,%
leftmargin=0cm,rightmargin=0cm,%
roundcorner=5pt,
topline=true,bottomline=true,rightline=false,leftline=false,%
backgroundcolor=gray!15,
frametitlebackgroundcolor=white,
frametitlerule=true,
frametitlerulewidth=1pt
}

\mdfdefinestyle{stebox5}{%
linecolor=steBoxLine,
linewidth=1pt,%
leftmargin=0cm,rightmargin=0cm,%
roundcorner=5pt,
topline=false,bottomline=false,rightline=false,leftline=true,%
backgroundcolor=gray!15
}

\newcommand{\steResearchQuestionBox}[1]
{   
    \begin{mdframed}[style=stebox5]
    {#1}
    \end{mdframed}
}

\usepackage{tabularx}
\usepackage{siunitx}
\usepackage{multirow}
\usepackage{colortbl}
\usepackage{longtable}
\usepackage{tabularray}
\usepackage{tabu}

\usepackage{dcolumn}

\usepackage{array}

\newcommand{\revised}[1]{{\color{black}#1}}

\newcolumntype{P}[1]{>{\centering\arraybackslash}p{#1}}
\newcolumntype{a}{>{\columncolor{headertable}}l}
\newcolumntype{d}[1]{D{.}{.}{#1}}

\journal{Journal of Systems and Software}

\begin{document}

\begin{frontmatter}


\title{From Challenge to Change: Design Principles for AI Transformations}

\tnotetext[mytitlenote]{Accepted for publication in the Journal of Systems and Software.}

\author[label1]{Theocharis Tavantzis\corref{cor1}}
\ead{thtah@cs.aau.dk}

\author[label1]{Stefano Lambiase}
\ead{stla@cs.aau.dk}

\author[label1]{Daniel Russo}
\ead{daniel.russo@cs.aau.dk.}

\author[label2]{Robert Feldt}
\ead{robert.feldt@chalmers.se}

\cortext[cor1]{Corresponding author}

\affiliation[label1]{
        organization={Department of Computer Science, Aalborg University},
        city={Aalborg},
        country={Copenhagen}}

\affiliation[label2]{
        organization={Department of Computer Science and Engineering, Chalmers University of Technology | University of Gothenburg},
        city={Gothenburg},
        country={Sweden}}

\begin{abstract}
The rapid rise of Artificial Intelligence (AI) technologies is reshaping Software Engineering (SE) practice, unlocking new opportunities while introducing human-centered challenges. Although prior research acknowledges behavioral and other non-technical factors affecting AI integration, most studies still emphasize technical concerns and offer limited insight into how teams adapt to and trust AI systems. This work proposes a Behavioral Software Engineering (BSE)-informed, human-centric framework to support SE organizations during early AI adoption. We employed a mixed-methods methodology to construct and refine the framework. A literature review of organizational change models established its theoretical foundation, and thematic analysis of interview data produced concrete, actionable steps. The resulting framework comprises nine dimensions: \textit{AI Strategy Design, AI Strategy Evaluation, Collaboration, Communication, Governance and Ethics, Leadership, Organizational Culture, Organizational Dynamics, and Up-skilling}, each supported by design principles and actionable steps. To collect preliminary practitioner feedback, we conducted a survey (N=105) and two expert workshops (N=4). Survey responses show that Up-skilling (\SI{15.2}{\percent}) and AI Strategy Design (\SI{15.1}{\percent}) received the highest \$100-method allocations, highlighting their perceived centrality in early AI initiatives. Findings suggest that organizations currently prioritize procedural aspects such as strategy design, while human-centered guardrails remain comparatively underdeveloped. Early feasibility checks of the workshops reinforced these patterns and highlighted the importance of grounding the framework in real-world practice. By identifying critical behavioral dimensions and offering actionable guidance, this contribution provides practitioners with a pragmatic roadmap for navigating the socio-technical complexity of early AI adoption and outlines future research directions for human-centric AI in SE.

\end{abstract}


\end{frontmatter}

\section{Introduction}
\label{sec1}

Over the past 70 years, programming languages and Software Engineering (SE) have undergone continuous and radical shifts~\cite{terragni2025futureAi-drivenSE}. The rise of AI technologies, especially with the advent of Large Language Models (LLMs), constitutes the latest revolution in the field, enabling capabilities previously unthinkable only a few years ago~\cite{fan2023llmforSE, du2024evaluatingLLMs}. Specifically, LLMs appear to influence various core tasks of the SE life-cycle, from requirements elicitation~\cite{ronanki2023chatGPT1RE} to code generation~\cite{hou2024largelanguagemodelssoftware} and maintenance~\cite{xiao2024genAI4pullRequests}.

In recent years, SE research has increasingly examined the impact of AI on the discipline~\cite{gupta2024genAIsystematicReview}. Studies have already underscored benefits such as increased automation, enhanced productivity, and opportunities for innovation~\cite{fan2023llmforSE, ozkaya2023nextFrontier, yoon2024intent}. Yet, much of the SE literature traditionally prioritizes technical and process-driven aspects, often overlooking the human and collaborative dimensions~\cite{capretz_humanSE_2014, perry_people_process_improvement_1994, storey2020SociotechnicalFramework}. This imbalance seems to be maintained or even amplified as AI radically influences the field~\cite{peretz_andersson_empirical_2022}, potentially sidelining the central role of human behavior and interaction to successful AI-enabled transformations~\cite{russo2024cphManifesto, shneiderman2020human_centredAI}.

Recent studies have shown that non-technical factors strongly influence AI adoption in SE organizations~\cite{russo2024navigatingComplexityOfgenAI, mit2025GenAIDivide, choudhuri2025trustAndBehaviorToGenAI}. For example, Russo~\cite{russo2024navigatingComplexityOfgenAI} found that compatibility of AI tools with existing workflows is the main driver of early AI adoption, while perceived usefulness and social influence were less decisive. To analyze these complexities, socio-technical frameworks~\cite{storey2020SociotechnicalFramework, hoda2021socio} and theories rooted in psychology, cognition, and social dynamics, such as Behavioral Software Engineering (BSE)~\cite{lenberg_bse_2014, lenberg_bse_slr_2015}, offer a valuable baseline to comprehend deeper the transformative processes in SE industry.

Relying on this perspective, previous research~\cite{tavantzis2024HumanCenteredAI} identified key BSE concepts, such as \textit{Emotions}, \textit{Personality}, and \textit{Politics}, that shape AI adoption as well as critical barriers, including \textit{Resistance to Change} and \textit{Ethical Concerns}. These findings underscore the need for adopting a human-centered approach in AI-driven transformations in SE. However, Tavantzis and Feldt~\cite{tavantzis2024HumanCenteredAI} note that existing change management models do not fully address the BSE-related challenges of AI adoption.

While some frameworks, such as Errida and Lotfi’s~\cite{errida2021DescriptiveOrgModel}, identify key factors like resistance management, issues unique to AI remain open, such as ethical concerns and over-reliance on AI. Furthermore, practitioners increasingly question the applicability of traditional change models in AI-driven transformations, expressing concerns that these frameworks may be insufficient or misaligned with the emerging organizational context.

Despite these contributions, little yet is known about how BSE perspectives can be used to guide AI adoption in practice. Hence, extending our earlier findings~\cite{tavantzis2024HumanCenteredAI}, this study aims to move from identifying challenges to offering practical guidance for SE companies. Specifically, this research seeks to propose a BSE-informed framework to mitigate the human and socio-technical complexities of \textbf{early-stage} AI-driven transformations in SE organizations.

\revised{The framework targets organizational and process design for early-stage AI adoption in SE organizations; it does not prescribe technical software/system design activities. This is made on purpose; While the framework does not map onto specific SE activities or phases of the SE life-cycle, it addresses the organizational readiness conditions that enable organizations to subsequently engage with structured SE processes. In terms of established SE process models such as ISO/IEC/IEEE 15288~\cite{iso15288}, the framework operates at a layer that corresponds most closely to organizational enabling processes related to infrastructure management, portfolio management, and human resource management, rather than with technical life-cycle processes.} Our focus is deliberately limited to these early phases, rather than considering later anchoring and sustaining activities, reflecting the current immaturity in AI integration across SE companies.

\revised{We define early-stage AI adoption as the initial organizational efforts to integrate AI into processes and workflows. At the organizational level, this stage is characterized by exploratory or pilot-level AI use, limited governance structures, emerging roles and responsibilities, limited internal AI expertise, and practices that have not yet been scaled or institutionalized across the organization. The present framework is intended for this formative stage, where organizations are still establishing the organizational, behavioral, and governance conditions for AI integration. Later-stage concerns, such as scaling AI use, anchoring change, institutionalizing AI-enabled practices, long-term sustainability, and continuous optimization, are outside the scope of this study and remain directions for future research.} This research decision allows us to generate timely and actionable insights for practitioners navigating the uncertainty and fluidity of early adoption, while leaving anchoring change initiatives as a future research direction.

To pursue this objective, we employed a mixed-methods methodology, combining theoretical frameworks in literature and practical insights. We began with a literature review to assess how existing change management research can mitigate the AI-risen challenges. Then, we enriched these insights using thematic analysis to an existing qualitative dataset, ensuring industrial perspectives from practitioners across diverse roles and organizations. Following a systematic way, we aimed to add practical actions to the design principles of the framework. \revised{These actionable steps are intended to support organizational decision-makers in navigating early-stage AI adoption and design their adoption strategies.} Finally, survey data and two early feasibility check sessions allowed us to preliminarily validate the framework and assess the perceived importance of its dimensions.

In a nutshell, this paper proposes a BSE-informed framework for early-stage AI-driven transformations, consisting of dimensions, design principles, and actionable steps \revised{with the aim to facilitate change management process in SE organizations}. The framework builds upon existing socio-technical perspectives by integrating behavioral dimensions such as resistance to change, ethical concerns, and over-reliance on AI. Our artifact suggests that while technical integration is valued, human aspects are decisive for adoption success. By offering actionable guidance and validated dimensions, this study aims to empower SE companies to navigate the uncertainty of early AI incorporation and provide a foundation for future research on human-centric organizational change.

The remainder of the paper is organized as follows. Section \ref{sec2} offers the theoretical background that is required for this research study by discussing key concepts and related publications. Section \ref{sec3} is devoted to the methodology we followed for the purposes of this work. Sections \ref{sec4} and \ref{sec5} present our results and discuss their connection with the existing literature. Finally, Section \ref{sec6} concludes the study, demonstrating the limitations and directions for future work.

\section{Background and Related Work}
\label{sec2}

This section provides an overview of the main areas of background work that inform our study, including Behavioral Software Engineering, organizational change, AI transformation, and the human aspects of AI transformation. Each subsection outlines prior research that frames and motivates our contribution. Accordingly, we position our contribution as an early-stage framework for AI transformation in SE, combining generic organizational change models and domain-specific frameworks, and focusing on the formative stages rather than end-to-end transformation.

\subsection{Behavioral Software Engineering}

Even though research community discussed the importance of human aspects in SE procedures since the early ages of the domain~\cite{weinberg_psychology_1971}, the majority of existing research and industrial initiatives focuses predominantly on the technical aspects of SE life-cycle~\cite{capretz_humanSE_2014, perry_people_process_improvement_1994, ferreira_spi_2011}. After 2014, Lenberg et al.~\cite{lenberg_bse_2014} and Graziotin et al.~\cite{graziotin2015UnderstandingAffect} necessitated the criticality of exploring the human dimension within SE processes, introducing the concepts of \textit{Behavioral Software Engineering (BSE)} and \textit{psychoempirical software engineering} respectively. Later on, in their joint study~\cite{graziotin2022PsychometricsinBSE}, agreed on the term of BSE as the \textit{study of cognitive, behavioral and social aspects of software engineering performed in three different levels of analysis, namely the individual, group, and organizational level}.

In 2015, Lenberg et al.~\cite{lenberg_bse_slr_2015} revealed a substantial gap in the research body of the newly-introduced field of BSE. Particularly, 42 out of the 55 identified BSE concepts had less than 10 publications, implying significant room for research contributions. Similarly, Storey et al.~\cite{storey2020SociotechnicalFramework} observed that most of the contributions, included in their work, are solution-oriented, despite the authors' claim to explore the human dimension. Hence, they argue that it is important to strike a balance between understanding human and social aspects and exploring technical innovation. Finally, Zolduoarrati et al.~\cite{zolduoarrati2024SecondaryStudies} provide a foundation for researchers and practitioners to understand the current state of the human aspects in SE field. Their tertiary study led to the identification of 16 categories, including workspace conditions, knowledge sharing, and trust as key factors influencing developer productivity.

\subsection{Organizational Change}

Organizational change is often driven by internal factors, such as the need to improve productivity, or external influences, like emerging technologies~\cite{beer_walton_organization_1987}. Even though it constitutes a field with persistent research interest, a shift towards the human aspects of change initiatives can be observed~\cite{lenberg_organizational_change_2017, understanding_org_change_2023, klotins2022continuous}.

In the literature, two types of change management models can be distinguished, the processual and the descriptive models~\cite{parry2024EmpiricalOrgChange}. The former define sequential steps for change, while the descriptive models determine the factors that impact the success of an organizational change endeavor. One of the most well-received processual models regarding is proposed by Kotter~\cite{kotter_leading_change_2012}.

Despite its wide acceptance and adoption, certain weaknesses and limitations of Kotter's model have been discussed within the research community. Appelbaum et al.~\cite{appelbaum2012back} discuss model's lack of flexibility, due to the sequential order of steps, and the unsuitability of the model in specific contexts, like changes with more technical orientation. Moreover, the authors highlight the lack of deep investigation of the human factors during the transition process that can affect the success of the whole endeavor, such as employees' resistance to change. A finding that is supported by recent research findings, as they have identified employees’ behavior and attitudes towards organizational change as key factors in the change process~\cite{lenberg_organizational_change_2017}. More recent additions in the research body, like Errida and Lotfi's descriptive framework~\cite{errida2021DescriptiveOrgModel}, include factors that address the human dimension of the change initiative, such as Resistance Management.

Lenberg et al.~\cite{lenberg_human_factors_2015} showed that a comprehensive understanding of the cognitive, behavioral, and social factors is vital for successful transformations within the field of Software Engineering. The thematic analysis resulted in the recognition of four themes as problems in neglecting the human factors in SE industry, with organizational change being one of these four issues. More specific to SE, Umarji and Seaman~\cite{umarji2005predictingAcceptanceOfSPI} explored how well-structured models in psychology literature, such as the Theory of Planned Behavior, can make the Software Process Improvement (SPI) smoother and more efficient. Despite the encouraging initial results, more empirical evaluation of the proposed framework is necessary by gathering quantitative data from industrial practitioners.

\subsection{AI Transformation}

Artificial Intelligence (AI) constitutes a substantial catalyst for organizational transformations, offering new capabilities to practitioners. Davila et al.~\cite{davila2024AIbasedAssistants} found that accelerating online searching, typing, and syntax recall are among the most frequent applications of AI tools, such as GitHub Co-pilot. In this process, the out-of-context replies and the non-functional code were noted as the main barriers related to the collaboration with AI-based assistants. The authors underscore the need to guide developers regarding the utilization of AI tools and the criticality in taking into account the ethical and legal issues that AI tools raise.

Bashir~\cite{bashir2024AI-centricRE} examined the use of NLP and LLMs to support Requirements Engineering in large-scale industrial systems, showing that models such as BERT can automate tasks like requirements elicitation but struggle with domain-specific adaptation. Similarly, Ajiga et al.~\cite{ajiga2024enhancingSEpractices} highlighted how AI is reshaping software development through automated code generation and testing, noting that thoughtful adoption can enhance productivity, quality, and innovation. Nonetheless, data security concerns and high implementation costs remain key barriers to effective AI integration.

AI readiness models \revised{and adoption frameworks} constitute an area of particular interest within \revised{academia and industry. Regulatory, such as EU AI Act~\cite{eu2025aiAct}, and industry guidance, such as Google's AI Principles~\cite{google2025AIprinciples}, provide a high-level overview of the principles and structures to the use and development of AI tools. Similarly, organizational readiness models~\cite{nortje2020AFF, johnk2021AIcomes_AIorgReadiness, holmstrom2022AItoDigitalTransformation, uren2023AIReadiness, bandara2025aiReasinessForPM, lavin2022TRLML} assess organizations' capacity to adopt AI technologies.

All these initiatives constitute a valuable foundation for the multi-faceted nature of AI technologies and indicate the different aspects that organizations need to consider. However, they either focus predominantly on technical aspects and overlook the human-centered angles (e.g.~\cite{holmstrom2022AItoDigitalTransformation, lavin2022TRLML}) or lack empirical validation regarding their applicability (e.g.~\cite{johnk2021AIcomes_AIorgReadiness}). For example, Holmstr\"om~\cite{holmstrom2022AItoDigitalTransformation} identified four dimensions in the proposed AI readiness framework, namely Technology, Activities, Boundaries, and Goals. These aspects often give less explicit treatment to behavioral and change-management mechanisms. On the other hand, the framework proposed by J\"ohnk et al.~\cite{johnk2021AIcomes_AIorgReadiness} is empirically grounded through qualitative expert interviews and preliminary validation, but it lacks a large-scale, in-depth quantitative validation.}

\subsection{Human Aspects in AI Transformation}

Although AI is advancing rapidly, there are still major gaps in interdisciplinary research of the field, especially regarding human factors. Peretz-Andersson and Torkar~\cite{peretz_andersson_empirical_2022} revealed via a systematic way the psychology-related research gap, as only four out of the fifty-two included studies delve into such topics. Since then, research contributions have focused more on the human and social dimension of the AI integration.

A considerable number of the challenges posed by AI originate from human-centered issues rather than technological limitations. In line with BSE’s emphasis on social aspects across different dimensions, Dwivedi et al.~\cite{dwivedi_AI_2021} identify potential job losses due to AI-driven automation, and ethical dilemmas, like algorithmic bias and lack of transparency in decision-making. Similarly, Dolata et al.~\cite{dolata2024freelancersGenerativeAI} featured the need to validate AI-generated outputs and clients' extraordinary expectations due to misconceptions of AI capabilities as two issues that are common with the emergence of generative AI. Stakeholders active collaboration and communication are crucial to mitigate ethical, social, and environmental challenges, aiming to align AI innovations with societal values and foster transparent and beneficial technological progress~\cite{polyportis2024navigating}.

Lately, studies started discussing that AI integration is more than a technical-oriented process. Russo~\cite{russo2024navigatingComplexityOfgenAI} showed that while usefulness and ease-of-use matter, compatibility with existing workflows is the dominant factor for GenAI incorporation in SE. This challenges traditional technology adoption theories, suggesting that effective AI integration requires smooth fit into the way teams already work, not just showcasing benefits and pushing adoption. A finding stressed in MIT’s 2025 business report~\cite{mit2025GenAIDivide}, discussing that 95\% of GenAI implementations still show no return. This is not because of the model's capabilities, but due to tools inability to integrate within existing workflows. Dell'Acqua et al. supported this view with their work~\cite{dellAcqua2023jaggedFrontier}. Particularly, they found that while generative AI can significantly boost productivity, it can mislead skilled professionals and lead to worse outcomes. Hence, AI integration strategy should focus on which tasks to incorporate AI, not just whether to adopt AI.

A recent study by Tavantzis and Feldt~\cite{tavantzis2024HumanCenteredAI} examines AI transformation through the lens of BSE. The authors conducted a qualitative study, utilizing thematic and narrative analysis, to explore the BSE concepts and challenges that shape AI integration. Their results demonstrate several sub-themes linked to twelve core BSE concepts, alongside six primary BSE-related challenges and their associated sub-challenges.

\begin{takehomebox}
    \textbf{Related Work Summary:} These insights highlight the inherent complexity of AI integration and \revised{reveal gaps in change management models, particularly around emerging risks like over-reliance on AI tools. Existing efforts provide a useful structure, but often lack deep empirical validation or overlook the human element of AI adoption. Our contribution, rooted in the BSE principles, aims to contribute to these gaps.}
\end{takehomebox}

\section{Methodology}
\label{sec3}

\begin{figure*}[htbp]
    \centerline{\includegraphics[width=1\linewidth]{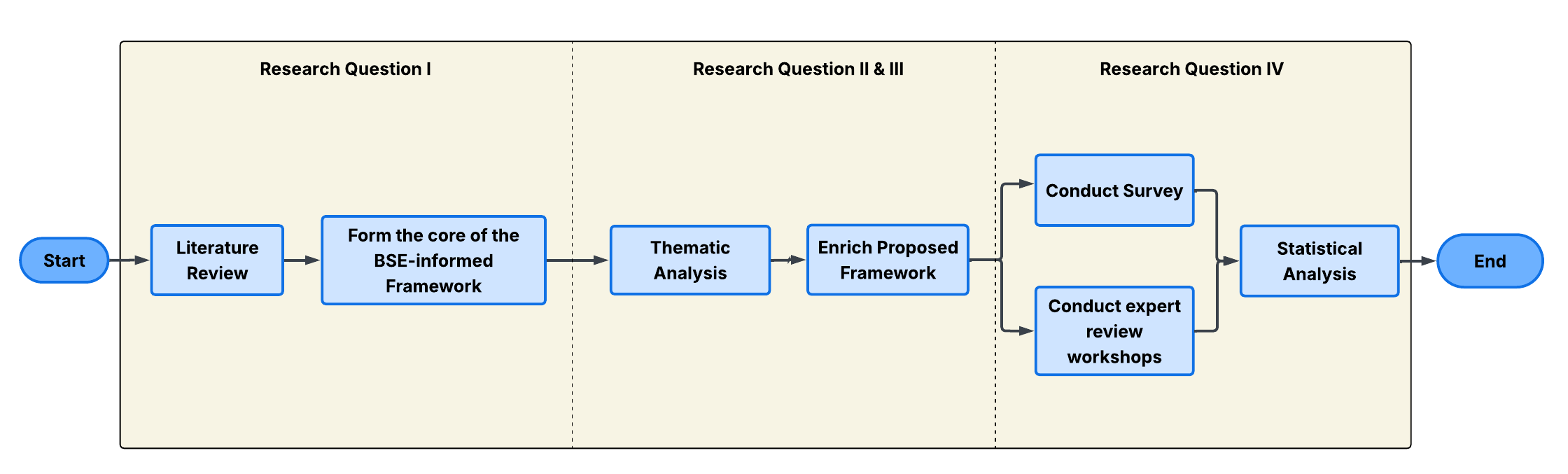}}
    \caption{Research Method Overview}
    \label{fig:research_method}
\end{figure*}

This section reports on the methodology of the study, describing the research objectives and research questions. As shown in Figure \ref{fig:research_method}, we employed a mixed-methods methodology to conduct this research.

\subsection{Objective and Research Questions}

The objective of this work is to \textbf{structure and propose an actionable BSE-informed framework} to support early-stage AI integration. Tavantzis and Feldt's findings~\cite{tavantzis2024HumanCenteredAI} informed the formalization of the framework. Specifically, they discuss the BSE-related challenges that AI incorporation raises. These challenges span various aspects, like \textit{Resistance to Change} and \textit{Skills in Era of AI}. Notably, study's participants — whose input informed these hurdles — work in organizations at the early stages of AI integration. Thus, their insights reflect this phase. Consequently, the proposed framework is deliberately tailored to address the challenges of the early-stage AI integration.

Beyond addressing the technical hurdles of early AI integration, the framework also seeks to support organizations in managing the behavioral and social challenges that accompany this process. To achieve this, we structured the study around four research questions to guide the research design choices and final output of the research~\cite{creswell2017ResearchDesign}. In particular:

\steResearchQuestionBox{\faQuestionCircle\ \textbf{RQ\textsubscript{1}}—\textit{What foundational steps can be derived from existing change management literature to support AI-driven organizational change in SE organizations?}}

RQ\textsubscript{1} aims to investigate whether existing organizational change models address the behavioral challenges of AI integration identified in prior research~\cite{tavantzis2024HumanCenteredAI}. Through a literature review, the study maps these challenges to the steps/factors of established change management frameworks, setting the foundation for further refinement subsequently.

\steResearchQuestionBox{\faQuestionCircle\ \textbf{RQ\textsubscript{2}}—\textit{How can a human-centric framework address the human and socio-technical complexities of AI-driven transformations in SE organizations?}}

Building on the foundation of RQ\textsubscript{1}, RQ\textsubscript{2} explores the behavioral challenges overlooked by current change management models. In this RQ, we i) identify these gaps, ii) employ thematic analysis on prior qualitative data to derive solutions, and iii) extend the framework to incorporate the BSE perspective, culminating in design principles (DPs) for effective AI-driven organizational change.

\steResearchQuestionBox{\faQuestionCircle\ \textbf{RQ\textsubscript{3}}—\textit{What tangible steps can be incorporated into the human-centric framework to ensure framework’s applicability?}}

RQ\textsubscript{3} relies on the critique that academic contributions are overly theoretical and lack practical guidance~\cite{ivanof2017theoryGapAndPractice}. To counter this, we synthesize insights from peer-reviewed and grey literature, the prior qualitative dataset, and our recommendations to develop actionable steps for each DP. This aims to enhance framework’s practical relevance \revised{for decision-makers and practitioners navigating early-stage AI adoption}, laying the foundation for RQ\textsubscript{4}.

\steResearchQuestionBox{\faQuestionCircle\ \textbf{RQ\textsubscript{4}}—\textit{How do SE practitioners perceive the relevance and clarity of the proposed framework's dimensions and design principles?}}

RQ\textsubscript{4} explores the perceived usefulness of the proposed framework for SE organizations. Given time constraints, a full Action Research study was unfeasible, so we conducted a preliminary assessment. This included a survey of SE professionals for quantitative feedback on the framework’s overview and expert review workshops for initial validation and detailed input on specific DPs, which we label as \textit{"early feasibility checks"}. Together, these methods provide early insights into the framework’s practical value and suggest directions for future work.

\subsection{Data Collection}

We adopted a multi-method data collection strategy to enhance the validity and reliability of the study. The approach comprised a literature review, a survey, and two early feasibility checks.

\subsubsection{Literature Review}

To answer RQ\textsubscript{1} and establish the groundwork of the proposed framework, we relied on earlier research materials. In this regard, we conducted a literature review to produce a structured, comprehensive, and unbiased synthesis of existing knowledge. The synthesis and analysis involved listing the different change management models and evaluating which of the identified BSE challenges can be addressed. The screening process of the literature relied on the Preferred Reporting Items for Systematic Reviews and Meta-Analyses (PRISMA) statement~\cite{moher2009PRISMA}.

\begin{figure*}[htbp]
    \centerline{\includegraphics[scale=0.5]{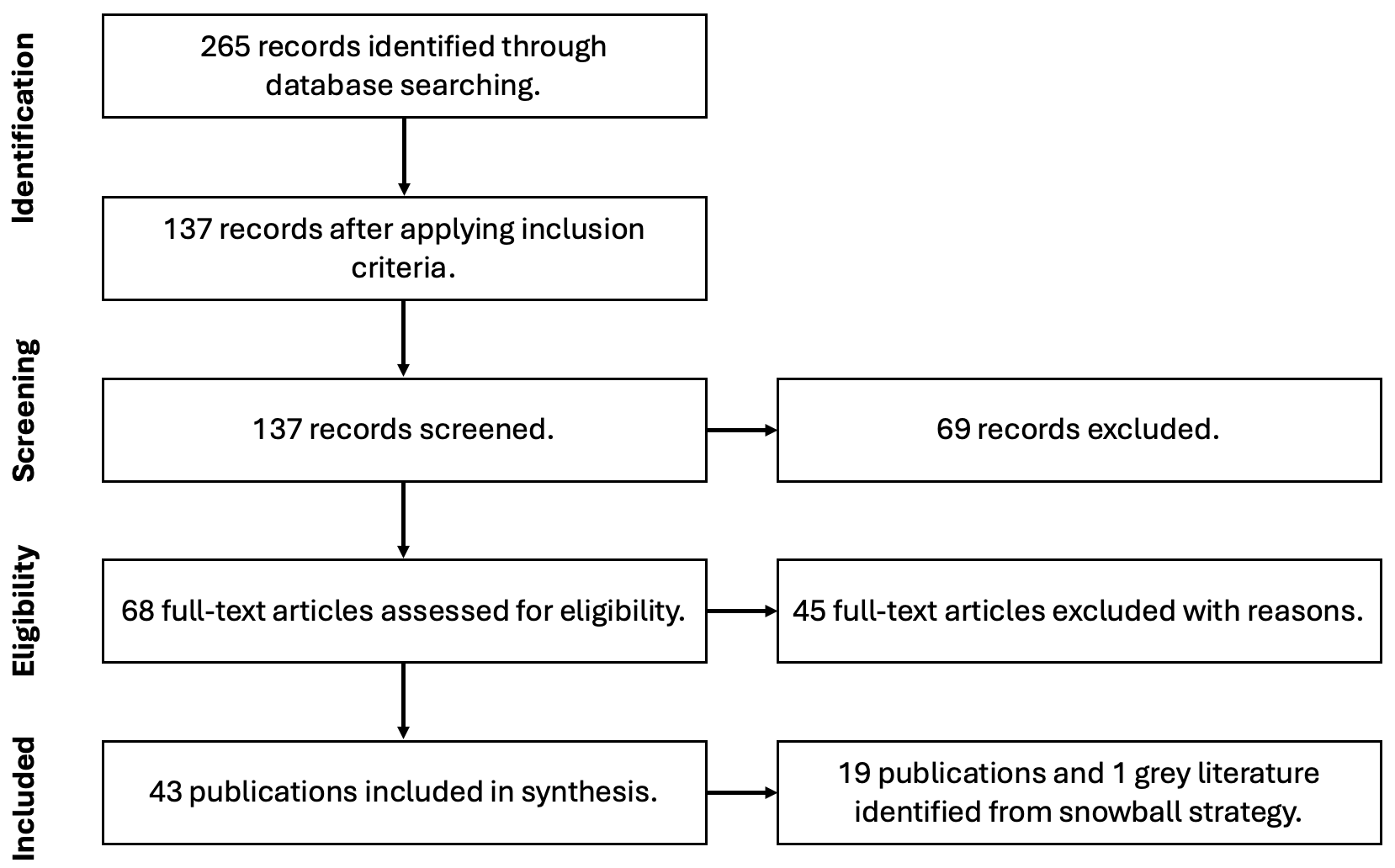}}
    \caption{PRISMA statement overview}
    \label{fig:prisma_statement}
\end{figure*}

PRISMA statement ensures transparency and comprehensive reporting in systematic reviews by providing a structured framework for detailed explanation of the studies included in the review. According to the methodology, the process consists of four discrete phases; \textit{the Identification phase, the Screening phase, the Eligibility phase, and the Included phase}. These phases extend from the process of searching for pertinent studies to the final stage of including the useful material found. The search took place in the scientific database of Scopus. Figure~\ref{fig:prisma_statement} presents an overview of the PRISMA statement as applied in this study.

\paragraph{Identification phase} The identification phase involved searching for relevant publications using keyword-based queries. The initial search string, \textit{TITLE-ABS-KEY(("Change Management" OR "Organizational Change") AND "model")}, returned 9,712 results. Through trial and error, the final search query, \textit{TITLE-ABS("Change Management model")}, yielded 265 studies. Then, we filtered these publications to include only journal and conference papers in English, within the fields of Business Management, Computer Science, Engineering, and Psychology. This limited the final set in 137 studies.

\revised{The search string was intentionally broad to capture foundational change management frameworks. This decision was motivated by our assumption that AI adoption shares core characteristics with other large-scale organizational change efforts, such as resistance to change, making these frameworks relevant to the context of our study. Thus, we assessed non-AI change management models for their capacity to address challenges identified in our previous study~\cite{tavantzis2024HumanCenteredAI}, on the basis that many of these challenges are common across large-scale organizational transformations (see Table~\ref{tab:mapping_results} for the detailed mapping).}

\paragraph{Screening phase} In this sub-step, we screened the studies retrieved from the previous step. To this end, we read all the titles and abstracts of the publications, removing the studies that were not relevant. In some cases, we read the full-text as well. This sub-step resulted in 68 remaining studies.

\paragraph{Eligibility phase} During this sub-step we filtered out the studies retrieved from the screening process by examining the full-texts and applying the following inclusion and exclusion criteria.

The \textbf{inclusion criteria} suggest; \textit{i) studies should include change management models or success factors, ii) studies should not include domain-specific change management models and can be useful for other domains as well, iii) studies should be peer-reviewed publications, either journal or conference papers}.

The \textbf{exclusion criteria} suggest studies are excluded when: \textit{i) they do not propose a new change management model or build upon an existing one, ii) they have a particular orientation that makes the model inapplicable for other domains, iii) they are on-going studies without having a well-defined outcome}.

\paragraph{Included phase} Finally, we applied a snowballing strategy by reviewing the references of the selected studies and identifying works that cited them. This process led to the inclusion of 20 additional publications, one of which was grey literature. As a result, the literature review comprised a total of 43 studies.

\subsubsection{Survey}

Survey constitutes an empirical method to collect data from large populations~\cite{kasunic2005designingSurvey}, making them particularly suitable for the SE field~\cite{ghazi2019surveyResearchInSE}. To minimize bias and ensure methodological rigor, the survey design followed the eight-step framework by Molleri et al.~\cite{molleri2016surveyGuidelines}, which includes: \textit{1. Define Research Objectives, 2. Identify Target Audience and Sampling Frame, 3. Design the Sample Plan, 4. Design the Survey Instrument, 5. Evaluate the Survey Instrument, 6. Collect and Analyze Survey Data, 7. Extract Conclusions, and 8. Document and Report the Survey}.

We identified SE practitioners as the key target audience to share their impressions regarding the proposed BSE-informed framework and the usefulness of particular dimensions. The survey included, predominantly, Likert scale questions and some open-ended questions to capture the impressions of the population. The distribution of the survey lasted four weeks and happened, mainly, through personal contacts in organizations and Prolific—a platform that facilitates collecting high-quality human-derived data \footnotemark[1]. Also, we employed snowball sampling, as personal connections to companies shared the link of the survey with other potential participants. This approach ensured that only people with related background responded to the survey, ensuring input's quality. The detailed survey structure and setting can be found on our Figshare replication package.\footnotemark[2].

The Research Ethics Committee at Aalborg University approved this research activity with reference number 1094922 in November 2025. In this context, all participants were older than 18 years and gave informed consent before participating in the study. Additionally, we notified them regarding their right to withdraw their participation at any point.

\footnotetext[1]{https://www.prolific.com}

\footnotetext[2]{https://doi.org/10.6084/m9.figshare.32051676}

\subsubsection{Early Feasibility Checks}

To preliminary assess the perceived relevance and clarity of the proposed framework, we conducted two expert review workshops. Relying on the output of RQ\textsubscript{3}, these sessions sought to gather experts’ impressions on the practicality of the proposed actions, serving as an initial form of validation. Rather than organizing large sessions, smaller, focused sessions were held to ensure that participants could provide relevant input based on their roles. For instance, steps under the \textit{Leadership} dimension may not apply to software developers, making targeted participation more effective. Participants were selected through personal contacts to ensure the necessary domain expertise.

Each session followed a five-phase structure. In the \textit{Introduction} phase, the research objectives, the BSE-related challenges, and the proposed framework were briefly presented. This was followed by a detailed explanation of the \textit{workshop-specific dimensions}, including their DPs and actionable steps. Then, \textit{Group discussions} took place regarding the applicability of these actions in real-world settings based on participants experiences. In the fourth phase, the \textit{Data Collection} happened through a survey distribution. Finally, the \textit{Closing phase} included a discussion of preliminary findings and an overview of the next steps. Each session lasted approximately one hour.

\subsection{Data Analysis}

The data analysis process involves both a qualitative and a quantitative approach. Specifically, we applied \textit{Thematic Analysis} to identify themes and structure the proposed framework. Then, we performed \textit{Statistical Analysis} to the quantitative data collected through the survey and the workshops to summarize and interpret the findings.

\subsubsection{Thematic Analysis}

We employed \textbf{Thematic analysis} to analyze the qualitative data gathered from semi-structured interviews of prior research, following the method outlined by Braun and Clarke~\cite{braun2006thematicAnalysis}. According to their methodology, thematic analysis comprises six discrete stages; namely, \textit{familiarizing yourself with your data, generating initial codes, searching for themes, reviewing themes, defining and naming themes, and producing the report}. This process enabled the identification, analysis, and patterns reporting across the interview transcripts.

Given the exploratory nature of the first part of the research, we adopted an inductive approach to derive codes and themes directly from participant experiences, without relying on pre-existing models. This decision is aligned with the objective of RQ\textsubscript{2}, aiming to enrich the proposed framework beyond traditional change management models. The goal was to provide a distinct contribution by capturing insights grounded in real-world contexts. Although the analysis was primarily conducted by one researcher, we followed a structured review process. After initial coding and theme development, two authors reviewed the results to refine the themes, ensure clarity and minimize potential interpretation bias, thereby enhancing the reliability of the findings.

\subsubsection{Statistical Analysis}

Complementing qualitative analysis with numerical evidence enhances the understandability and credibility of the findings. To analyze the quantitative data from the survey and workshops, we performed \textbf{descriptive statistics}, such as mean values and standard deviation, to summarize key trends, assess the perceived importance of framework's dimensions, and measure professionals' agreement. We utilized different ways of visualization, like bar charts, to clearly exhibit the results of participants' impressions on the usefulness of the BSE-informed framework.

Finally, we employed the \textbf{Scott-Knott Effect Size Difference (ESD) test}~\cite{jelihovschi2014scottKnott}, a hierarchical clustering algorithm designed to identify statistically distinct homogeneous groups using mean values. Due to its simplicity and accuracy, Scott-Knott has become a widely adopted method in hierarchical clustering analysis~\cite{jelihovschi2014scottKnott}. In the context of this study, the algorithm could support recognizing clusters with dimensions of significant importance, as perceived by SE professionals.

\begin{table*}[h]
    \centering
    \begin{tabularx}{\linewidth}{@{}l X l X@{}}
        \toprule
        \multicolumn{4}{c}{\textbf{Change Management Models}} \\
        \midrule
        \textbf{ID} & \textbf{Reference} & \textbf{ID} & \textbf{Reference} \\
        \midrule
        S01 & ADKAR model \cite{abdallah2016adkarModel}                              & S23 & Leppitt \cite{leppitt2006challengingTheCodeOfChange} \\
        S02 & Armenakis et al. \cite{armenakis2007orgChangeRecipientsBeliefScale}    & S24 & Lewin's model \cite{schein2010organizational} \\
        S03 & Bamford \& Daniel \cite{bamford2005caseStudyOfChangeManagement}        & S25 & Li \& Yodmongkol \cite{li2019ChangeManagementAntibiotic} \\
        S04 & Beckhard \& Harris \cite{beckhard1987organizationalTransitions_ManagingComplexChange} & S26 & McKinsey 7-S Model \cite{kocaoglu2019mcKinsey7sModel} \\
        S05 & Bolboli \& Reiche \cite{bolbori2013businessExcellence}                 & S27 & Mento et al. \cite{mento2002changeManagementProcess} \\
        S06 & Bridges' Transition Model \cite{bridges2003bridgesTransitionModel}     & S28 & Merino-Barbancho \& Fico \cite{merino_barnacho2025understandingChangeManagement} \\
        S07 & Bullock \& Batten \cite{bullock1985reviewAndSynthesisOfOrgChange}      & S29 & Mukheli \& Naidoo \cite{mukheli2023changeManagementInSouthAfrica} \\
        S08 & Burke \& Litwin \cite{burke1992aCausalModelOfOrgChange}                & S30 & Narciso \& Allison \cite{narciso2014overcomingStructuralResistanceinSPI} \\
        S09 & Calvert \cite{calvert2006change_management_model_for_ERP}              & S31 & Owad et al. \cite{owad2023leanSixSigmaAndKotterframework} \\
        S10 & Carnall \cite{carnall2007managingChangeInOrganizations}                & S32 & Phillips \& Klein \cite{phillips2023changeManagementfromTheory} \\
        S11 & Congruence Model \cite{nadler1980congruenceModel}                      & S33 & Pulido \& Taherdoost \cite{pulido2024changeManagementInDigitalTransformation} \\
        S12 & Cummings \& Worley \cite{cummings2013organizationDevelopmentAndChange} & S34 & Rajamanoharan \& Collier \cite{rajamanoharan2006sixSigmaImplementation} \\
        S13 & Errida \& Lotfi \cite{errida2021DescriptiveOrgModel}                   & S35 & Rohman \& Subriadi \cite{rohman2020changeManagementModelforIS} \\
        S14 & GE's Change Acceleration \cite{vonderlinn2009cap}                      & S36 & Spyropoulou et al. \cite{spyropoulou2021businessExcellenceinLargeOrganizations} \\
        S15 & Hayes \cite{hayes2014theTheoryAndPracticeOfChangeManagement}           & S37 & Sulistiyani \& Susanto \cite{sulistiyani2018changeManagementForeGov} \\
        S16 & Hourani et al. \cite{hourani2019proposed7Emodel}                       & S38 & Tampoe \cite{tampoe1990drivingOrgChange} \\
        S17 & Jick \cite{jick1993implementingChange}                                 & S39 & Toherhi \& Recardo \cite{toherhi2012nineBlunders} \\
        S18 & Judson \cite{judson1991changingBehaviorInOrganizations}                & S40 & Victor \& Franckeiss \cite{victor2002theFiveDimensionsOfChange} \\
        S19 & Kanter et al. \cite{kanter1992challengesOfOrgChange}                   & S41 & Wan et al. \cite{wan2020UNchangeManagement} \\
        S20 & Kirmizi \& Kocaoglu \cite{kirmizi2022ERP}                              & S42 & Wheelan-Berry \& Somerville \cite{whelan-berry2010linkingChangeDrivers} \\
        S21 & Knoster \cite{knoster2000frameworkForSystemsChange}                    & S43 & Worley \& Mohrman \cite{worley2014isChangeManagementObsolete} \\
        S22 & Kotter's model \cite{kotter_leading_change_2012}                       &     & \\
        \bottomrule
    \end{tabularx}
    \caption{List of organizational change management literature.}
    \label{tab:changeManagementLiterature}
\end{table*}

\begin{table*}[h]
    \centering
    \begin{tabularx}{\linewidth}{@{}l X X@{}}
        \toprule
        \textbf{Domain} & \textbf{Processual} & \textbf{Descriptive} \\
        \midrule
        \textbf{Business Excellence}
            & Bolboli \& Reiche \cite{bolbori2013businessExcellence}
            & Spyropoulou et al. \cite{spyropoulou2021businessExcellenceinLargeOrganizations} \newline
              Rajamanoharan \& Collier \cite{rajamanoharan2006sixSigmaImplementation} \\
        \cmidrule(l){1-3}
        \textbf{Digital Transformation}
            & Kirmizi \& Kocaoglu \cite{kirmizi2022ERP} \newline
              Pulido \& Taherdoost \cite{pulido2024changeManagementInDigitalTransformation}
            & Calvert \cite{calvert2006change_management_model_for_ERP} \\
        \cmidrule(l){1-3}
        \textbf{Education}
            & ---
            & Rohman \& Subriadi \cite{rohman2020changeManagementModelforIS} \\
        \cmidrule(l){1-3}
        \textbf{Generic}
            & ADKAR model \cite{abdallah2016adkarModel} \newline
              Bridges' Transition Model \cite{bridges2003bridgesTransitionModel} \newline
              Bullock \& Batten \cite{bullock1985reviewAndSynthesisOfOrgChange} \newline
              Hayes \cite{hayes2014theTheoryAndPracticeOfChangeManagement} \newline
              Jick \cite{jick1993implementingChange} \newline
              Judson \cite{judson1991changingBehaviorInOrganizations} \newline
              Kanter et al. \cite{kanter1992challengesOfOrgChange} \newline
              Kotter's model \cite{kotter_leading_change_2012} \newline
              Lewin's model \cite{schein2010organizational} \newline
              Mento et al. \cite{mento2002changeManagementProcess} \newline
              Tampoe \cite{tampoe1990drivingOrgChange} \newline
              Toherhi \& Recardo \cite{toherhi2012nineBlunders} \newline
              Wheelan-Berry \& Somerville \cite{whelan-berry2010linkingChangeDrivers} \newline
              Victor \& Franckeiss \cite{victor2002theFiveDimensionsOfChange} \newline
              Worley \& Mohrman \cite{worley2014isChangeManagementObsolete}
            & Armenakis et al. \cite{armenakis2007orgChangeRecipientsBeliefScale} \newline
              Beckhard \& Harris \cite{beckhard1987organizationalTransitions_ManagingComplexChange} \newline
              Burke \& Litwin \cite{burke1992aCausalModelOfOrgChange} \newline
              Carnall \cite{carnall2007managingChangeInOrganizations} \newline
              Congruence Model \cite{nadler1980congruenceModel} \newline
              Cummings \& Worley \cite{cummings2013organizationDevelopmentAndChange} \newline
              Errida \& Lotfi \cite{errida2021DescriptiveOrgModel} \newline
              GE's change acceleration \cite{vonderlinn2009cap} \newline
              Knoster \cite{knoster2000frameworkForSystemsChange} \newline
              Leppitt \cite{leppitt2006challengingTheCodeOfChange} \newline
              McKinsey 7-S Model \cite{kocaoglu2019mcKinsey7sModel} \newline
              Phillips \& Klein \cite{phillips2023changeManagementfromTheory} \\
        \cmidrule(l){1-3}
        \textbf{Healthcare}
            & Li \& Yodmongkol \cite{li2019ChangeManagementAntibiotic} \newline
              Owad et al. \cite{owad2023leanSixSigmaAndKotterframework}
            & Bamford \& Daniel \cite{bamford2005caseStudyOfChangeManagement} \newline
              Merino-Barbancho \& Fico \cite{merino_barnacho2025understandingChangeManagement} \\
        \cmidrule(l){1-3}
        \textbf{Public Sector}
            & Sulistiyani \& Susanto \cite{sulistiyani2018changeManagementForeGov}
            & Mukheli \& Naidoo \cite{mukheli2023changeManagementInSouthAfrica} \newline
              Wan et al. \cite{wan2020UNchangeManagement} \\
        \cmidrule(l){1-3}
        \textbf{Software Engineering}
            & Narciso \& Allison \cite{narciso2014overcomingStructuralResistanceinSPI}
            & Hourani et al. \cite{hourani2019proposed7Emodel} \\
        \bottomrule
    \end{tabularx}
    \caption{Change management models taxonomy by domain and approach.}
    \label{tab:changeManagementModelsTaxonomy}
\end{table*}

\section{Findings}
\label{sec4}

This section presents the results derived from a multi-stage analysis addressing RQ\textsubscript{1}, RQ\textsubscript{2}, RQ\textsubscript{3}, and RQ\textsubscript{4}. It includes findings from the literature review, the thematic analysis, and the systematic development of actionable steps to supplement each identified DP. Then, it discusses the insights of the survey and the expert review workshops.

\subsection{RQ\textsubscript{1} — Literature Review Results}

The PRISMA statement methodology resulted in the inclusion of 43 publications in the literature review (See Figure \ref{fig:prisma_statement}).  Table \ref{tab:changeManagementLiterature} exhibits alphabetically the included change management models, together with their unique id numbers. This section is devoted to evaluate whether the existing change management literature addresses the identified BSE-related challenges.

\subsubsection{Publications Profiles}

\begin{figure*}[h]
    \centerline{\includegraphics[scale=0.55]{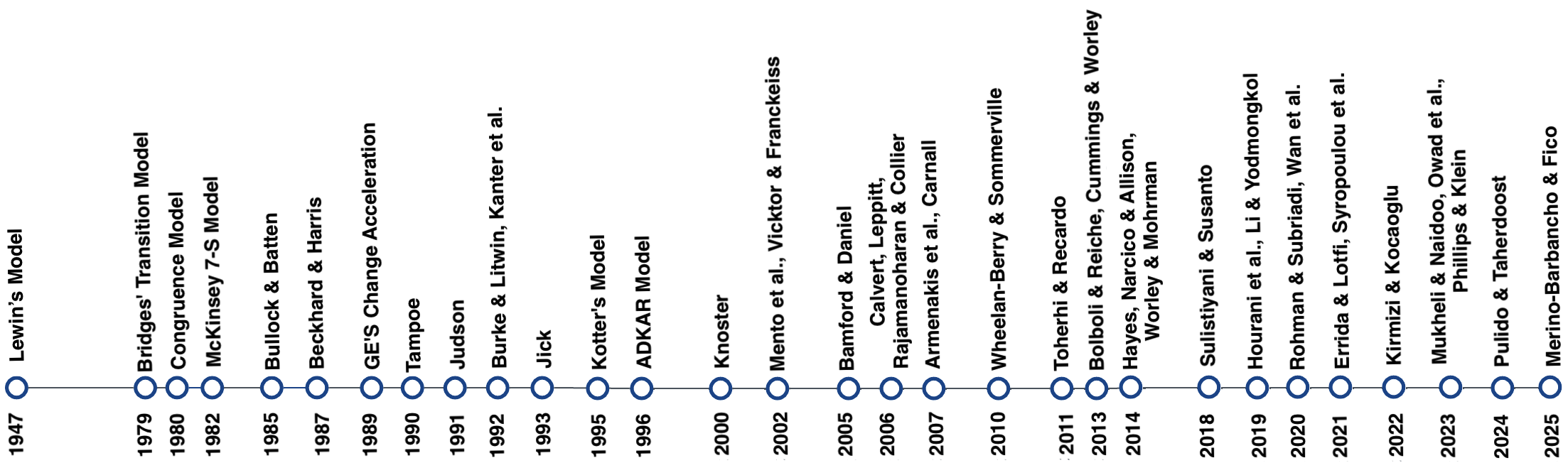}}
    \caption{Timeline of Change Management Models}
    \label{fig:studies_timeline}
\end{figure*}

The publication dates of the selected models span from 1947 to 2025. Although thirty of the forty-three models were introduced after the year 2000, some of the most influential frameworks in existing literature, such as those by Kotter's and Lewin's models, originated prior to 2000. Figure \ref{fig:studies_timeline} illustrates the chronological distribution of the reviewed change management models, indicating key trends and offering valuable insights. Interestingly, the models are fairly evenly distributed across the timeline, reflecting a consistent interest in exploring and comprehending change management processes since the field's early years.

For the purposes of the study we employed a dual taxonomy approach to classify the selected change management models (see Table \ref{tab:changeManagementModelsTaxonomy}).

Consistent with prior research \cite{parry2024EmpiricalOrgChange}, the first classification criterion distinguishes models as either \textbf{processual}, which delineate a series of sequential steps, and \textbf{descriptive} models, which emphasize critical success factors for change. Processual models typically offer a structured step-by-step process, often covering stages such as initiation and planning, while descriptive models highlight key influential elements, such as leadership and communication, that affect change outcomes.

The second classification criterion pertains to \textbf{contextual applicability}, differentiating between generic models designed for uniform use across industries and domain-specific models developed for particular sectors, such as healthcare or the public sector. The identified domains encompass: \textit{Business Excellence, Digital Transformation, Education, Generic, Healthcare, Public Sector, and Software Engineering}.

\subsubsection{Comparison with identified BSE-related challenges}

The objective of RQ\textsubscript{1} is to start forming the human-centric framework,  assessing how traditional change management literature can address the identified AI-risen hurdles. The mapping strategy relied on two underlying rationales: i) \textit{steps or factors of the included models should address a potential hardship and} ii) \textit{variations of the same step should be grouped to enhance readability}. For example, the step \textit{"Create a sense of urgency"} appeared with different phrasings, such as \textit{"Create a burning platform"} \cite{toherhi2012nineBlunders}. To avoid such duplicates and redundancy, these variations were grouped under the umbrella step \textit{"Create a sense of urgency"}. As for the strategy we followed, the first author conducted the mapping process, while the remaining authors reviewed the outcomes and proposed refinements where needed.

Table \ref{tab:mapping_results} (see Appendices) presents the results of the mapping procedure. As can be observed, combining existing organizational change models can mitigate the substantial majority of the identified BSE-related challenges. Specifically, the mapping process addressed all but three of the BSE-related sub-challenges, namely: \textit{"Considerate use of external AI with company’s data", "Inter-generational Collaboration", and "Lack of skills due to AI"}.

For the rest challenges, there is at least one step that can mitigate the negative consequences that they could cause. In particular:

\begin{itemize}
    \item \textbf{Change Management Strategy}: This challenge highlights the pivotal role of \textbf{Communication} within AI adoption both internally and externally to the organization. Established models emphasize the necessity of transparency and the definition of well-defined communication channels~\cite{rohman2020changeManagementModelforIS}, along with a comprehensive stakeholder engagement strategy~\cite{leppitt2006challengingTheCodeOfChange, kocaoglu2019mcKinsey7sModel, sulistiyani2018changeManagementForeGov, hourani2019proposed7Emodel, worley2014isChangeManagementObsolete}. These initiatives can foster the acceptance and understanding of organization's AI vision~\cite{errida2021DescriptiveOrgModel, kotter_leading_change_2012, abdallah2016adkarModel, bamford2005caseStudyOfChangeManagement,
    li2019ChangeManagementAntibiotic, bolbori2013businessExcellence,
    mento2002changeManagementProcess, mukheli2023changeManagementInSouthAfrica,
    calvert2006change_management_model_for_ERP,
    phillips2023changeManagementfromTheory, pulido2024changeManagementInDigitalTransformation, rohman2020changeManagementModelforIS, spyropoulou2021businessExcellenceinLargeOrganizations,
    tampoe1990drivingOrgChange, jick1993implementingChange, judson1991changingBehaviorInOrganizations, victor2002theFiveDimensionsOfChange, kanter1992challengesOfOrgChange, wan2020UNchangeManagement, kirmizi2022ERP, worley2014isChangeManagementObsolete}.

    \item \textbf{Ethical Concerns}: The outcome of the mapping process revealed that prior models cannot fully address the ethical considerations that AI raises. However, by encouraging experimentation through pilot projects~\cite{pulido2024changeManagementInDigitalTransformation, vonderlinn2009cap}, companies can determine the balance to mitigate the risks associated with sensitive data.
    
    \item \textbf{Organizational Readiness for AI Adoption}: Organizational readiness for AI integration is shaped by six distinct sub-challenges. Considering the domain knowledge, such as regulations~\cite{leppitt2006challengingTheCodeOfChange, burke1992aCausalModelOfOrgChange}, has the potential to enhance institutions' preparedness. Furthermore, cultivating a shared sense of urgency~\cite{kotter_leading_change_2012, leppitt2006challengingTheCodeOfChange, bolbori2013businessExcellence, owad2023leanSixSigmaAndKotterframework, pulido2024changeManagementInDigitalTransformation, jick1993implementingChange, toherhi2012nineBlunders, kanter1992challengesOfOrgChange, kirmizi2022ERP} regarding AI adoption and creating a shared need~\cite{errida2021DescriptiveOrgModel, abdallah2016adkarModel, armenakis2007orgChangeRecipientsBeliefScale, vonderlinn2009cap, kanter1992challengesOfOrgChange, worley2014isChangeManagementObsolete} can help reduce the diverse perspectives across organizational levels. However, navigating the political dynamics inherent within AI transition necessitates securing political support~\cite{mento2002changeManagementProcess, cummings2013organizationDevelopmentAndChange, jick1993implementingChange, kanter1992challengesOfOrgChange}, actively engaging diverse perspectives and considerations~\cite{phillips2023changeManagementfromTheory}, and identifying the right timing for implementation~\cite{wan2020UNchangeManagement}. Lastly, it is important that change teams are interdisciplinary~\cite{calvert2006change_management_model_for_ERP, tampoe1990drivingOrgChange}, as the multi-faceted nature of AI demands expertise from multiple disciplines, while solutions should also be tailored to each organization's context~\cite{errida2021DescriptiveOrgModel, leppitt2006challengingTheCodeOfChange, worley2014isChangeManagementObsolete}.
    
    \item \textbf{Resistance to Change}: The output of the literature review suggests that training~\cite{errida2021DescriptiveOrgModel, kotter_leading_change_2012, abdallah2016adkarModel, kocaoglu2019mcKinsey7sModel, mukheli2023changeManagementInSouthAfrica, narciso2014overcomingStructuralResistanceinSPI, calvert2006change_management_model_for_ERP, phillips2023changeManagementfromTheory, pulido2024changeManagementInDigitalTransformation, rajamanoharan2006sixSigmaImplementation, kanter1992challengesOfOrgChange, wan2020UNchangeManagement, worley2014isChangeManagementObsolete} and coaching~\cite{errida2021DescriptiveOrgModel, spyropoulou2021businessExcellenceinLargeOrganizations} the employees are vital actions to minimize feelings of resistance. Constant communication~\cite{mento2002changeManagementProcess, jick1993implementingChange, kanter1992challengesOfOrgChange} seems crucial to understand the complementary role of AI technologies and lessen the concerns of job insecurity. Finally, encouraging innovation~\cite{pulido2024changeManagementInDigitalTransformation, vonderlinn2009cap} and providing incentives and rewards~\cite{errida2021DescriptiveOrgModel, abdallah2016adkarModel, kanter1992challengesOfOrgChange, knoster2000frameworkForSystemsChange} can motivate and increase experimentation with AI technologies, in order to establish a deeper understanding of their capacity.
    
    \item \textbf{Skills in Era of AI}: Initiatives around practitioners' education are the main components to address the challenges of this particular area. Continuous Learning~\cite{rajamanoharan2006sixSigmaImplementation, vonderlinn2009cap, wan2020UNchangeManagement, worley2014isChangeManagementObsolete}, AI Training~\cite{errida2021DescriptiveOrgModel, kotter_leading_change_2012, abdallah2016adkarModel, kocaoglu2019mcKinsey7sModel, mukheli2023changeManagementInSouthAfrica, narciso2014overcomingStructuralResistanceinSPI, calvert2006change_management_model_for_ERP, phillips2023changeManagementfromTheory, pulido2024changeManagementInDigitalTransformation, rajamanoharan2006sixSigmaImplementation, kanter1992challengesOfOrgChange, wan2020UNchangeManagement, worley2014isChangeManagementObsolete} and Coaching~\cite{errida2021DescriptiveOrgModel, spyropoulou2021businessExcellenceinLargeOrganizations} constitute key efforts to keep up with the constant and rapid AI progress and demands. Nevertheless, concerns regarding the potential absence of fundamental skills due to over-reliance on AI continue to be unaddressed.
    
    \item \textbf{Strategic Adoption of AI}: A systematic approach to AI integration is crucial to ensure a smooth transition. In this regard, ongoing assessment~\cite{errida2021DescriptiveOrgModel, li2019ChangeManagementAntibiotic, bolbori2013businessExcellence, mento2002changeManagementProcess, bullock1985reviewAndSynthesisOfOrgChange, phillips2023changeManagementfromTheory, pulido2024changeManagementInDigitalTransformation, hayes2014theTheoryAndPracticeOfChangeManagement, kanter1992challengesOfOrgChange} and measurement of the incorporation process, using predefined KPIs~\cite{bolbori2013businessExcellence, merino_barnacho2025understandingChangeManagement, spyropoulou2021businessExcellenceinLargeOrganizations}, where applicable, can facilitate tracking the progress towards transition goals. For example, organizations can evaluate their efforts in achieving an optimal balance between automation and human involvement~\cite{errida2021DescriptiveOrgModel, li2019ChangeManagementAntibiotic, bolbori2013businessExcellence, mento2002changeManagementProcess, bullock1985reviewAndSynthesisOfOrgChange, phillips2023changeManagementfromTheory, pulido2024changeManagementInDigitalTransformation, hayes2014theTheoryAndPracticeOfChangeManagement, kanter1992challengesOfOrgChange}, as well as in leveraging internal versus external AI solutions~\cite{bolbori2013businessExcellence, merino_barnacho2025understandingChangeManagement, spyropoulou2021businessExcellenceinLargeOrganizations}. Training~\cite{errida2021DescriptiveOrgModel, kotter_leading_change_2012, abdallah2016adkarModel, kocaoglu2019mcKinsey7sModel, mukheli2023changeManagementInSouthAfrica, narciso2014overcomingStructuralResistanceinSPI, calvert2006change_management_model_for_ERP, phillips2023changeManagementfromTheory, pulido2024changeManagementInDigitalTransformation, rajamanoharan2006sixSigmaImplementation, kanter1992challengesOfOrgChange, wan2020UNchangeManagement, worley2014isChangeManagementObsolete} and coaching~\cite{errida2021DescriptiveOrgModel, spyropoulou2021businessExcellenceinLargeOrganizations} possess a vital role in supporting understanding of when and how AI applications can be more appropriate and impactful. Finally, establishing short-term objectives~\cite{errida2021DescriptiveOrgModel, kotter_leading_change_2012, leppitt2006challengingTheCodeOfChange, mento2002changeManagementProcess, narciso2014overcomingStructuralResistanceinSPI, phillips2023changeManagementfromTheory, pulido2024changeManagementInDigitalTransformation} and performing gap analyses between the current and desired states~\cite{narciso2014overcomingStructuralResistanceinSPI, spyropoulou2021businessExcellenceinLargeOrganizations, sulistiyani2018changeManagementForeGov} can help translate the long-term vision of AI adoption into a clear, actionable, and purpose-driven strategy.
\end{itemize}

The majority of the traditional, well-established change management models encompass steps related to anchoring change and integrating lessons learned during the change process. The identified BSE-related challenges \cite{tavantzis2024HumanCenteredAI}, based on which we structure the framework, do not discuss such concerns. The interviewees, whose responses shaped these challenges, represent organizations at the beginning of their AI integration journey. Therefore, their perspectives do not yet extend to the later phases of AI transformation, where long-term sustainability and reinforcement of change become critical.

Accordingly, the proposed BSE-informed framework maintains an early-stage focus within the broader process of AI adoption. This focus does not suggest that sustaining change is unimportant; rather, it emphasizes the necessity of establishing a robust foundation before reaching that stage. A successful initial integration phase can enable smoother progression and reinforce organizational efforts in subsequent stages of transformation.

\subsubsection{Summary of Literature Review — RQ\textsubscript{1} output}
\label{sec4-literature-review}

The summary of the literature review constitutes the core of the proposed BSE-informed framework, fulfilling the objective of RQ\textsubscript{1}. This foundation was formed as follows into dimensions and DPs.

\paragraph{\textbf{Agile Culture} — emphasizes adaptability and experimentation to navigate uncertainty and enable effective AI integration}
\begin{itemize}
    \item \textbf{Encourage Experimentation through pilot projects}: \revised{In early stages, unawareness and misconceptions are common; therefore experimentation can facilitate understanding of the real capabilities.}
    \item \textbf{Set short-term wins}: \revised{Trust in early phases is fragile, therefore bounded pilots with visible benefits reduce resistance and create momentum.}
\end{itemize}

\paragraph{\textbf{AI Strategy Design} — aims to define a coherent vision, aligning AI initiatives with organizational objectives, and considering contextual constraints}
\begin{itemize}
    \item \textbf{Be aware of parameters within the external environment, such as regulations}: \revised{External parameters, such as regulations, impose constraints; therefore systematic monitoring of these factors enhance strategic responsiveness and institutional compliance.}
    \item \textbf{Design tailored solutions, considering the unique nature of the domain}: \revised{Each organizational context has unique requirements, for example regarding data privacy; therefore each tool has to conform to the domain-specific demands.}
\end{itemize}

\paragraph{\textbf{AI Strategy Evaluation} — ensures continuous monitoring and assessment of AI integration progress}
\begin{itemize}
    \item \textbf{Continuously evaluate the integration process}: \revised{AI progress is continuous and rapid; therefore strategy evaluation and adaptation are necessary.}
    \item \textbf{Define and monitor KPIs for integration goals}: \revised{Measurable indicators translate abstract objectives, such as the balance between automation and human presence, into observable outcomes; therefore establishing KPIs enables data-driven evaluation and adaptive management.}
\end{itemize}

\paragraph{\textbf{AI Up-skilling} — stresses the necessary initiatives to prepare employees for AI-embedded solutions}
\begin{itemize}
    \item \textbf{Provide AI Training for employees}: \revised{The rise of AI introduce new skills for practitioners, like prompt and context engineering; therefore they need to enrich their skill set to fully-leverage the potential of emerging tools.}
    \item \textbf{Provide AI Mentoring and Coaching sessions}: \revised{Effective AI adoption depends not only on technical skills, but also on understanding, confidence, and ethical awareness; therefore mentoring and coaching sessions are important to build these qualities.}
    \item \textbf{Institutionalize continuous learning initiatives}: \revised{AI evolves rapidly and new tools continually emerge; therefore establishing continuous learning mechanisms is pivotal to keep up with AI rapid evolution and sustain competence.}
\end{itemize}

\paragraph{\textbf{Communication} — aims to clear and continuous information flow between stakeholders of the AI integration process}
\begin{itemize}
    \item \textbf{Structure a clear communication strategy with all stakeholders}: \revised{Transparency and consistency in communication could mitigate uncertainty; therefore a well-defined communication strategy is vital to foster trust and collaboration.}
    \item \textbf{Establish communication channels with all stakeholders}: \revised{Transparent stakeholder engagement relies on systematic communication; therefore introducing both formal and informal channels is vital to enhance collaboration quality and organizational cohesion.}
    \item \textbf{Engage with all stakeholders}: \revised{Effective AI integration requires stakeholders being onboard; therefore it is essential to ensure continuous engagement with them.}
    \item \textbf{Listen to employees' concerns}: \revised{Constant AI development brings uncertainty and anxiety in employees regarding their future; therefore listening and addressing their concerns fosters trust and psychological safety during AI integration.}
\end{itemize}

\paragraph{\textbf{Leadership} — provides direction, alignment, and motivation for AI transformation}
\begin{itemize}
    \item \textbf{Establish a clear and shared need}: \revised{Convincing stakeholders about the benefits of AI incorporation enables broad acceptance and embracement of the efforts; therefore cultivating this need will enhance trust in AI adoption endeavors.}
    \item \textbf{Establish a sense of urgency}: \revised{The lack of a clear definition of the risks and benefits of timely transformation leads to delays and inactive periods; therefore cultivating a sense of urgency accelerates decision-making and impels commitment.}
    \item \textbf{Form a cross-functional AI team}: \revised{AI constitutes a multifaceted technology involving technical, ethical and legal aspects; therefore it is crucial to form interdisciplinary teams to be responsible for the AI strategy design.}
    \item \textbf{Encourage AI adoption and experimentation by rewarding desired behaviors and achievements}: \revised{Practitioners are more likely to embrace AI experimentation when their efforts and ideas are visibly valued; therefore establishing reward mechanisms could reinforce commitment and reduce resistance.}
\end{itemize}

\paragraph{\textbf{Organizational Dynamics} — deals with individuals (internal politics) or situations (timing, resistance to change) that could enable or hinder AI integration efforts}
\begin{itemize}
    \item \textbf{Be aware of the organizational situation, which might indicate internal pressures or resistance}: \revised{Internal dynamics, such as resistance to change, may lead to significant delays and cancellations of projects; therefore diagnosing cases of tension and resistance is essential to design targeted interventions.}
    \item \textbf{Secure the political support of key players}: \revised{Organizational change success is influenced by stakeholders' interests; therefore obtaining political support of key individuals ensures commitment and sustains the defined strategy.}
\end{itemize}

\begin{takehomebox}
    \textbf{RQ\textsubscript{1} take-away:} AI integration appears to follow patterns seen in previous change initiatives. Particularly, combining traditional change management models addresses most identified BSE-related challenges, leaving only a few unaddressed. This provides a solid foundation for the proposed BSE-informed, early-stage framework.
\end{takehomebox}

\subsection{RQ\textsubscript{2} — Thematic Analysis Results}

To re-analyze the qualitative data collected during previous research study \cite{tavantzis2024HumanCenteredAI}, we employed thematic analysis. The nature of the former work was exploratory, which was reflected in the involved organizations and participants. Specifically, institutions of different domains (N=4) and professionals (N=10) that hold various roles participated in the semi-structured interviews. More detailed demographic information is provided in the study. The detailed protocol and setting of the interviews can be found on GitHub.\footnotemark[3] 

\footnotetext[3]{https://github.com/tavantzish/Appendix-Unpacking\_AI\_integration}

Despite the distinguished orientation between the previous and the present study, interviewees shared and elaborated on lessons learned and success factors that they could recognize throughout the AI integration endeavors of their organizations. Concentrating, predominantly, on these parts and interpreting them through the lens of this study, we generated the initial codes. Then, the codes were grouped and formed the themes to avoid introducing duplicates. Given that the outcome aims to be a set of DPs, the themes were expressed as DPs for organizations to take their first steps towards AI integration. Finally, we grouped the themes into broader dimensions, which will constitute the higher layer of the proposed framework.

\subsubsection{Identified themes grouped by dimensions}

Following the aforementioned systematic way for the analysis of the interview transcripts, we identified the following dimensions and themes. Several design principles identified through thematic analysis align with those derived from the literature review (Section~\ref{sec4-literature-review}). Where this is the case, only the supporting practitioner quote is provided, and the reader is referred to Section~\ref{sec4-literature-review} for the full rationale.

\paragraph{\textbf{AI Strategy Design}} AI technologies bring new opportunities and capabilities for organizations. Consequently, there is a growing desire to integrate AI. This dimension refers to the DPs that are required to ensure a robust strategy for AI adoption.

\begin{enumerate}[label=\roman*]
    \item \textit{Determine projects' prioritization strategy}: \revised{The fast pace of AI progress leads to multiple ideas for implementation; therefore prioritization is necessary to handle the increasing number of projects.}
    
    \begin{quote}
    \textbf{P03:} \textit{I try to be organized, I try to use planning tools, and prioritize my ideas. I need to be clear on the ones that I'm focusing on right now, the ones that I'm expecting to move forward, and the ones that can wait.}
    \end{quote}
    
    \item \textit{Develop AI strategy in line with the latest AI advances}: \revised{AI advancements usually overlap with designing integration strategies; therefore aligning them is vital to avoid unnecessary adjustments and delays.}

    \begin{quote}
    \textbf{P06:} \textit{From the developers’ perspective, it is important to start with the latest available model, instead of trying and testing the older models and then adopting the newer models. We should stick to the plan without changing the timelines. So, it's about better planning.}
    \end{quote}
    
    \item \textit{Consider user-friendliness during AI strategy design}: \revised{Perceived ease-of-use is a baseline requirement for adoption; therefore delineating a user-friendly solution should be top-priority of the design phase}.

    \begin{quote}
    \textbf{P02:} \textit{The one is to make it easy to try… So, we build a really easy co-pilot studio chatbot that works with data from the web.}
    \end{quote}
    
    \item \textit{Recognize a business need, before starting an AI integration project}: \revised{AI does not improve productivity and efficiency in all contexts; therefore it is critical to identify an existing business need that AI can address, rather than start adopting AI excessively.}

    \begin{quote}
    \textbf{P01:} \textit{What we should focus on is being business-driven, to make sure that there is a business need. I think when we look at projects now, sometimes we have to take a step back and ask do we really need AI?}
    \end{quote}
    
\end{enumerate}

\paragraph{\textbf{Communication}} The rapid AI advancements lead to constant changes and misconception among individuals. Hence, this dimension underscores the pivotal role of communication for AI integration.
        \begin{enumerate}[label=\roman*]
            \item \textit{Communicate the change constantly}: \revised{Getting people on-board might be challenging and hinder the adoption progress; therefore continuous communication with all stakeholders, internally and externally to the organization is necessary.}

            \begin{quote}
            \textbf{P01:} \textit{I think the key lesson that we have learned is to over-communicate. AI and what an AI transformation means to a large organization is so complex and it's not enough to say it once. You have to over-communicate, you have to ingrain into people's minds what the plan is, what the strategy is, and keep spreading the word.}
            \end{quote}
            
            \item \textit{Communicate the need for AI}: \revised{Elucidating the motivation that turns AI transition into a pivotal shift for organizations can get stakeholders contribute to the change effort with their resources; therefore it is imperative to invest on inspiring and convincing all stakeholders.}

            \begin{quote}
            \textbf{P09:} \textit{Then you need to ensure that people see it the same way, because it doesn't matter if I think it's a good idea to implement AI for a specific area and I can't get my stakeholders to actually agree on this. Communication is a key.}
            \end{quote}
            
        \end{enumerate}

\paragraph{\textbf{Collaboration}} Given the interdisciplinary nature of AI, collaboration is a critical element for smooth AI adoption. Hence, this dimension aims to ensure collaborative practices and engagement within organizations.
        \begin{enumerate}[label=\roman*]
            \item \textit{Institutionalize knowledge and experience-sharing practices, such as internal forums and CoPs}: \revised{Sharing knowledge and past experiences can lead to patterns or anti-patterns identification in organization's workflow, which can save resources, like time and money; therefore internal forums and Communities of Practice are initiatives that can aid this purpose.}

            \begin{quote}
            \textbf{P01:} \textit{It's about knowledge-sharing, for everyone to understand the potential of what we're trying to do and how to do it, having internal forums and CoPs.}
            \end{quote}
            
            \item \textit{Form interdisciplinary teams}: See Section \ref{sec4-literature-review} for rationale—DP: \textit{Form a cross-functional AI team}.

            \begin{quote}
            \textbf{P03:} \textit{I had some interns who were actually software developers, and they could teach me a lot. It was like we had sort of a little team. If I could, I would have done it one year earlier, because then they could have helped me all along my learning journey. I think it is crucial to network and communicate with different backgrounds, because then things can happen much easier and faster.}
            \end{quote}
            
        \end{enumerate}
        
\paragraph{\textbf{Governance and Ethics}} AI rise brings ethical challenges for organizations. Therefore, Governance and Ethics are foundational pillars for responsible AI integration — what is allowed and how organizations shall verify it. The following DPs aim to ensure responsible AI governance.
        \begin{enumerate}[label=\roman*]
            \item \textit{Determine clear AI application guidelines}: \revised{AI use brings risks and ethical implications. For example, practitioners could query models using sensitive data; therefore it is advisable to establish internal guidelines regarding the utilization of AI solutions.}

            \begin{quote}
                \textbf{P04:} \textit{We have a general framework for using external solutions. Some colleagues used chatGPT, especially when we didn't have yet available the internal tool. But then, they were very mindful about what to ask, to remove proprietary data and keep it generic, so that they can still have their answer and convert it back to something with proprietary information in it.}
            \end{quote}
            
            \item \textit{Monitor work quality to examine cases of over-reliance on AI}: \revised{Over-reliance on AI may impact negatively critical thinking and creativity of individuals; therefore it is pivotal to introduce regular assessments and oversight to ensure that AI complements individuals' and team's efforts.}

            \begin{quote}
                \textbf{P03:} \textit{I guess it would be like if somebody discovers this is not a good output or this is incorrect. Then that would be an issue, and it should be I guess the management team to check if happening and address it.}
            \end{quote}
            
            \item \textit{Evaluate and criticize the quality of AI outputs}: \revised{AI outputs might encompass biases and not be relevant to particular contexts; therefore assessing AI outputs should be an integral part of the AI incorporation workflow.}

            \begin{quote}
            \textbf{P10:} \textit{I'm thinking that maybe we shouldn't trust AI so much. I mean, in that project I had, people trusted whatever the output that was generated, but it wasn’t always that accurate.}    
            \end{quote}
            
        \end{enumerate}
        
\paragraph{\textbf{Leadership}} Given the multifaceted nature of AI, Leadership initiatives constitute a decisive pillar for seamless AI adoption. The following DPs outline key leadership practices that enable effective, responsible, and purposeful AI integration.
        \begin{enumerate}[label=\roman*]
            \item \textit{Set short-term goals}: See Section \ref{sec4-literature-review} for rationale—DP: \textit{Set short-term goals}.

            \begin{quote}
            \textbf{P05:} \textit{I think that would be difficult (to convince external stakeholders). It would be a long time work and it will be by slowly adding things and achieving aims... And eventually they will build trust.}    
            \end{quote}
            
            \item \textit{Establish a sense of urgency for AI adoption}: See Section \ref{sec4-literature-review} for rationale—DP: \textit{Establish a sense of urgency}.

            \begin{quote}
                \textbf{P07:} \textit{We need to be making these changes now because these organizations will be much more efficient than we will. So, we really need to be able to communicate it.}
            \end{quote}
            
            \item \textit{Define collective ownership}: See Section \ref{sec4-literature-review} for rationale—DP: \textit{Establish a shared need}.

            \begin{quote}
                \textbf{P03:} \textit{I think feeling the ownership makes it easier to implement a change in the organization. People feel that it's something that's coming from us, rather than only from me. It’s something we do together.}
            \end{quote}
            
            \item \textit{Leverage talents for product and process innovation}: \revised{Junior professionals and recent graduates bring fresh perspectives and mindsets; therefore integrating their ideas in strategy design could increase the creativity, innovation and experimentation capacity.}

            \begin{quote}
                \textbf{P04:} \textit{Maybe even chase some students who just finished studying AI and see what they can contribute and say okay, where do you think we can inject AI into our products to make it more efficient and cheaper, without substituting human presence.}
            \end{quote}
            
            \item \textit{Provide reliable internal AI solutions}: \revised{Using external solutions with proprietary data brings risks of data privacy; therefore companies should provide internal AI solutions to enable experimentation.}

            \begin{quote}
                \textbf{P05:} \textit{And then the availability is necessary as well. We have an AI set up within our network. So we can use it for company confidential documents, since a lot of the material we work with we don't want to put it in the cloud and have it over to Google.}
            \end{quote}
            
            \item \textit{Promote collaboration with academia for research on AI}: \revised{Collaborations with academia can lead to innovative, interdisciplinary, empirical research and knowledge sharing; therefore engaging with universities supports responsible and trustworthy AI integration.}

            \begin{quote}
                \textbf{P05:} \textit{I think managers within our department, we have also been active again, doing research on AI, trying to find collaboration with universities.}
            \end{quote}
            
            \item \textit{Institutionalize skill verification activities}: \revised{Increased AI capabilities might allow people to overlook enriching their skill set and knowledge background; therefore skill verification activities should be integral part of AI incorporation.}

            \begin{quote}
                \textbf{P06:} \textit{...being over-reliant on GPT or AI makes you personally inefficient. So, let's say without them, your skills will be zero in the future. I think companies should have some kind of skills verification.}
            \end{quote}
            
            \item \textit{Hire AI experts to facilitate the design of the strategy}: \revised{Inadequate technical expertise can lead to misaligned investments and unforeseen barriers; therefore recruiting experienced AI professionals early in the design phase mitigates risks and enhances organizational readiness.}
            
            \begin{quote}
                \textbf{P08:} \textit{We have a new AI expert in our HR team. So, we need more of these types of profiles who have the know-how. I think it's always important to have these people on-board.}
            \end{quote}
            
        \end{enumerate}
        
\paragraph{\textbf{Organizational Culture}} Shifts in organizational culture appear to be salutary for AI incorporation. As a continuously evolving technology, AI requires a proactive culture, relying on experimentation and multiple iterations. This dimension aims to emphasize this need.
        \begin{enumerate}[label=\roman*]
            \item \textit{Foster a culture of experimentation}: See Section \ref{sec4-literature-review} for rationale—DP: \textit{Encourage Experimentation through pilot projects}.

            \begin{quote}
                \textbf{P07:} \textit{I think to be able to have hands-on experience. Nobody taught us how to use PowerPoint, which we tend to use a lot. We just learn how to do it by experimenting and it's the same with AI. Try it, experiment, see what it can do.}
            \end{quote}
            
            \item \textit{Adopt Agile principles}: \revised{Working in iterations and avoiding rigid, long-term plans is in line with the dynamic nature of AI; therefore incorporating agile principles and an iterative approach is beneficial.}

            \begin{quote}
                \textbf{P01:} \textit{I think the nature of working with AI in our group leads to be more agile, because things are happening so fast both in the AI environment and in our company.}
            \end{quote}
            
            \item \textit{Foster an innovation-driven and proactive culture}: \revised{AI integration demands continuous adaptation and openness to technological change; therefore cultivating a proactive and innovation-driven culture enables organizations to anticipate disruptions and sustain competitive advantage.}

            \begin{quote}
                \textbf{P04:} \textit{Like I said, our customers are not asking for it. It would be good that a product manager would have an idea of what to implement, where to go, where to use it... Now it seems like we're just waiting until an application pops up or we get forced to use it somehow, rather than looking for the opportunities to create even a competitive advantage over our competitors. Generally, a more self-initiative approach.}
            \end{quote}
        \end{enumerate}

\paragraph{\textbf{Up-skilling}} AI technologies introduce new skills, like prompt engineering. This dimension is related to initiatives that companies need to take to ensure that practitioners possess the required skills in the era of AI.
        \begin{enumerate}[label=\roman*]
            \item \textit{Support the AI training of employees}: \revised{AI brings new skills and it is vital to ensure practitioners' robust knowledge foundations; therefore companies should introduce up-skilling initiatives, such as training programs, coaching sessions, and workshops.}

            \begin{quote}
            \textbf{P09:} \textit{The key here is to enlighten people, try to remove the fear. That's one of the things we did during our workshops, where we tried to educate people on what they can do and communicate that we're not trying to replace anybody, we’re trying to make them more efficient and up-skill them.}
            \end{quote}
            
            \item \textit{Calibrate understanding of AI capabilities and limitations}: \revised{Understanding the capabilities of AI tools is essential for individuals to form realistic expectations regarding their performance; therefore cases of over-reliance on AI tools can be reduced.}

            \begin{quote}
            \textbf{P05:} \textit{The more you use the tool, the less over-reliant you should be. I think you should have realistic expectations of AI and know the capabilities. For me, over-reliance is when you expect it to solve a problem, and it doesn't. But if you're trusting it for the 90\% of your work and it solves the 100\%, then it's not over-reliance.}
            \end{quote}
            
            \item \textit{Introduce cross-generational mentoring sessions}: \revised{Constant advancements lead to variations in the educational background of practitioners of different generations; therefore cross-generational mentoring sessions could increase the productivity and quality of collaboration between junior and more experienced individuals.}

            \begin{quote}
            \textbf{P08:} \textit{I think there are big differences in how people are being educated now vs 10 years ago vs 20 years ago. So, we need to be prepared that those entering the workplace are going to have a completely different expectations. And creating these links, maybe with cross-generational mentoring, between the different parts of our workforce, I think, will be very important.}
            \end{quote}
            
        \end{enumerate}

Table \ref{tab:thematic_analysis_outcome} exhibits the DPs that emerged from the thematic analysis, grouped by potential dimensions of the human-centric framework. Moreover, the table encompasses a column \textit{"Frequency"}, showing the number of participants that shared that particular DP. The frequency values can act as an indication of each DP's relative importance. Higher frequency values are indicative of recommendations that may have already been adopted by different organizations. In contrast, lower frequency values may signify insights that are more domain-specific.

\begin{table*}[h!]
    \centering
    \begin{tabularx}{\linewidth}{@{}l X c@{}}
        \toprule
        \textbf{Dimension} & \textbf{Theme — Design Principle} & \textbf{Frequency} \\
        \midrule
        \multirow{4}{*}{\textbf{AI Strategy Design}}
            & Determine projects' prioritization strategy                  & 2 \\
            & Develop AI strategy in line with the latest AI advances      & 1 \\
            & Consider user-friendliness during AI strategy design         & 1 \\
            & Recognize a business need before starting an AI integration project & 7 \\
        \cmidrule(l){1-3}
        \multirow{2}{*}{\textbf{Communication}}
            & Communicate the change constantly                            & 7 \\
            & Communicate the need for AI                                  & 1 \\
        \cmidrule(l){1-3}
        \multirow{2}{*}{\textbf{Collaboration}}
            & Institutionalize knowledge and experience-sharing practices  & 4 \\
            & Form interdisciplinary teams                                 & 2 \\
        \cmidrule(l){1-3}
        \multirow{3}{*}{\textbf{Governance \& Ethics}}
            & Determine clear AI application guidelines                    & 7 \\
            & Monitor work quality to examine cases of over-reliance on AI & 1 \\
            & Evaluate and criticize the quality of AI outputs             & 2 \\
        \cmidrule(l){1-3}
        \multirow{8}{*}{\textbf{Leadership}}
            & Set short-term wins                                          & 2 \\
            & Establish a sense of urgency for AI adoption                 & 2 \\
            & Define collective ownership                                  & 2 \\
            & Leverage talents for product and process innovation          & 1 \\
            & Provide reliable internal AI solutions                       & 1 \\
            & Promote collaboration with academia for research on AI       & 1 \\
            & Institutionalize skill verification activities               & 1 \\
            & Hire AI experts to facilitate the design of the strategy     & 1 \\
        \cmidrule(l){1-3}
        \multirow{3}{*}{\textbf{Organizational Culture}}
            & Foster a culture of experimentation                          & 6 \\
            & Adopt Agile principles                                       & 1 \\
            & Foster an innovation-driven and proactive culture            & 2 \\
        \cmidrule(l){1-3}
        \multirow{3}{*}{\textbf{Up-skilling}}
            & Support the AI training of employees                         & 9 \\
            & Calibrate understanding of AI capabilities and limitations & 1 \\
            & Introduce cross-generational mentoring sessions              & 1 \\
        \bottomrule
    \end{tabularx}
    \caption{Thematic analysis outcome: proposed design principles grouped by dimension, with frequency of mention across participants.}
    \label{tab:thematic_analysis_outcome}
\end{table*}

During RQ\textsubscript{1}, existing change management models did not address three identified sub-challenges. Table \ref{tab:thematic_analysis_results_mapping} demonstrates the mapping between the DPs that emerged from thematic analysis and these hurdles.

\begin{table*}[h!]
    \centering
    \begin{tabularx}{\linewidth}{@{}l l X@{}}
        \toprule
        \textbf{BSE-related sub-challenge} & \textbf{Dimension} & \textbf{Design Principle} \\
        \midrule
        \multirow{2}{*}{\textbf{\makecell[l]{Considerate use of external \\ AI with company's data}}}
            & Governance \& Ethics & Determine clear AI application guidelines \\
            & Leadership           & Provide reliable internal AI solutions \\
        \cmidrule(l){1-3}
        \textbf{Inter-generational collaboration}
            & Up-skilling           & Introduce cross-generational mentoring sessions \\
        \cmidrule(l){1-3}
        \multirow{2}{*}{\textbf{Lack of skills due to AI}}
            & Governance \& Ethics & Monitor work quality to examine cases of over-reliance on AI \\
            & Leadership           & Institutionalize skill verification activities \\
        \bottomrule
    \end{tabularx}
    \caption{Mapping between the unaddressed BSE-related sub-challenges and the design principles that emerged from the thematic analysis.}
    \label{tab:thematic_analysis_results_mapping}
\end{table*}

\subsubsection{Summary of Thematic Analysis — RQ\textsubscript{2} output}

In addition to addressing the residual BSE-related sub-challenges, the thematic analysis resulted in the identification of twenty-one more themes that have the potential to enrich the core of the proposed framework. A comparison of the outcomes of RQ\textsubscript{1} and RQ\textsubscript{2} reveals similar DPs, like \textit{"Create sense of urgency for AI adoption"}. To avoid introducing duplicates and redundancy, we consolidated such steps.

Figure \ref{fig:proposed_framework} (see Appendices) illustrates the dimensions of the proposed framework and their key DPs. The light blue boxes represent the DPs that arose by the literature review, while with light yellow are the ones that emerged from the thematic analysis. Finally, the light green boxes symbolize the DPs that appeared in both processes.

To improve clarity and readability, we clustered the nine dimensions of the framework into three overarching levers, namely \textbf{Direction}, \textbf{People}, and \textbf{Guardrails}. This grouping reflects the distinct roles these dimensions play in organizational transformation. \textit{Direction} captures the strategic and evaluative mechanisms that guide change—Strategy Design and Evaluation. \textit{People} encompasses the human and cultural drivers of organizational change, including Leadership, Organizational Culture, Collaboration, and Up-skilling. \textit{Guardrails} represent the structural, ethical and political constraints that shape how initiatives are governed and sustained over time —Governance and Ethics and Organizational Dynamics. Together, these three levers integrate strategic intent, organizational capabilities, and human and behavioral dynamics, offering a more integrated view of how AI transformation unfolds in practice.

\begin{takehomebox}
    \textbf{RQ\textsubscript{2} take-away:} Thematic analysis substantially enriched the foundation established in RQ\textsubscript{1}. It introduced levers absent from conventional models, such as mechanisms for skill verification. Hence, the framework was extended to incorporate the BSE-informed perspective that is missing from prior change models. 
\end{takehomebox}

\subsection{RQ\textsubscript{3} — Applicability of the framework}

In preliminary exploratory discussions, industry experts acknowledged that the importance of components such as \textit{Communication} is already well recognized. Therefore, RQ\textsubscript{3} seeks to explore this area more deeply by proposing actionable steps for each DP, thereby enhancing the study's relevance to organizational contexts. \revised{In other words, the DPs represent mid-level organizational design orientations that guide the overall direction of the organization, rather than discrete actions or intended outcomes. The actionable steps listed alongside each DP serve as concrete means to operationalize these orientations.}

To achieve this, we utilized the outcomes of thematic analysis, existing peer-reviewed research, and grey literature. In certain instances, we provided our own recommendations. This systematic approach forms a more robust evaluation of the framework. While several actionable steps are grounded in empirical research, others are derived from practitioners' interviews or authors' experience. We therefore distinguish evidence-backed components from those that require future empirical study to confirm their effectiveness.

The following tables (Table \ref{tab:tangible_steps_AI_strategy_design} - Table \ref{tab:tangible_steps_upskilling}) present the actionable steps that support the applicability of the proposed DPs. These actions are categorized by each dimension of the framework, thus improving readability and avoiding the introduction of a single, overwhelming table. The final column, \textit{"References"}, cites the key resources that informed the development of each tangible step.

\revised{Table \ref{tab:tangible_steps_AI_strategy_design} presents the actionable steps for the \textbf{\textit{AI strategy design}} dimension. Forming interdisciplinary teams with clear roles and responsibilities is beneficial to address the multifaceted challenges associated with AI adoption, as shared by P01 and P03. For example, engaging people of legal department ensures adherence to both national and ethical standards. Finally, involving domain experts in the AI strategy design, even without much experience with AI technologies, could facilitate considering the uniqueness of the field.

Modica et al. \cite{modica2010portfolioPriotization} explain the pivotal role of determining prioritization criteria and their weights. Then, stakeholders assign their ranks for the projects, forming the decision matrix. Moreover, P05 explained how \textit{cooperating with academia can facilitate forming the AI strategy}, while P8 mentioned that \textit{hiring AI experts} has positively affected their progress.

P01 noted that the development of AI solutions often prioritizes technical aspects, neglecting user testing and feedback mechanisms, such as user acceptance testing. This insight aligns with Venkatesh and Davis \cite{venkatesh1996perceivedEaseOfUse}, who emphasize that hands-on experience is crucial for assessing ease of use. Therefore, involving users throughout the design process is essential for ensuring user-friendliness. Additionally, P01 highlighted the importance of conducting \textit{"business case analysis to identify existing business needs before integrating AI technologies"}. This involves pinpointing organizational pain points and understanding market needs through targeted surveys and interviews~\cite{harvard2022identifyBusinessNeeds}.}

\begin{table*}[h!]
    \centering
    \begin{tabularx}{\linewidth}{@{}X X l@{}}
        \toprule
        \textbf{Design Principle} & \textbf{Actionable Steps} & \textbf{References} \\
        \midrule
        Be aware of parameters within the external environment, such as regulation
            & \revised{Bring together people with different educational backgrounds, such as IT, business, AI, and legal backgrounds}
            & \revised{P01, P03} \\
            & \revised{Establish clear roles within the team} & \\
        \cmidrule(l){1-3}
        Design tailored solutions, considering the unique nature of the domain
            & Involve domain experts in the design of the AI strategy
            & Authors' suggestion \\
        \cmidrule(l){1-3}
        \multirow{2}{=}{Determine projects' prioritization strategy}
            & Define prioritization criteria and their weight
            & \multirow{2}{*}{\makecell[l]{Modica et al.\\\cite{modica2010portfolioPriotization}}} \\
            & Collect stakeholders' ranks & \\
        \cmidrule(l){1-3}
        \multirow{2}{=}{Develop AI strategy in line with the latest AI advances}
            & Establish connections with academia to track and contribute to AI research advances
            & P05 \\
            & Hire AI experts to design and update the strategy
            & P08 \\
        \cmidrule(l){1-3}
        Consider user-friendliness during AI strategy design
            & Include users in the loop, conducting user acceptance tests
            & \makecell[l]{P01,\\Venkatesh \& Davis\\\cite{venkatesh1996perceivedEaseOfUse}} \\
        \cmidrule(l){1-3}
        \multirow{2}{=}{Recognize a business need before starting an AI integration project}
            & Conduct interviews and questionnaires to capture organization's pain points and/or market needs
            & \makecell[l]{Harvard Business\\School \cite{harvard2022identifyBusinessNeeds}} \\
            & Perform business case analysis
            & P01 \\
        \bottomrule
    \end{tabularx}
    \caption{Actionable steps for the design principles of the \textit{AI Strategy Design} dimension.}
    \label{tab:tangible_steps_AI_strategy_design}
\end{table*}

Table \ref{tab:tangible_steps_AI_strategy_evaluation} exhibits the tangible steps related to the \textbf{\textit{AI Strategy Evaluation}} dimension. Remenyi and Sherwood-Smith \cite{remenyi1999continuousEvaluation} proposed a systematic approach for the continuous evaluation of information systems. This process involves establishing an evaluation team composed of users, managers, engineers, and developers. Their responsibilities include defining business success criteria and designing a multi-metric evaluation framework to assess them.

Subsequently, regular evaluation sessions should be scheduled to analyze these metrics and implement necessary adjustments. The multi-metric framework may incorporate Key Performance Indicators (KPIs) aligned with organizational priorities. For example could encompass: (i) strategic aspects, such as \textit{the rate of AI Projects aligned with organizational objectives} — assesses how well AI integration supports organizational goals, (ii) operational aspects, such as \textit{the rate of rework of AI-Generated Content} — measures output quality and human correction burden, and (iii) human-centric aspects, such as \textit{Automation-to-Human Oversight Ratio} — evaluates the balance between automated generation and human review or approval, and \textit{Compliance or Ethical Review Incidents Related to AI} — tracks adherence to data governance and ethical standards.

\begin{table}[h!]
    \centering
    \begin{tabularx}{\linewidth}{@{}X X l@{}}
        \toprule
        \textbf{Design Principle} & \textbf{Actionable Steps} & \textbf{References} \\
        \midrule
        \multirow{2}{=}{Continuously evaluate the integration process}
            & Create an evaluation team comprising different roles
            & \multirow{4}{*}{\makecell[l]{Remenyi \&\\Sherwood-Smith\\\cite{remenyi1999continuousEvaluation}}} \\
            & Define clear business success criteria & \\
        \cmidrule(l){1-2}
        \multirow{2}{=}{Define and monitor KPIs for integration goals}
            & Determine a multi-metric system, including KPIs & \\
            & Set regular evaluation sessions to monitor and adapt the strategy & \\
        \bottomrule
    \end{tabularx}
    \caption{Actionable steps for the design principles of the \textit{AI Strategy Evaluation} dimension.}
    \label{tab:tangible_steps_AI_strategy_evaluation}
\end{table}

Table \ref{tab:tangible_steps_collaboration} outlines the DPs and necessary steps to ensure \textbf{\textit{Collaboration}} during the AI integration process. As P01 and P03 stated, AI constitutes a technology that spans multiple domains. Thus, it is essential to bring together individuals from various backgrounds, like IT, business or legal, and establish clear roles within their workflow. This will increase the quality of their cooperation, comparing to relying on communication channels between different departments, working on their own projects.

Moreover, sharing knowledge and experience is vital to improve team effectiveness and avoid repeating mistakes that waste resources, such as money and time. Earlier research findings \cite{abu-shanab2014knowledge_sharing_practices} indicate a positive correlation between knowledge-sharing practices and organizational learning, the latter being a prerequisite for preserving competitive advantage. P01 suggested Communities of Practice (CoPs) and internal forums as ways to facilitate knowledge and experience-sharing, while Calantone et al. \cite{calantone2002learning} proposed that organizations should document and share lessons learned from past projects.

\begin{table}[h!]
    \centering
    \begin{tabularx}{\linewidth}{@{}X X p{0.25\linewidth}@{}}
        \toprule
        \textbf{Design Principle} & \textbf{Actionable Steps} & \textbf{References} \\
        \midrule
        \multirow{2}{=}{Form a cross-functional team}
            & Bring together people with different educational backgrounds, such as IT, business, AI, and legal backgrounds
            & P01, P03 \\
            & Establish clear roles within the team & \\
        \cmidrule(l){1-3}
        \multirow{2}{=}{Institutionalize knowledge and experience-sharing practices}
            & Set up internal forums and Communities of Practice
            & P01 \\
            & Share lessons learned from past projects
            & Calantone et al. \cite{calantone2002learning} \\
        \bottomrule
    \end{tabularx}
    \caption{Actionable steps for the design principles of the \textit{Collaboration} dimension.}
    \label{tab:tangible_steps_collaboration}
\end{table}

Table \ref{tab:tangible_steps_communication} demonstrates the actionable DPs for the \textit{\textbf{Communication}} dimension. Kujala et al. \cite{kujala2022stakeholdersEngagement} underscores that one-way or two-way information flows form strategic stakeholder engagement. For instance, the strategy could include newsletters, presentations, and written reports as a one-way flow of information, or anonymous surveys, interviews, and stakeholder observations as means of gathering feedback through a two-way communication flow. This enables stakeholders, including employees, to articulate their perspectives and express concerns.

Establishing a well-defined and transparent communication strategy requires the identification and documentation of both formal and informal communication channels. To support direct and ongoing interactions among stakeholders, organizations may utilize platforms such as Slack by creating dedicated channels for AI-related topics, thereby enhancing the dissemination of relevant information. In addition to specifying communication methods, a comprehensive communication plan should also encompass the communication objectives, identify the target audience, and outline implementation and evaluation procedures, including timelines and feedback mechanisms \cite{van_gemert1999designingCommunicationPlan}. These elements are essential considerations for organizations when designing their AI communication plan.

\begin{table}[h!]
    \centering
    \begin{tabularx}{\linewidth}{@{}X X l@{}}
        \toprule
        \textbf{Design Principle} & \textbf{Actionable Steps} & \textbf{References} \\
        \midrule
        Engage with stakeholders
            & \revised{Initiate stakeholder communication through one-way information flows, such as newsletters, as a first step towards engagement}
            & Kujala et al.~\cite{kujala2022stakeholdersEngagement} \\
            & \revised{Initiate two-way information flows, e.g. anonymous surveys, with primary stakeholders}  & \\
        \cmidrule(l){1-2}
        Listen to employees' concerns
            & Use two-way information flows, such as anonymous surveys, interviews, and observations
            & \\
        \cmidrule(l){1-3}
        Establish constant communication channels with all stakeholders
            & Create AI-specific channels in communication platforms, such as Slack
            & Authors' suggestion \\
        \cmidrule(l){1-3}
        \multirow{4}{=}{Structure a clear communication strategy with all stakeholders}
            & Determine communication goals
            & \multirow{4}{*}{\makecell[l]{Van Gemert\\\& Woudstra\\\cite{van_gemert1999designingCommunicationPlan}}} \\
            & Determine means of communication & \\
            & Determine target groups / stakeholders & \\
            & Determine implementation and evaluation strategy & \\
        \bottomrule
    \end{tabularx}
    \caption{Actionable steps for the design principles of the \textit{Communication} dimension.}
    \label{tab:tangible_steps_communication}
\end{table}

Table \ref{tab:tangible_steps_governance} contains the applicable DPs for the \textit{\textbf{Governance and Ethics}} dimension of the proposed framework. As multiple interviewees stated, organizations should offer rules and directions on the utilization of AI solutions, with a focus on external ones. These instructions should explain situations that the use of AI solutions is not permitted and guidelines to anonymize proprietary information, before prompting to AI products.

The University of Toronto Library \cite{uoft2025AIevaluation} has published a road-map for the evaluation of AI tools and their generated outputs. The output assessment pertains to the significance of linking AI-generated outcomes to real-world examples, ensuring their alignment with reality, and comparing model's performance with human efforts. This indicates the necessity for maintaining human in the loop, ensuring the supervision and reliability of AI-generated outputs. Similarly, organizations could introduce AI oversight roles. The purpose of such roles would be to monitor and review AI-assisted work, performing audits to certify that people work with AI tools in a critical manner, avoiding over-reliance on GenAI.

\begin{table}[h!]
    \centering
    \begin{tabularx}{\linewidth}{@{}X X p{0.25\linewidth}@{}}
        \toprule
        \textbf{Design Principle} & \textbf{Actionable Steps} & \textbf{References} \\
        \midrule
        \multirow{2}{=}{Determine clear AI application guidelines}
            & Explain situations where utilizing AI solutions is not permitted
            & \multirow{2}{*}{\makecell[l]{P01, P04,\\P06--P10}} \\
            & Describe when and how to anonymize proprietary and sensitive data, before using AI solutions
            & \\
        \cmidrule(l){1-3}
        \multirow{2}{=}{Evaluate and criticize the quality of AI outputs}
            & Compare AI-generated results with real-world examples
            & \multirow{2}{*}{\makecell[l]{University of\\Toronto Library\\\cite{uoft2025AIevaluation}}} \\
            & Keep human in the loop and compare their judgments with AI outputs
            & \\
        \cmidrule(l){1-3}
            Monitor work quality to examine cases of over-reliance on AI
            & Introduce AI oversight roles to audit AI-assisted work
            & Authors' suggestion \\
        \bottomrule
    \end{tabularx}
    \caption{Actionable steps for the design principles of the \textit{Governance and Ethics} dimension.}
    \label{tab:tangible_steps_governance}
\end{table}

Table \ref{tab:tangible_steps_leadership} exhibits the actionable steps to execute the DPs that are associated with the \textit{\textbf{Leadership}} dimension. In their work, Nazir et al. \cite{nazir2024collectiveOwnership} emphasized a research gap concerning collective ownership, concentrating on the ways to cultivate this sense. However, based on psychological theory, they proposed that the formation of collective ownership feelings can be achieved through the following three mechanisms: i) developing a common understanding of the project and its goals, ii) ensuring active collaboration within the team, and iii) cultivating a sense of control on their tasks.

Creating a sense of urgency regarding AI integration aligns with the initial stage of Kotter’s change management model~\cite{kotter_leading_change_2012}. In a subsequent work \cite{kotter2008senseOfUrgency}, Kotter outlined four key steps to cultivate this urgency: \textit{i) bring the outside in (recognize opportunities in the market), ii) behave with urgency every day (yesterday's success does not prevent tomorrow's failure), iii) find opportunity in crises, and iv) deal with no-nos (people that are not open-minded)}. Similarly, articulating a shared and well-defined need for AI adoption is critical for facilitating organizational change. Conducting industry benchmark analyses, identifying organizational pain points that AI can address, and transparently sharing these findings can aid instilling this shared need.

Agarwal \cite{agarwal1998rewardSystems} investigated four organizational reward strategies, among which implementing team-based incentives, like performance bonuses for proposing and executing innovative AI initiatives can foster collaboration and encourage experimentation with emerging technologies. While skill verification activities own a pivotal role in recruitment processes, Alami et al. \cite{alami2024bestPracticesInRecruitment} listed best practices in online hiring for Software Engineering roles. Rather than relying solely on automated platforms, they advocate for incorporating \textit{task-oriented and targeted questions} to assess technical skills, \textit{AI detection} methods to filter automated responses, and \textit{qualitative manual reviews} to ensure authenticity. Thereby, they underscore the importance of human involvement. Such skill assessment practices can be integrated into employee development programs, not just hiring.

As P05 highlighted, collaboration between industry and academia is salutary for the former. For example, organizations gain insights to the latest advancements in AI research and can establish connections with emerging talents capable of driving innovation. Joint research initiatives, in particular, offer long-term strategic value. In a similar vein, events like Hackathons can enhance firms branding and serve as effective recruitment tools for attracting skilled professionals.

Dalrymple et al. \cite{dalrymple2024guaranteedsafeaiframework} proposed the Guaranteed Safe (GS) AI framework to ensure the safety and reliability of AI systems. This framework consists of four components: \textit{the Formal Safety Specification}, which defines strict safety constraints for AI systems, \textit{the World Model}, a structured representation for predicting and mitigating risks, \textit{the Verifier}, which guarantees compliance with the safety specifications, and \textit{the Deployment Infrastructure}, which serves as a monitoring and control mechanism during system operation. By adhering such structured methodologies, organizations can enhance the trustworthiness of their AI deployments. Additionally, incorporating transparent means of communication, like FAQs, regarding the use of data can further reinforce system's reliability.

Finally, Laoyan \cite{laoyan2025importanceOfSettingShortTermGoals} stated the value of setting short-term goals as a mechanism for building project momentum. The process begins with clearly defining measurable goals, followed by breaking them down into smaller, actionable steps. Assigning specific timelines to each task helps ensure continuous and steady progress and reduces the risk of procrastination.

\begin{table*}[h!]
    \centering
    \begin{tabularx}{\linewidth}{@{}X X l@{}}
        \toprule
        \textbf{Design Principle} & \textbf{Actionable Steps} & \textbf{References} \\
        \midrule
        \multirow{3}{=}{Define collective ownership}
            & Develop a common understanding of the project and its goal
            & \multirow{3}{*}{\makecell[l]{Nazir et al.\\\cite{nazir2024collectiveOwnership}}} \\
            & Ensure active collaboration of the team members & \\
            & Cultivate a feeling of control in team members' tasks & \\
        \cmidrule(l){1-3}
        \multirow{4}{=}{Establish a sense of urgency}
            & Bring the outside in
            & \multirow{4}{*}{\makecell[l]{Kotter\\\cite{kotter2008senseOfUrgency}}} \\
            & Behave with urgency every day & \\
            & Find opportunity in crises & \\
            & Deal with the no-nos & \\
        \cmidrule(l){1-3}
        \multirow{3}{=}{Establish a clear and shared need}
            & Perform industry benchmark analysis
            & \multirow{3}{*}{Authors' suggestions} \\
            & Identify pain points that AI can address & \\
            & Share the results of the analysis & \\
        \cmidrule(l){1-3}
        Encourage AI adoption and experimentation by rewarding desired behaviors and achievements
            & Institutionalize a reward system
            & Agarwal \cite{agarwal1998rewardSystems} \\
        \cmidrule(l){1-3}
        Hire AI experts to facilitate the design of the AI strategy
            & \textit{Further break-down is not feasible}
            & --- \\
        \cmidrule(l){1-3}
        \multirow{2}{=}{Institutionalize skill verification activities}
            & Involve human in the hiring process to have manual assessment of individuals' skills
            & Alami et al. \cite{alami2024bestPracticesInRecruitment} \\
            & Include skill assessments in employee development plans
            & Authors' suggestion \\
        \cmidrule(l){1-3}
        \multirow{2}{=}{Leverage talents for product and process innovation}
            & Develop partnerships with universities
            & P05 \\
            & Organize events, such as hackathons
            & Authors' suggestion \\
        \cmidrule(l){1-3}
        Promote collaboration with academia for research on AI
            & Create joint research projects
            & P05 \\
        \cmidrule(l){1-3}
        \multirow{2}{=}{Provide reliable internal AI solutions}
            & Follow structured protocols and frameworks, such as the GS AI framework
            & \makecell[l]{Dalrymple et al.\\\cite{dalrymple2024guaranteedsafeaiframework}} \\
            & Accompany the internal AI solution with a FAQ endpoint
            & Authors' suggestion \\
        \cmidrule(l){1-3}
        \multirow{3}{=}{Set short-term wins}
            & Define the goal
            & \multirow{3}{*}{\makecell[l]{Laoyan\\\cite{laoyan2025importanceOfSettingShortTermGoals}}} \\
            & Break the goal into steps & \\
            & Set a timeline & \\
        \bottomrule
    \end{tabularx}
    \caption{Actionable steps for the design principles of the \textit{Leadership} dimension.}
    \label{tab:tangible_steps_leadership}
\end{table*}

Table \ref{tab:tangible_steps_organizational_culture} displays the tangible steps that ensure the applicability of the \textit{\textbf{Organizational Culture}} dimension. Agility can be defined as the ability to sense and respond to change~\cite{tallon2019search_for_organizational_agility}. According to Adebayo \cite{adebayo2022AgileMindset}, successful Agile adoption is associated with four key factors: \textit{leadership buy-in}, \textit{cross-functional collaboration}, \textit{continuous learning}, and \textit{employee empowerment}. Top management should embrace Agile principles, such as by ensuring flexibility and teams empowerment. Promoting cross-departmental collaboration fosters creativity and problem-solving capabilities. In parallel, continuous feedback mechanisms drive both product and process improvements, ensuring adaptability. Finally, encouraging collective ownership enhances innovation and accountability, strengthening an Agile mindset.

Kumar \cite{kumar2024fosteringInnovation} introduced a framework identifying drivers of innovation. Central to this framework is the establishment of a psychologically safe environment, where employees feel autonomous to experiment, collaborate, and exercise creativity. Endorsement of such initiatives, along with the implementation of incentive systems, can significantly improve the levels of creativity. Nevertheless, it is crucial to define organization’s risk capacity to maintain balance between exploitation of current capabilities and exploration of novel opportunities.

On a similar note, Thomke in his report to Harvard Business School \cite{thomke2024buildingExperimentationCulture}, highlighted the attributes of a robust experimentation culture, encompassing continuous learning, structured reward systems, leadership support, and employee empowerment. Complementary, Pulido and Taherdoost \cite{pulido2024changeManagementInDigitalTransformation} underscored the importance of embracing experimentation through the initiation and funding of pilot projects.

\begin{table*}[h!]
    \centering
    \begin{tabularx}{\linewidth}{@{}p{0.25\linewidth} X l@{}}
        \toprule
        \textbf{Design Principle} & \textbf{Actionable Steps} & \textbf{References} \\
        \midrule
        \multirow{4}{=}{Adopt Agile principles}
            & Support agile principles from top management
            & \multirow{4}{*}{\makecell[l]{Adebayo\\\cite{adebayo2022AgileMindset}}} \\
            & Encourage cross-functional collaboration & \\
            & Encourage continuous learning and establish feedback mechanisms & \\
            & Promote employees' empowerment and ownership & \\
        \cmidrule(l){1-3}
        \multirow{5}{=}{Foster an innovation-driven and proactive culture}
            & Cultivate a psychologically safe environment
            & \multirow{5}{*}{\makecell[l]{Kumar\\\cite{kumar2024fosteringInnovation}}} \\
            & Ensure seniors' support & \\
            & Ensure the appropriate knowledge and technical expertise & \\
            & Introduce a reward system & \\
            & Determine the risk capacity & \\
        \cmidrule(l){1-3}
        \multirow{5}{=}{Foster a culture of experimentation}
            & Promote and fund pilot projects
            & \makecell[l]{Pulido \& Taherdoost\\\cite{pulido2024changeManagementInDigitalTransformation}} \\
            & Introduce a reward system
            & \multirow{4}{*}{\makecell[l]{Thomke\\\cite{thomke2024buildingExperimentationCulture}}} \\
            & Support continuous learning & \\
            & Ensure seniors' support & \\
            & Support employees' empowerment & \\
        \bottomrule
    \end{tabularx}
    \caption{Actionable steps for the design principles of the \textit{Organizational Culture} dimension.}
    \label{tab:tangible_steps_organizational_culture}
\end{table*}

\revised{Table \ref{tab:tangible_steps_organizational_dynamics} illustrates the actionable steps of the \textit{\textbf{Organizational Dynamics}} dimension. Wan et al. \cite{wan2020UNchangeManagement} suggest that determining a transparent and constant communication strategy is essential for staying aware of the organizational timing and the pressures that might take place during the change process. As a concrete implementation of this principle, establishing dedicated AI-specific channels in platforms, such as Slack, could foster engagement between different roles. Several organizational situations are particularly suited to such channels, including communicating policy or tool updates, distributing shared resources like prompts or context files, surfacing unexpected AI outputs, and facilitating dialogue between peers and top management about AI-related concerns. This could enable knowledge-sharing across roles to capture concerns and mitigate resistance.

Several change management models, such as Kotter's \cite{kotter_leading_change_2012}, necessitated the support of influential people, to reduce conflicting views regarding the upcoming change. In this regard, Jick \cite{jick1993implementingChange} introduced a methodology to identify key individuals or groups whose support is critical. This approach encompasses \textit{defining the necessary mass to achieve the change}, \textit{developing a strategy to secure their commitment}, and \textit{establishing a monitoring system to track the progress}.}

\begin{table*}[tb]
    \centering
    \begin{tabularx}{\linewidth}{@{}X X l@{}}
        \toprule
        \textbf{Design Principle} & \textbf{Actionable Steps} & \textbf{References} \\
        \midrule
        Be aware of the organizational situation, which might indicate internal pressures or resistance
            & \revised{Create AI-specific channels in communication platforms for updates, announcements and knowledge-sharing}
            & Authors' suggestion \\
        \cmidrule(l){1-3}
        \multirow{4}{=}{Secure the political support of key players}
            & Identify influential individuals or groups
            & \multirow{4}{*}{\makecell[l]{Jick \cite{jick1993implementingChange}}} \\
            & Define the necessary mass to support the change & \\
            & Develop a plan to gain the commitment of the critical mass & \\
            & Track the progress & \\
        \bottomrule
    \end{tabularx}
    \caption{Actionable steps for the design principles of the \textit{Organizational Dynamics} dimension.}
    \label{tab:tangible_steps_organizational_dynamics}
\end{table*}

Finally, Table \ref{tab:tangible_steps_upskilling} manifests the tangible steps responsible for \textit{\textbf{Up-skilling}} dimension. P08 explained in their interview that different generations have quite various educational foundations. Hence, cross-generational sessions, pairing younger employees with richer AI background with senior staff for mentorship may improve their collaboration.

Many interview participants emphasized the importance of AI training initiatives, such as hands-on workshops and both mandatory and optional training programs, in establishing a shared understanding of AI technologies. Similarly, offering tailored training that accounts for individuals' personality and attitudes toward AI can enhance learning effectiveness. Complementing these efforts with demonstrations of real-world AI use cases can provide a more holistic perspective on the capabilities and limitations of AI systems.

P07 elaborated on the value of hiring AI coaches to guide employees in effectively utilizing AI tools and technologies. Mughairi and Karim \cite{mughairi2020coachingAndMentoring} concluded that effective coaching and mentoring require clear objectives, strong interpersonal relationships, and structured communication. Setting measurable goals for individual performance and team growth, along with consistent and transparent feedback, helps cultivate trust between employees and their coaches or mentors. Hence, this approach can create supportive environments that promote both personal and business growth.

Lastly, London and Smither \cite{london1999empowerContinuousLearning} explored strategies to nurture a culture of continuous learning culture within organizations. Key enablers include constructive feedback, skill-based payments, and self-management training. Continuous feedback loops allow individuals to identify areas for development more accurately, while career centers and self-assessment tools support setting personal development goals. By rewarding skill acquisition, companies can motivate employees to embrace ongoing learning and remain adaptable to the emerging challenges.

\begin{table*}[h!]
    \centering
    \begin{tabularx}{\linewidth}{@{}p{0.3\linewidth} X l@{}}
        \toprule
        \textbf{Design Principle} & \textbf{Actionable Steps} & \textbf{References} \\
        \midrule
        Introduce cross-generational mentoring sessions
            & Pair younger employees with AI knowledge and experience with senior staff for mentorship
            & P08 \\
        \cmidrule(l){1-3}
        \multirow{3}{=}{Provide AI training for employees}
            & Set up hands-on workshops
            & P02, P03, P08 \\
            & Set up mandatory and optional programs
            & P01, P04, P06, P07, P09 \\
            & Set up tailored training programs
            & Authors' suggestion \\
        \cmidrule(l){1-3}
        \multirow{4}{=}{Provide AI mentoring and coaching sessions}
            & Hire AI experts as coaches
            & P07 \\
            & Set measurable goals
            & \multirow{3}{*}{\makecell[l]{Mughairi \& Karim\\\cite{mughairi2020coachingAndMentoring}}} \\
            & Establish continuous and transparent feedback loops & \\
            & Cultivate trust between employees and coaches / mentors & \\
        \cmidrule(l){1-3}
        \multirow{3}{=}{Institutionalize continuous learning initiatives}
            & Establish continuous and transparent feedback loops
            & \multirow{3}{*}{\makecell[l]{London \& Smither\\\cite{london1999empowerContinuousLearning}}} \\
            & Introduce skill-based payment system & \\
            & Develop self-management awareness & \\
        \cmidrule(l){1-3}
        \multirow{2}{=}{\revised{Calibrate understanding of AI capabilities and limitations}}
            & Provide AI training initiatives
            & \multirow{2}{*}{Authors' suggestions} \\
            & Demonstrate real-world AI case studies & \\
        \bottomrule
    \end{tabularx}
    \caption{Actionable steps for the design principles of the \textit{Up-skilling} dimension.}
    \label{tab:tangible_steps_upskilling}
\end{table*}

\begin{takehomebox}
    \textbf{RQ\textsubscript{3} take-away:} This RQ developed the lower level of the early-stage BSE-informed framework by adding actionable steps to each DP. This aimed to improve the framework's applicability and form the baseline for RQ\textsubscript{4}.
\end{takehomebox}

\subsection{Survey Results}

We surveyed SE professionals to collect their impressions regarding the high-level overview of the proposed framework, encompassing the dimensions and the respective DPs. The survey distribution lasted five weeks and resulted in 140 responses. We removed 35 responses (\SI{25}{\percent}) because they were incomplete or failed our predefined authenticity checks. This exclusion rate is within the anticipated range in studies conducted via Prolific.

\subsubsection{Demographic Information}

The objective of the survey was exploratory, involving individuals that hold various positions of different backgrounds and have different years of experience. Of the 105 respondents, \SI{81}{\percent} were male, \SI{18}{\percent} were female, and \SI{1}{\percent} were no-binary. The participants are located in eight distinct countries, with the most representations coming from the United Kingdom (\SI{40}{\percent}), Sweden (\SI{20}{\percent}), and the United States of America (\SI{18}{\percent}).

Our sample prominently featured individuals holding software developer (\SI{48}{\percent}) and team lead (\SI{14}{\percent}) roles, making up \SI{62}{\percent} of all respondents. Other significant positions included Data Analysts / Data Engineers / Machine Learning Engineers (\SI{10}{\percent}), DevOps Engineers (\SI{7}{\percent}), and high leadership roles, such as CEO and CIO (\SI{7}{\percent}). The remaining respondents held System Architecture (\SI{4}{\percent}), Quality Assurance (\SI{4}{\percent}), and other roles (\SI{6}{\percent}).

In terms of work tenure, there was almost equal distribution across the different groups of work experience. While entry-level (0-5 years of experience) and mid-senior (11-20 years of experience) positions were the most frequent groups across the sample—both appeared in the \SI{30}{\percent} of the responses— there was significant representation of mid- (5-10 years of experience) and senior-level (20+ years of experience) individuals—\SI{22}{\percent} and \SI{18}{\percent} respectively.

\subsubsection{Current AI integration landscape}

The first part of the survey was devoted to the challenges that SE organizations face during AI incorporation. \textbf{Ethical considerations} are central to current AI integration efforts, with \textit{Risk aversion due to sensitive data} and \textit{Considerate use of external AI solutions} being mentioned 49 and 41 times, respectively. \textbf{Challenges regarding organizational processes} also appeared prominently. Particularly, \textit{Keeping up with AI advancements} and \textit{Identify cases where AI is effective and where it is not} were cited by 34 and 31 respondents. Additionally, \textbf{Organizational Politics} emerged as a key factor, with \textit{Handling differing views on AI transformation} appearing in 21 responses. Interestingly, only one new challenge was added to the existing list, indicating that the initial group of BSE-related challenges~\cite{tavantzis2024HumanCenteredAI} effectively captured the main obstacles of early-stage AI integration.

The average rating of existing organizational initiatives to address these challenges was 3.05 out of 5, indicating room for improvement and highlighting the need for a structured framework to support them. While the 63.8\% of the respondents identified the existence of an internal AI integration strategy, almost 13\% were uncertain about the existence of such a strategy. While the percentage is relatively low, this uncertainty underscores the need for clearer and more consistent communication. Among those aware of their organization's strategy, the mean satisfaction score was 3.61. This score suggests a generally positive experience, but also indicating opportunities for organizations to incorporate employees' feedback and tailor their strategy based on their insights.

\subsubsection{Overall impressions for the proposed framework}

The respondents rated the proposed framework as relatively easy to understand, with an average score of 3.17 out of 5. The primary concern affecting comprehension was its complexity, stemming from multiple dimensions and numerous DPs per dimension. However, given the inherently multi-dimensional nature of AI integration, a certain level of complexity is anticipated, as some respondents conveyed. Participants recommended the potential classification of dimensions and DPs, sharing their concern though that these multiple layers could increase the complexity of the artifact.

Importantly, only four respondents indicated that the framework is ineffective. In fact, \SI{28.6}{\percent} believed it could fully help organizations overcome AI-related challenges, while \SI{67.6}{\percent} saw it as partially helpful. This finding supports the view that there is no one-size-fits-all solution and organizations should tailor the framework to their specific contexts.

Participants' feedback also pertained to the applicability of the framework, reinforcing the need for prioritizing the dimensions and DPs, and supplementing actionable steps linked to each DP. Although such steps were included in the broader proposal, they were excluded from the survey to ensure quality in the responses. Additionally, a few respondents highlighted the importance of keeping the framework up-to-date and considering technical aspects as well—like technical infrastructure—noting that rapid progress demands regular updates to remain relevant and effective.

\subsubsection{\$100 method — Dimensions prioritization}

The final section of the survey was devoted to the perceived importance of dimensions for participants, using the \$100 method. Table \ref{tab:survey_descriptive_statistics} shows the summary of the descriptive statistical analysis. Notably, budget prioritization reflects practitioners’ current pain points, not necessarily the long-term determinants of successful AI transformation. Because many respondents are early in their AI adoption journey, their ability to identify long-term success factors may be limited.

The \$100-method allocations are compositional, as allocations are bounded and sum to a fixed total. While mean values are commonly reported in exploratory analyses, \SI{95}{\percent} bootstrap confidence intervals are also provided to indicate sampling variability. Given the exploratory nature of the study, we consider the use of descriptive statistics and the Scott–Knott ESD algorithm appropriate.

\begin{table*}[h]
    \centering
    \sisetup{table-format=2.1, detect-weight=true}
    \begin{tabular}{@{}l S S c S[table-format=2.0] S[table-format=1.0] S[table-format=2.0]@{}}
        \toprule
        \textbf{Dimension} & {\textbf{Mean}} & {\textbf{Std. Dev.}} & \textbf{CI \SI{95}{\percent}} & {\textbf{Median}} & {\textbf{Min}} & {\textbf{Max}} \\
        \midrule
        AI Strategy Design      & 15.1 & 7.7 & [13.6, 16.6] & 15 & 0 & 40 \\
        AI Strategy Evaluation  & 11.4 & 5.6 & [10.3, 12.5] & 10 & 0 & 30 \\
        Collaboration           &  9.6 & 5.5 & [8.5, 10.6]  & 10 & 0 & 30 \\
        Communication           & 10.6 & 5.7 & [9.5, 11.7]  & 10 & 0 & 30 \\
        Governance \& Ethics    & 11.6 & 7.8 & [10.1, 13.1] & 10 & 0 & 50 \\
        Leadership              & 11.0 & 8.1 & [9.5, 12.6]  & 10 & 0 & 63 \\
        Organizational Culture  &  8.4 & 5.4 & [7.4, 9.4]   &  9 & 0 & 30 \\
        Organizational Dynamics &  7.1 & 4.4 & [6.2, 7.9]   &  5 & 0 & 30 \\
        Up-skilling              & 15.2 & 9.6 & [13.3, 17.0] & 12 & 0 & 50 \\
        \bottomrule
    \end{tabular}
    \caption{Descriptive statistics of survey data: dimension prioritization scores from the \$100-method.}
    \label{tab:survey_descriptive_statistics}
\end{table*}

\textbf{Up-skilling} and \textbf{AI Strategy Design} emerged as the most critical dimensions for successful AI integration, with mean scores of 15.2 and 15.1. This suggests that SE practitioners place strong emphasis on developing internal capacity and establishing clear strategic planning for early-stage AI integration. \textbf{Governance and Ethics} ranked third in perceived importance (11.6), followed closely by \textbf{AI Strategy Evaluation} (11.4) and \textbf{Leadership} (11.0), reflecting growing awareness of oversight and accountability mechanisms once AI initiatives are underway.

Surprisingly, \textbf{Organizational Dynamics} was rated the least important dimension, despite \textit{Differing internal views on AI adoption} being a frequently cited challenge. This discrepancy indicates that political, cultural, and structural aspects across organizations are often under-recognized or insufficiently discussed. Although some respondents expressed extreme positions—as shown by the wide minimum and maximum values in Table~\ref{tab:survey_descriptive_statistics}—the relatively narrow \SI{95}{\percent} confidence intervals suggest a moderate level of consensus around the relative importance of each dimension.

\begin{figure*}[h]
    \centerline{
    \includegraphics[width=0.95\linewidth]{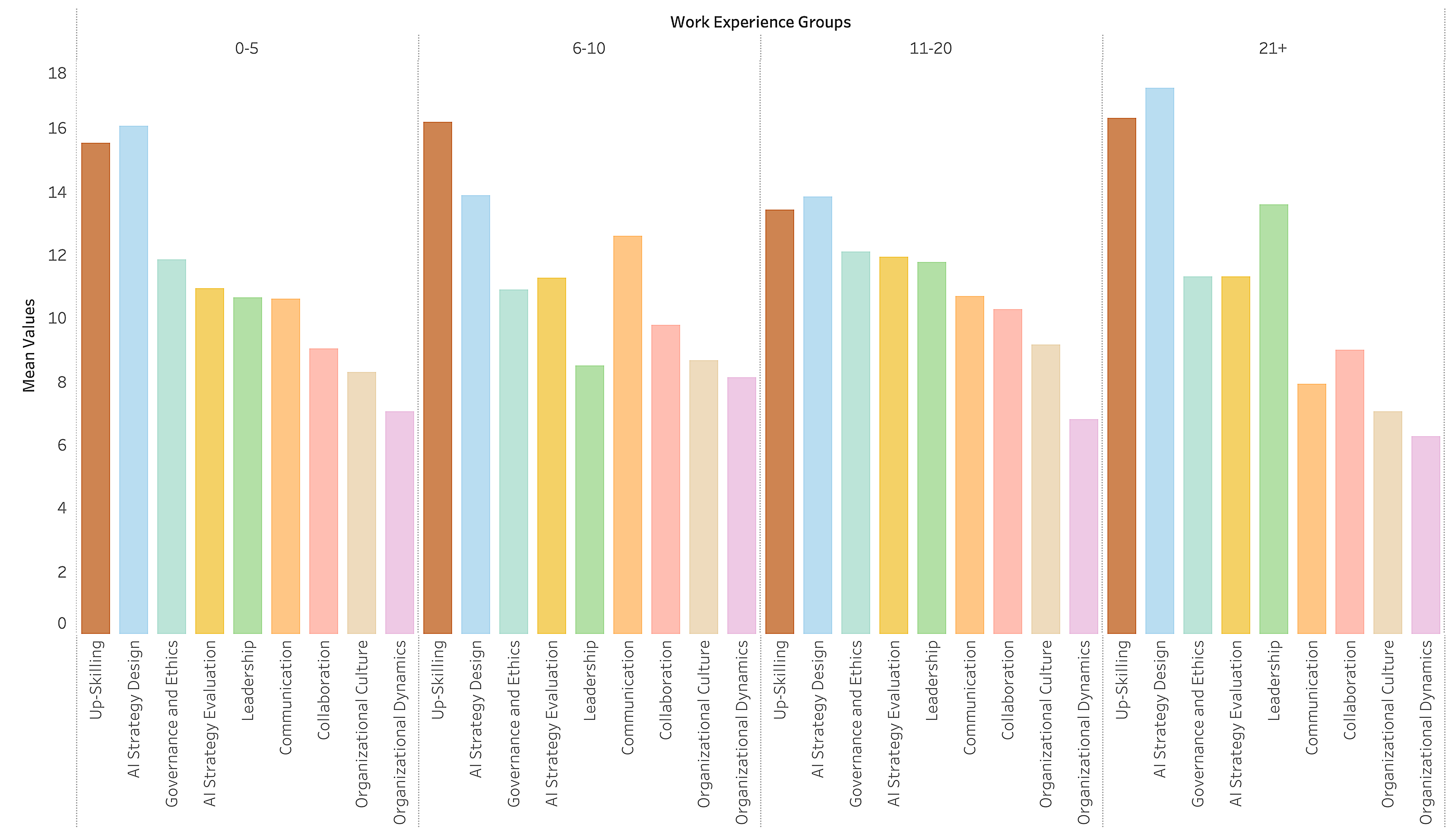}}
    \caption{Dimensions' prioritization per group of working experience}
    \label{fig:survey_meanValuesGrouping}
\end{figure*}

Figure \ref{fig:survey_meanValuesGrouping} illustrates mean prioritization scores for the dimensions of the proposed framework, grouped by respondents’ years of professional experience (0–5, 6–10, 11–20, and 21+ years). Overall, two dimensions —\textit{AI Strategy Design} and \textit{Up-skilling}— receive consistently high mean scores across experience groups, while other dimensions show greater variation.

Early-career respondents (0–5 years) rate \textit{AI Strategy Design} and \textit{Up-skilling} highest with a notable difference to the remaining dimensions, indicating a strong preference for structured planning and capacity building. Respondents with 6–10 years emphasize \textit{Communication}, \textit{Collaboration}, and \textit{Organizational Dynamics} more strongly than other groups, which may reflect increased exposure to cross-team coordination and change-management responsibilities at mid career. The 11–20 year group largely follows the overall ranking pattern (this group comprised the largest share of respondents together with the group with 0-5), while practitioners with 21+ years place relatively greater importance on \textit{Leadership}, consistent with more senior strategic roles.

\subsubsection{Scott-Knott Algorithm}

We employed the Scott-Knott Effect Size Difference (ESD) test to cluster the framework's dimensions according to their mean scores derived from the \$100 method. This analysis provides practitioners with a shared understanding of the relative importance of the dimensions by grouping them accordingly.

\begin{figure*}[h]
    \centerline{\includegraphics[width=1\linewidth]{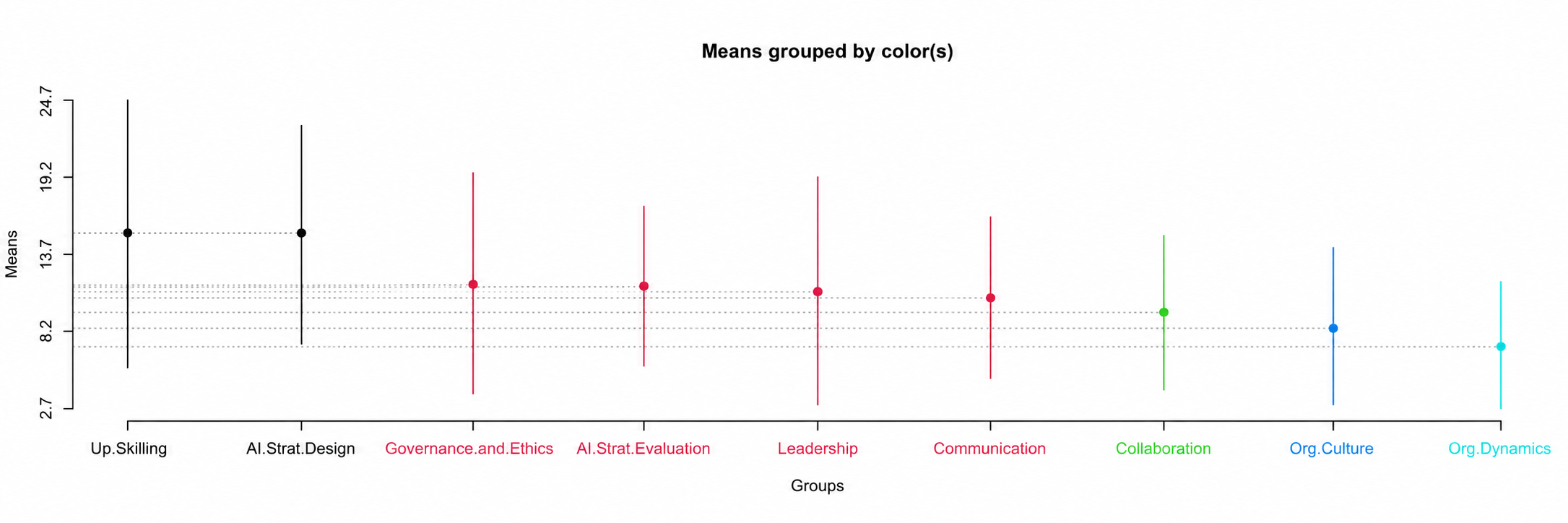}}
    \caption{Statistically distinct clusters of framework dimensions based on Scott-Knott ESD test results}
    \label{fig:scottKnott}
\end{figure*}

As illustrated in Figure \ref{fig:scottKnott}, the dimensions are organized into \textbf{five distinct groups}. The first group includes \textit{AI Strategy Design} and \textit{Up-skilling}, which represent the highest-priority dimensions. The second group comprises four dimensions—\textit{Governance and Ethics, AI Strategy Evaluation, Leadership and Communication}—followed by three single-dimension groups: \textit{Collaboration}, \textit{Organizational Culture}, and \textit{Organizational Dynamics}.

\subsection{Early Feasibility Check Results}

We conducted two expert review workshops to supplement the survey data and complete the preliminary evaluation of the proposed framework. The objective of these sessions was to collect professionals' input regarding the relevance and clarity of the recommended steps within the industrial context.

\subsubsection{Demographic Information}

The sessions included individuals actively involved in AI strategy decision-making within their organizations, enabling collection of in-depth insights across all framework dimensions. Three participants represented the Automotive industry, while one came from the Aerospace sector. Table \ref{tab:workshop_org_descriptive_information} presents descriptive information about their organizations.

\begin{table}[h!]
    \centering
    \begin{tabularx}{\linewidth}{@{}l X l l@{}}
        \toprule
        \textbf{Org. ID} & \textbf{Participant IDs} & \textbf{Company Size} & \textbf{Domain} \\
        \midrule
        Org01 & Pt01, Pt02, Pt03 & 102{,}000 & Automotive \\
        Org02 & Pt04             &  24{,}500 & Aerospace \\
        \bottomrule
    \end{tabularx}
    \caption{Descriptive information of the organizations represented in the workshop sessions.}
    \label{tab:workshop_org_descriptive_information}
\end{table}

Table \ref{tab:workshop_descriptive_information} provides descriptive information of the sessions' participants. The group was diverse across several dimensions. Participants represented two experienced age groups, offering perspectives shaped by their experiences. Gender diversity further enriched the discussions. The participants also held different roles, including managerial positions with both technical and business expertise, which expanded the depth of insights gathered. Due to work obligations, only three professionals participated in the data collection.

\begin{table*}[h!]
    \centering
    \begin{tabular}{@{}l l l l l@{}}
        \toprule
        \textbf{Participant ID} & \textbf{Age Group} & \textbf{Gender} & \textbf{Current Role} & \textbf{Work Experience} \\
        \midrule
        Pt01 & 45--54 & Male   & Manager                              & $>10$ years \\
        Pt02 & 35--44 & Female & Expert Product Strategy Manager      & $>10$ years \\
        Pt03 & 35--44 & Male   & Technology \& Strategy Leader for AI & $>10$ years \\
        Pt04 & 35--44 & Male   & Group Manager                        & $>10$ years \\
        \bottomrule
    \end{tabular}
    \caption{Descriptive information of the workshop participants.}
    \label{tab:workshop_descriptive_information}
\end{table*}

\subsubsection{Early Feasibility Check Overview}

Table \ref{tab:workshop_general_insights} summarizes the preliminary feedback from the workshop participants on the clarity and applicability of the proposed framework. Overall, the clarity of the DPs was rated highly, with nearly all dimensions receiving mean scores of 4 or above out of 5. The exception was Organizational Culture, which received a slightly lower score (3.67).

In contrast, the applicability of the proposed steps for each dimension received lower scores. While certain dimensions, such as Communication, were seen as highly actionable, most were rated closer to the midpoint value of 3. This suggests that although the framework is generally clear, there is room for improvement and further refinement to improve practical implementation. Notably, the applicability scores obtained in the workshops reflect role-specific feasibility judgments rather than general usability assessments.

\begin{table}[h!]
    \centering
    \sisetup{table-format=1.2}
    \begin{tabular}{@{}l S S@{}}
        \toprule
        \textbf{Dimension} & {\textbf{DPs Clarity}} & {\textbf{Steps Applicability}} \\
        \midrule
        AI Strategy Design      & 4.33 & 3.67 \\
        AI Strategy Evaluation  & 4.00 & 4.00 \\
        Collaboration           & 4.00 & 3.33 \\
        Communication           & 4.67 & 4.67 \\
        Governance \& Ethics    & 4.33 & 4.00 \\
        Leadership              & 4.33 & 3.67 \\
        Organizational Culture  & 3.67 & 3.00 \\
        Organizational Dynamics & 4.33 & 3.67 \\
        Up-skilling              & 4.33 & 3.00 \\
        \bottomrule
    \end{tabular}
    \caption{General insights from the early feasibility check: mean clarity and applicability scores of the design principles across framework dimensions (1--5 scale).}
    \label{tab:workshop_general_insights}
\end{table}

An interesting trend appears when comparing the applicability scores (see Table \ref{tab:workshop_general_insights}) with the mean \$100-method allocations across dimensions (see Table \ref{tab:survey_descriptive_statistics}). Figure \ref{fig:early-stage_paradox} (see Appendices) visualizes this pattern by mapping the perceived importance of each framework's dimension against its perceived implementation difficulty. While SE practitioners acknowledge the high applicability and long-term importance of dimensions such as Governance and Ethics, they still allocated comparatively little budget to it.

This divergence can be interpreted in multiple ways. One possibility is what we term the early-stage paradox: \textit{SE practitioners recognize the necessity and feasibility of enablers like Governance and Ethics, yet concentrate early efforts on visible, short-term, high-impact areas rather than on building a sustained integration foundation}. However, an alternative explanation is that organizations are consciously prioritizing speed, experimentation, and capability building before investing in safety-oriented or governance-related structures—potentially accruing technical debt in the process. Rather than a paradox per se, this pattern may reflect strategic sequencing or limited awareness of future governance needs.

\textbf{AI Strategy Design} emerged as a high-priority dimension, receiving a mean score of 4.33 regarding the clarity of its DPs (see Figure \ref{tab:workshop_general_insights}). However, their applicability was rated slightly lower (3.67), suggesting that while some steps are actionable, others may lack effectiveness in practice.

Among the recommended actions, only \textit{forming a cross-functional team} was unanimously deemed applicable. Other useful steps included \textit{involving domain experts, conducting interviews,} and \textit{analyzing business cases}. Conversely, \textit{defining prioritization criteria} and \textit{collecting stakeholders' ranks} were neither experienced nor considered applicable by the participants.

On the other hand, no step was unanimously considered as inapplicable, but opinions varied. For instance, some valued \textit{interviews for identifying organizational needs}, while others found them ineffective. One participant also noted the difficulty of \textit{hiring AI experts with sufficient domain knowledge} and following a structured methodology for projects' prioritization.

\textbf{AI Strategy Evaluation} showed similar patterns to the AI Strategy Design dimension, with both clarity and applicability receiving a mean score of 4, implying well-defined and feasible DPs.

All suggested steps of this dimension were considered actionable by at least one professional.\textit{ Creating a diverse evaluation team} received unanimous support, while \textit{setting regular metrics} was frequently cited as practical. The appearance of all the proposed steps in the responses reflects the alignment with industrial needs.

However, some concerns emerged. A few participants warned that KPIs might mislead teams if not cautiously defined, potentially increasing complexity and causing conflicting success criteria. These thoughts were discussed during the sessions, with \textit{defining business success criteria} and \textit{developing a multi-metric system with KPIs} being viewed by some as impractical in real-world settings.

The \textbf{Communication} dimension received high ratings for both DPs' clarity and applicability, with a mean score of 4.67 in each case. This suggests that participants found the DPs both well-articulated and practical in industrial contexts.

Most of the proposed actions were widely endorsed, with five out of seven steps considered both useful and feasible. The only exceptions were \textit{creating AI-specific channels} and \textit{defining implementation and evaluation strategies}, which did not receive unanimous support. Participants emphasized the central role of communication in successful AI adoption, both across the organization and within teams.

Only two actions were flagged as potentially ineffective. One practitioner expressed concerns about using two-way information flows, particularly in the form of surveys, citing participants fatigue and disengagement. Similarly, while strategies are important, one professional felt that they should keep communication process simple to avoid introducing unnecessary complexity. It was noted that successful communication is often intuitive, \textit{"people can feel and know when communication is working"}.

Experts rated highly the DPs of the \textbf{Collaboration} dimension for their clarity, with a mean score of 4, even though their applicability received the lower score of 3.33. This indicates some challenges in implementing the proposed steps.

Among the recommended actions, only \textit{setting up internal forums and Communities of Practice} was unanimously viewed as both feasible and valuable. Other steps, like \textit{bringing together people from different backgrounds} and \textit{sharing lessons learned from past projects}, were seen as partly actionable but received fewer agreements. One participant emphasized that effective collaboration requires actually \textit{working together}, noting that without this, communication tends to break down over time.

Some steps were also identified as impractical by few participants. Specifically, \textit{Establishing clear roles within the team} and \textit{sharing lessons learned} were noted as problematic. Practitioners argued that overly rigid roles can be counterproductive, and that formalized knowledge-sharing mechanisms like "white books" or structured sessions often fail in practice. Instead, informal approaches, such as Communities of Practice, seem to be more beneficial for maintaining and sharing knowledge.

Experts assigned high clarity score (4.33) regarding the DPs of the \textbf{Governance and Ethics} dimension and slightly lower for their applicability (4). Despite these scores, participants raised concerns about real-world implementation.

All proposed steps were acknowledged as feasible by at least one participant, even though none received unanimous agreement. The most frequently cited actions included \textit{anonymizing proprietary data} before AI use, \textit{comparing AI outputs with real-world examples}, and \textit{keeping humans in the loop}. These steps belong to different DPs, suggesting that further refinement may be required to ensure their full implementation.

Only one action was deemed impractical, and only by a single participant. Still, broader concerns emerged. Some noted the challenge of effectively monitoring governance measures, while others warned that fear and misconceptions around AI can lead to excessive control mechanisms. This could hinder AI integration efforts.

\textbf{Leadership} emerged as the ''richest'' dimension, containing the largest number of DPs and actionable steps. Despite its complexity, practitioners rated the DPs highly for their clarity (4.33). However, their applicability score received a lower value (3.67), implying a gap between theoretical recommendations and practical feasibility.

Only 15 out of the 21 proposed actions were considered applicable by at least one participant, and none received unanimous support. Notably, most steps were endorsed by a single respondent. Professionals acknowledged that Leadership dimension spans various aspects within AI adoption process, explaining the anticipated complexity.

Several steps were explicitly identified as impractical. Notably, all actions inspired by Kotter’s framework for establishing urgency were considered infeasible. Practitioners, also, criticized the step \textit{institutionalizing a reward system}, arguing that it could lead to counterproductive behaviors. Instead, they emphasized the importance of fostering genuine belief and trust in the value of AI, rather than relying on financial incentives.

The \textbf{Organizational Culture} dimension stood out with the lowest ratings across all evaluated aspects. DPs clarity received a mean score of 3.67, while their applicability scored just 3, indicating difficulties in interpreting and implementing the proposed DPs.

Experts offered limited but valuable feedback. Only two steps, namely \textit{supporting agile principles from top management} and \textit{determining risk capacity}, were identified as actionable. The rest recommendations were neither endorsed nor rejected. Instead, participants noted that cultural traits are inherently difficult to measure. They emphasized that while the recommendations may be theoretically sound, effective implementation requires adaptation to each organization's specific context and industry.

The \textbf{Organizational Dynamics} dimension pertains to complex factors, influencing how organizations operate and adapt. Despite its conceptual depth, participants rated the clarity of its DPs relatively high with the mean score of 4.33. However, the applicability score was relatively lower receiving the score of 3.67, highlighting challenges in translating theory into practice.

Only a few steps were recognized as actionable, each by a single participant, reflecting diverse experiences and limited consensus. Two actions received no support. Specifically, \textit{defining the necessary mass to support change} and \textit{creating AI-specific communication channels} deemed impractical. Participants stressed the difficulty of \textit{securing political support from key players}, acknowledging it as critical in theory but highly complex in practice. These insights point to the need for deeper investigation in this area to propose more targeted DPs.

\textbf{Up-skilling} constitutes the final dimension of the framework, thus one of the most critical ones, according to the survey respondents. Practitioners assigned the high DPs' clarity score of 4.33.  However, their practical implementation received a lower score of 3.

Most of the proposed actions were viewed as feasible by at least one participant, and none were deemed completely impractical. Notably, \textit{cross-generational mentoring} and \textit{diverse AI training formats}, including workshops and tailored programs, were emphasized as meaningful. Furthermore, experts stressed that such efforts should involve active collaboration rather than relying on passive learning formats, reinforcing the need for dynamic and engaging initiatives.

\begin{takehomebox}
    \textbf{RQ\textsubscript{4} take-away:} Overall, the early-stage BSE-informed framework appears largely actionable but still needs empirical validation. The emphasis SE practitioners place on Up-skilling and AI Strategy Design reflects an industry preference for procedural rather than human-centered principles.
\end{takehomebox}

\section{Discussion}
\label{sec5}

This section is devoted to discussing the relevance and connection of our findings with existing literature. Also, it includes the limitations and threats to validity that the current research setting raises.

\subsection{Findings' interpretation}

The objective of this study was to propose a BSE-informed framework of actionable DPs, \revised{targeting organizational adoption strategies and practices}, to enable SE organizations for smoother and more effective early-stage AI integration. In this regard, the structuring process relied on identified BSE-related challenges of our former work \cite{tavantzis2024HumanCenteredAI}. To answer the raised research questions, we designed a mixed-methods methodology.

Table \ref{tab:findings-summary} summarizes the key findings emerged from our study, together with their implications. This structured approach aims to guide both SE practitioners and researchers in leveraging the benefits of the proposed framework.

\begin{table*}[t]
    \centering
    \begin{tabularx}{\textwidth}{@{}>{\raggedright\arraybackslash}X >{\raggedright\arraybackslash}X@{}}
        \toprule
        \textbf{Finding} & \textbf{Implication} \\
        \midrule
        AI transformation resembles prior change initiatives.
            & SE practitioners can approach AI integration using established practices from earlier software change initiatives. \\\addlinespace
        Up-skilling is the most important dimension in the framework.
            & Up-skilling initiatives could mitigate the already identified confusion when developing ML solutions. \\\addlinespace
        The \textit{Governance and Ethics} dimension emerged solely from the thematic analysis.
            & Responsible and ethical mechanisms are essential within AI-embedded workflows. \\\addlinespace
        The framework aligns with the EU AI Act, emphasizing human oversight.
            & Human presence and monitoring should be part of designing AI-embedded workflows. \\\addlinespace
        The framework partly covers each organization's needs.
            & AI strategists should adapt the framework to each domain's unique context to fully leverage its benefits. \\\addlinespace
        Communication gaps were identified, as almost \SI{15}{\percent} of respondents were unsure whether an AI strategy exists.
            & SE organizations should invest more in their internal communication strategy. \\\addlinespace
        Clusters of dimensions were identified based on their perceived importance for smooth AI integration.
            & These clusters indicate aspects that AI strategists could focus on for early-stage integration. \\\addlinespace
        Early feasibility check showed that, while the steps are clear, some received limited perceived applicability.
            & There is a need for more empirical studies and collaborations between academia and industry.\\
        \bottomrule
    \end{tabularx}
    \caption{Summary of the study's key findings and their implications for research and practice.}
    \label{tab:findings-summary}
\end{table*}

\begin{itemize}
    \item[\faHandORight] \textbf{AI as Software Integration}: AI adoption in SE is best treated as disciplined software integration rather than a break with prior practice.

    The synthesis of 43 change management models addressed all but three BSE sub-challenges. No single model suffices, yet a composite approach is effective; this underpins the proposed early-stage, BSE-informed framework as a pragmatic scaffold for action. Evidence from the review and the interviews converges on viewing AI adoption as structured integration into existing socio-technical systems, not as a radical departure. This stance is consistent with prior work cautioning against AI hype and encouraging adoption only where there is a clear, situated need~\cite{floridi2024whyAIhypeIsTechBubble, mattmann2024AIhypeVsReality}. Given their alignment with many identified challenges, these models remain broadly relevant. However, in light of today’s fast-paced change environments, further research is needed to assess how well established models translate to contemporary organizational contexts, particularly those involving emerging technologies like GenAI.

    Documented confusion around developing ML-enabled products~\cite{nahar2023metasummarychallengesbuildingproducts} indicates that capability building should precede scale-up. A focus on \emph{Up-skilling} can reduce coordination and cognitive barriers by establishing shared baselines in skills, vocabulary, and expectations. Survey responses further indicate that there is no one-size-fits-all solution, confirming the view that \textbf{Compatibility}—a close fit with current workflows, practices, and tools—emerges as a decisive condition for effective integration~\cite{russo2024navigatingComplexityOfgenAI, mit2025GenAIDivide, dellAcqua2023jaggedFrontier}. The immediate implication is to tailor the framework to local constraints and domain knowledge rather than apply generic templates, aligned with prior research~\cite{nadler1980congruenceModel}.

    \item[\faHandORight] \textbf{Governance and Oversight: Human-Centric Controls}. Early AI efforts require explicit governance mechanisms that safeguard transparency, accountability, and human decision rights.

    Thematic analysis extended the literature-derived framework with human-centric controls previously underrepresented; the \emph{Governance and Ethics} dimension originates entirely from this analysis. This addition reflects the argument that AI introduces ethical risks—opacity, misplaced trust, context drift—that are not fully captured by traditional change models~\cite{Eitel-Porter2021ethicalAI}. The design principles in this dimension align with the EU AI Act’s requirements for transparency and human oversight (Article 13)~\cite{eu2025aiAct}. Naturally, the framework situates current endeavors at Level~6 in Parasuraman et al.’s automation taxonomy, where systems propose actions and humans retain veto authority~\cite{parasuraman2000automationModel}. This allocation clarifies accountability while keeping assurance activities tractable during early adoption.

    \item[\faHandORight] \textbf{Industry Evidence: Priorities, Communication, and Feasibility}. SE professionals prioritize procedural aspects over guardrails, and in some cases fail to communicate the AI strategy clearly.

    The survey indicated a communication gap: Almost \SI{15}{\percent} of respondents were unsure whether their company had an AI strategy. Existing research stresses that both message and medium are crucial to achieve alignment~\cite{goodman2004mediumAndTheMessage}. SE organizations should therefore use interactive channels and internal forums to anchor intent, scope, and boundaries.

    \$100-method allocations placed \emph{Up-skilling} and \emph{AI Strategy Design} at the top, while human-centric safeguards were comparatively underfunded. The framework aims to address this skew by embedding governance and responsible-use mechanisms as first-class elements rather than afterthoughts. Scott–Knott ESD grouped the dimensions into five clusters; \emph{Up-skilling} and \emph{AI Strategy Design} formed the highest-priority cluster, while \emph{Organizational Dynamics} ranked lowest. These clusters offer a preliminary empirical basis for sequencing when resources are constrained.
    
    Early feasibility check (two workshop sessions, four participants) viewed the steps as clear but uneven in immediate applicability; only a subset was unanimously actionable. Although the sample is small, this pattern mirrors arguments to ground interventions in practitioners’ lived experiences~\cite{ivanof2017theoryGapAndPractice}. Follow-up studies should identify recurring patterns and anti-patterns and test minimally viable interventions in situ.
    
    Overall, the framework shows strong perceived usefulness as a guide for early AI integration. Feasibility appears to be high overall, as the recommended steps are generally implementable, though some DPs—such as securing political sponsorship and establishing effective communication channels—remain contested; however, we consider them as boundary conditions rather than inherent weaknesses. In early phases, investments in \emph{Strategy Design} and \emph{Up-skilling} appear to drive perceived usefulness more than investments in \emph{Governance and Ethics}. To reduce over-reliance risks without depressing adoption, we recommend skill verification and dedicated oversight roles that institute lightweight checks while preserving delivery pace.
    
\end{itemize}

\subsection{Threats to Validity}

This section illustrates the threats to validity of the study, according to Wohlin et al.~\cite{wohlin2012experimentation}, and the way we mitigated them.

\subsubsection{Construct Validity}

Several risks to construct validity arose from conducting the research methodology. The systematic mapping of RQ\textsubscript{1} relied on the interpretive judgment of the first author, which could potentially introduce interpretability bias in the process. To mitigate this, we followed a systematic mapping process and discussed ambiguous mappings.

Another risk to construct validity was the reuse of an existing qualitative dataset to refine the proposed framework, informing a new contribution with relatively old data. For this reason, having the data collected at the end of 2024 and from organizations at the beginning of their AI integration strategy, we emphasize the early-stage focus of the framework. \revised{Specifically, the framework particularly addresses early-stage AI-driven transformation, as the included organizations were at the beginning of their AI integration journey. While the dimensions themselves may remain relevant across adoption, the specific design principles captured here, such as establishing communication channels with stakeholders and across the organization, reflect early adoption priorities and may give way to different principles in later phases. Similarly, actions related to anchoring change, such as sustaining adoption and managing long-term organizational adaptation, are not yet represented, constituting a possible direction for future work.}

Although the primary analysis was performed by the first author, two authors reviewed the final themes and the coding process. In this way, we tried to minimize any bias in codes and themes recognition. A remaining risk is that the original data collection had a different research orientation and was not designed to elicit all aspects of the current framework.

Additionally, the design and execution of the quantitative data collection via the survey and workshops constitutes a risk to construct validity, as biases like question framing may occur. To mitigate this, we followed established methodologies, while combining qualitative and quantitative methods enhanced findings' reliability. Moreover, social desirability bias posed a challenge to construct validity, as participants might have given socially acceptable responses. During the process, we emphasized the confidentiality of participants' responses to encourage honesty. To address potential cultural bias of the sample, we included participants from varied national and cultural backgrounds.

\subsubsection{External Validity}

External validity is constrained by the characteristics of the sample. Specifically, the SE practitioners that participated in this work form a relatively small sample, representing organizations from a narrow domain in early stages of AI adoption. Mature, large-scale AI-native organizations may face different challenges not fully captured in this study. Cultural and regulatory differences across regions may also affect generalizability, although we attempted to mitigate this through geographically diverse sampling. 

The rapid evolution of AI technologies poses a time-related external validity risk. Accordingly, parts of the framework may become outdated as tooling, regulation, and organizational practices evolve. To mitigate this, the framework focuses on early stage AI integration and emphasizes structural and behavioral change mechanisms rather than proposing a holistic and tool-specific recommendation. Still, SE practitioners should adapt the framework with caution, considering domain-specific challenges, organizational maturity, and evolving AI regulations.

\subsubsection{Internal Validity}

Internal validity threats primarily refer to bias in data collection. Since few companies have reached a mature stage within their AI incorporation strategies, selection bias during the identification of survey and workshop participants poses a potential threat to internal validity. In order to capture the complexity of AI Transformation and address this risk, we included organizations of various domains and participants with diverse backgrounds.

Selection bias may also have affected our survey sample, as respondents could disproportionately represent early adopters or individuals with strong interest in AI, potentially overstating the perceived importance of personal career development factors—such as Up-skilling—and not organizational necessities.

Furthermore, AI technologies and their adoption process is a research area with a lot of concern and misconceptions. This state further complicates the process of collecting reliable data to deduce major and realistic conclusions. By combining qualitative and quantitative methods, we aimed to enhance the reliability of the final outcome, namely the actionable BSE-informed framework.

Finally, during the mapping process between BSE challenges and change management models' steps we did not follow a formal inter-rater reliability process. This absence constitutes a limitation, and disagreements were instead resolved through consensus discussions.

\section{Conclusions}
\label{sec6}

This study employed a mixed-methods approach to investigate human-centered AI integration. Building on our earlier work identifying BSE-related challenges in AI adoption~\cite{tavantzis2024HumanCenteredAI}, this study proposes a BSE-informed framework to support early-stage AI incorporation in SE-heavy contexts. Findings show that while practitioners recognize the importance of human and ethical dimensions, they tend to prioritize procedural activities, like strategy design, over governance and ethics. This indicates a gap between the recognized importance of these human-centered principles and their practical adoption in workflows.

Our contributions include (i) a practical framework tailored to early-stage AI adoption in SE contexts, (ii) a baseline for developing a more complete, end-to-end AI integration model, and (iii) directions for future research.

We position this framework as going beyond classic change models by embedding human-centered and ethical dimensions rooted in BSE theory, which are often underrepresented in traditional approaches. Given these findings and framework are grounded in early-stage, SE-dominant settings, future work should include further evaluation of the framework across diverse industries to enhance its generalizability. Although experts challenged certain steps, the present work does not modify the framework solely based on this limited input (N=4). Instead, we identify this divergence as an area for future research to explore how classical change steps translate into AI-driven contexts.

Similarly, applying Action Research may help uncover adoption patterns and anti-patterns in practice. Moreover, exploring hierarchical groupings and prioritization strategies could improve usability. Finally, a longitudinal study would support evolving the framework into a scalable, end-to-end solution.

\revised{\section*{Data Availability}

More information about the data extracted from the literature review, the thematic analysis process, and the survey protocol is available in our Figshare replication package (\url{https://doi.org/10.6084/m9.figshare.32051676}).}

\section*{Generative AI Use Statement}

ChatGPT and Deep-L Write\footnote{https://www.deepl.com/en/write} were utilized to improve the phrasing of some parts of the text, originally written by the authors.

\section*{Authors' Contributions}

Theocharis Tavantzis led the execution of the study, including data collection, data analysis, and reporting of the findings. He additionally contributed to the research design.

Stefano Lambiase complemented the data collection and analysis activities, and contributed to writing and revising the manuscript.

Daniel Russo contributed to the data collection during the design and distribution of the survey and actively participated in the manuscript review.

Robert Feldt was primarily responsible for the research design and the organization of the quantitative data collection process, and also reviewed the manuscript.

All authors reviewed and approved the final version of the manuscript prior to submission.

\bibliographystyle{elsarticle-num}
\bibliography{references}

@article{terragni2025futureAi-drivenSE,
    author = {Terragni, Valerio and Vella, Annie and Roop, Partha and Blincoe, Kelly},
    title = {{The Future of AI-Driven Software Engineering}},
    year = {2025},
    issue_date = {June 2025},
    publisher = {Association for Computing Machinery},
    address = {New York, NY, USA},
    volume = {34},
    number = {5},
    issn = {1049-331X},
    doi = {10.1145/3715003},
    journal = {ACM Trans. Softw. Eng. Methodol.},
    month = may,
    articleno = {120},
    numpages = {20}
}

@inproceedings{du2024evaluatingLLMs,
    author = {Du, Xueying and Liu, Mingwei and Wang, Kaixin and Wang, Hanlin and Liu, Junwei and Chen, Yixuan and Feng, Jiayi and Sha, Chaofeng and Peng, Xin and Lou, Yiling},
    title = {{Evaluating Large Language Models in Class-Level Code Generation}},
    year = {2024},
    isbn = {9798400702174},
    publisher = {Association for Computing Machinery},
    address = {New York, NY, USA},
    doi = {10.1145/3597503.3639219},
    booktitle = {Proceedings of the IEEE/ACM 46th International Conference on Software Engineering},
    articleno = {81},
    numpages = {13},
    location = {Lisbon, Portugal},
    series = {ICSE '24}
}

@misc{hou2024largelanguagemodelssoftware,
      title={{Large Language Models for Software Engineering: A Systematic Literature Review}}, 
      author={Xinyi Hou and Yanjie Zhao and Yue Liu and Zhou Yang and Kailong Wang and Li Li and Xiapu Luo and David Lo and John Grundy and Haoyu Wang},
      year={2024},
      eprint={2308.10620},
      archivePrefix={arXiv},
      primaryClass={cs.SE}
}

@inproceedings{ronanki2023chatGPT1RE,
    author = { Ronanki, Krishna and Berger, Christian and Horkoff, Jennifer },
    booktitle = { 2023 49th Euromicro Conference on Software Engineering and Advanced Applications (SEAA) },
    title = {{Investigating ChatGPT's Potential to Assist in Requirements Elicitation Processes}},
    year = {2023},
    volume = {},
    ISSN = {},
    pages = {354-361},
    doi = {10.1109/SEAA60479.2023.00061},
    publisher = {IEEE Computer Society},
    address = {Los Alamitos, CA, USA},
    month =sep
}

@article{xiao2024genAI4pullRequests,
    author = {Xiao, Tao and Hata, Hideaki and Treude, Christoph and Matsumoto, Kenichi},
    title = {{Generative AI for Pull Request Descriptions: Adoption, Impact, and Developer Interventions}},
    year = {2024},
    issue_date = {July 2024},
    publisher = {Association for Computing Machinery},
    address = {New York, NY, USA},
    volume = {1},
    number = {FSE},
    doi = {10.1145/3643773},
    journal = {Proc. ACM Softw. Eng.},
    month = jul,
    articleno = {47},
    numpages = {23}
}

@article{gupta2024genAIsystematicReview,
    title = {{Generative AI: A systematic review using topic modelling techniques}},
    journal = {Data and Information Management},
    volume = {8},
    number = {2},
    pages = {100066},
    year = {2024},
    note = {Systematic Review and Meta-analysis in Information Management Research - Part II},
    issn = {2543-9251},
    doi = {https://doi.org/10.1016/j.dim.2024.100066},
    author = {Priyanka Gupta and Bosheng Ding and Chong Guan and Ding Ding}
}

@inproceedings{fan2023llmforSE,
  author    =   {Fan, Angela and Gokkaya, Beliz and Harman, Mark and Lyubarskiy, Mitya and Sengupta, Shubho and Yoo, Shin and Zhang, Jie M.},
  booktitle =   {2023 IEEE/ACM International Conference on Software Engineering: Future of Software Engineering (ICSE-FoSE)}, 
  title     =   {{Large Language Models for Software Engineering: Survey and Open Problems}}, 
  year      =   {2023},
  pages     =   {31-53},
  doi       =   {10.1109/ICSE-FoSE59343.2023.00008}
}

@article{ozkaya2023nextFrontier,
  author    =   {Ozkaya, Ipek},
  journal   =   {IEEE Software}, 
  title     =   {{The Next Frontier in Software Development: AI-Augmented Software Development Processes}}, 
  year      = {2023},
  volume    =   {40},
  number    =   {4},
  pages     =   {4-9},
  doi={10.1109/MS.2023.3278056}
}

@article{russo2024cphManifesto,
    title = {{Generative AI in Software Engineering Must Be Human-Centered: The Copenhagen Manifesto}},
    journal = {Journal of Systems and Software},
    volume = {216},
    pages = {112115},
    year = {2024},
    issn = {0164-1212},
    doi = {https://doi.org/10.1016/j.jss.2024.112115},
    author = {Daniel Russo et al.}
}

@article{shneiderman2020human_centredAI,
    author = {Ben Shneiderman},
    title = {{Human-Centered Artificial Intelligence: Reliable, Safe \& Trustworthy}},
    journal = {International Journal of Human--Computer Interaction},
    volume = {36},
    number = {6},
    pages = {495--504},
    year = {2020},
    publisher = {Taylor \& Francis},
    doi = {10.1080/10447318.2020.1741118}
}

@inproceedings{yoon2024intent,
  title={{Intent-driven mobile gui testing with autonomous large language model agents}},
  author={Yoon, Juyeon and Feldt, Robert and Yoo, Shin},
  booktitle={2024 IEEE Conference on Software Testing, Verification and Validation (ICST)},
  pages={129--139},
  year={2024},
  organization={IEEE}
}

@article{capretz_humanSE_2014,
    author = {Capretz, L. F.},
    title = {{Bringing the Human Factor to Software Engineering}},
    journal = {IEEE Software},
    year = {2014},
    volume = {31},
    number = {2},
    doi = {10.1109/MS.2014.30}
}

@article{perry_people_process_improvement_1994,
    author = {Perry, Dewayne E. and Staudenmayer, Nancy A. and Votta, Lawrence G.},
    title = {{People, Organizations, and Process Improvement}},
    journal = {IEEE Software},
    year = {1994},
    volume = {11},
    number = {4},
    pages = {36--45}
}

@article{storey2020SociotechnicalFramework,
    author = {Storey, Margaret-Anne and Ernst, Neil A. and Williams, Courtney and Kalliamvakou, Eirini},
    title = {{The who, what, how of software engineering research: a socio-technical framework}},
    year = {2020},
    issue_date = {Sep 2020},
    publisher = {Kluwer Academic Publishers},
    address = {USA},
    volume = {25},
    number = {5},
    issn = {1382-3256},
    doi = {10.1007/s10664-020-09858-z},
    journal = {Empirical Softw. Engg.},
    pages = {4097--4129},
    numpages = {33},
}

@article{russo2024navigatingComplexityOfgenAI,
    author = {Russo, Daniel},
    title = {{Navigating the Complexity of Generative AI Adoption in Software Engineering}},
    year = {2024},
    issue_date = {June 2024},
    publisher = {Association for Computing Machinery},
    address = {New York, NY, USA},
    volume = {33},
    number = {5},
    issn = {1049-331X},
    doi = {10.1145/3652154},
    journal = {ACM Trans. Softw. Eng. Methodol.},
    month = jun,
    articleno = {135},
    numpages = {50}
}

@inproceedings{choudhuri2025trustAndBehaviorToGenAI,
    author = {Choudhuri, Rudrajit and Trinkenreich, Bianca and Pandita, Rahul and Kalliamvakou, Eirini and Steinmacher, Igor and Gerosa, Marco and Sanchez, Christopher and Sarma, Anita},
    title = {{What Guides Our Choices? Modeling Developers' Trust and Behavioral Intentions Towards GenAI}},
    year = {2025},
    isbn = {9798331505691},
    publisher = {IEEE Press},
    doi = {10.1109/ICSE55347.2025.00087},
    booktitle = {Proceedings of the IEEE/ACM 47th International Conference on Software Engineering},
    pages = {1691--1703},
    numpages = {13},
    location = {Ottawa, Ontario, Canada},
    series = {ICSE '25}
}

@article{hoda2021socio,
  title={Socio-technical grounded theory for software engineering},
  author={Hoda, Rashina},
  journal={IEEE Transactions on Software Engineering},
  volume={48},
  number={10},
  pages={3808--3832},
  year={2021},
  publisher={IEEE}
}

@inproceedings{lenberg_bse_2014,
    author = {Lenberg, Per and Feldt, Robert and Wallgren, Lars G\"oran},
    title = {{Towards a Behavioral Software Engineering}},
    booktitle = {Proceedings of the 7th International Workshop on Cooperative and Human Aspects of Software Engineering},
    year = {2014},
    pages = {48--55},
    publisher = {ACM}
}

@article{lenberg_bse_slr_2015,
    author = {Lenberg, Per and Feldt, Robert and Wallgren, Lars G\"oran},
    title = {{Behavioral Software Engineering: A Definition and Systematic Literature Review}},
    journal = {Journal of Systems and Software},
    year = {2015}
}

@inproceedings{tavantzis2024HumanCenteredAI,
    author = { Tavantzis, Theocharis and Feldt, Robert },
    booktitle = { 2025 IEEE/ACM 18th International Conference on Cooperative and Human Aspects of Software Engineering (CHASE) },
    title = {{ Unpacking Organizational Change in AI Transformations of Software Engineering }},
    year = {2025},
    volume = {},
    ISSN = {},
    pages = {149-160},
    doi = {10.1109/CHASE66643.2025.00026},
    publisher = {IEEE Computer Society},
    address = {Los Alamitos, CA, USA},
    month =apr
}

@techreport{mit2025GenAIDivide,
  title        = {{State of AI in Business 2025: The GenAI Divide}},
  author       = {Challapally, Aditya and Pease, Chris and Raskar, Ramesh and Chari, Pradyumna},
  institution  = {MIT Project NANDA},
  year         = {2025},
  month        = {July},
  note         = {Preliminary Findings from AI Implementation Research by Project NANDA},
  url          = {https://iceberg.mit.edu/index.html}
}

@techreport{dellAcqua2023jaggedFrontier,
  author       = {Dell'Acqua, Fabrizio and McFowland III, Edward and Mollick, Ethan R. and Lifshitz-Assaf, Hila and Kellogg, Katherine and Rajendran, Saran and Krayer, Lisa and Candelon, Fran\c{c}ois and Lakhani, Karim R.},
  title        = {{Navigating the Jagged Technological Frontier: Field Experimental Evidence of the Effects of AI on Knowledge Worker Productivity and Quality}},
  institution  = {Harvard Business School Technology \& Operations Management Unit / The Wharton School},
  type         = {Working Paper},
  number       = {24-013},
  year         = {2023},
  doi          = {10.2139/ssrn.4573321},
  note         = {Available at SSRN},
  month        = sep,
  date         = {2023-09-15}
}

@book{kotter_leading_change_2012,
    author      = {Kotter, John P.},
    title       = {{Leading Change}},
    year        = {2012},
    publisher   = {Harvard Business Press},
    address     = {USA}
}

@article{lenberg_organizational_change_2017,
    author      = {Lenberg, Per and Wallgren Tengberg, Lars G\"oran and Feldt, Robert},
    title       = {{An Initial Analysis of Software Engineers' Attitudes Towards Organizational Change}},
    journal     = {Empirical Software Engineering},
    year        = {2017},
    volume      = {22},
    pages       = {2179--2205},
    doi         = {10.1007/s10664-016-9482-0}
}

@inproceedings{lenberg_human_factors_2015,
    author      = {Lenberg, Per and Feldt, Robert and Wallgren, Lars G\"oran},
    title       = {{Human Factors Related Challenges in Software Engineering -- An Industrial Perspective}},
    booktitle   = {2015 IEEE/ACM 8th International Workshop on Cooperative and Human Aspects of Software Engineering (CHASE)},
    year        = {2015},
    pages       = {43--49},
    publisher   = {IEEE/ACM}
}

@article{peretz_andersson_empirical_2022,
    author = {Peretz-Andersson, Einav and Torkar, Richard},
    title  = {{Empirical AI Transformation Research: A Systematic Mapping Study and Future Agenda}},
    journal = {e-Informatica Software Engineering Journal},
    year = {2022},
    volume = {16},
    number = {1},
    pages = {220108},
    doi = {10.37190/e-Inf220108}
}

@article{beer_walton_organization_1987,
    author = {Beer, Michael and Walton, Anna Elise},
    title = {{Organization Change and Development}},
    journal = {Annual Review of Psychology},
    year = {1987},
    volume = {38},
    number = {1},
    pages = {339--367}
}

@article{errida2021DescriptiveOrgModel,
  author    = {Errida, Abdelouahab and Lotfi, Bouchra},
  title     = {{The determinants of organizational change management success: Literature review and case study}},
  journal   = {International Journal of Engineering Business Management},
  year      = {2021},
  volume    = {13},
  doi       = {10.1177/18479790211016273}
}

@book{weinberg_psychology_1971,
    author = {Weinberg, Gerald M.},
    title = {{The Psychology of Computer Programming}},
    year = {1971},
    volume = {932633420},
    publisher = {Van Nostrand Reinhold},
    address = {New York}
}

@article{ferreira_spi_2011,
	author = {Ferreira, Mar\'ilia Guterres and Wazlawick, Raul Sidnei},
	title = {{Software process improvement: A organizational change that need to be managed and motivated}},
	year = {2011},
	journal = {World Academy of Science, Engineering and Technology},
	volume = {50},
	pages = {296 -- 304},
}

@inproceedings{graziotin2015UnderstandingAffect,
    author = {Graziotin, Daniel and Wang, Xiaofeng and Abrahamsson, Pekka},
    title = {{Understanding the affect of developers: theoretical background and guidelines for psychoempirical software engineering}},
    year = {2015},
    isbn = {9781450338189},
    publisher = {Association for Computing Machinery},
    address = {New York, NY, USA},
    doi = {10.1145/2804381.2804386},
    pages = {25--32},
    numpages = {8},
}

@article{graziotin2022PsychometricsinBSE,
  author = {Graziotin, Daniel and Lenberg, Per and Feldt, Robert and Wagner, Stefan},
  title = {{Psychometrics in Behavioral Software Engineering: A Methodological Introduction with Guidelines}},
  journal = {ACM Transactions on Software Engineering and Methodology},
  year = {2022},
  volume = {31},
  number = {1},
  pages = {1--36},
  doi = {10.1145/3469888}
}

@article{zolduoarrati2024SecondaryStudies,
    title = {{Secondary studies on human aspects in software engineering: A tertiary study}},
    journal = {Journal of Systems and Software},
    volume = {200},
    pages = {111654},
    year = {2023},
    issn = {0164-1212},
    doi = {https://doi.org/10.1016/j.jss.2023.111654},
    author = {Elijah Zolduoarrati and Sherlock A. Licorish and Nigel Stanger}
}

@inproceedings{understanding_org_change_2023,
    author = {Lima, Rayfran and Albuquerque, Luis and Ayres, Marcelo and Silva, Suelen and Oran, Ana and Fran\c{c}a, C\'{e}sar},
    title = {{Understanding an organizational change and development intervention applied in a Global Software Industry: A case study: A Case Study}},
    year = {2023},
    isbn = {9798400707872},
    publisher = {Association for Computing Machinery},
    address = {New York, NY, USA},
    doi = {10.1145/3613372.3613388},
    booktitle = {Proceedings of the XXXVII Brazilian Symposium on Software Engineering},
    pages = {164--173},
    numpages = {10},
    location = {Campo Grande, Brazil},
    series = {SBES '23}
}

@standard{iso15288,
  title        = {{ISO/IEC/IEEE 15288:2023 Systems and software engineering --- System life cycle processes}},
  author       = {{International Organization for Standardization} and {International Electrotechnical Commission} and {Institute of Electrical and Electronics Engineers}},
  year         = {2023},
  institution  = {{ISO/IEC/IEEE}},
  address      = {Geneva, Switzerland},
  type         = {International Standard},
  number       = {ISO/IEC/IEEE 15288:2023}
}

@inproceedings{klotins2022continuous,
  author       = {Klotins, Eriks and Gorschek, Tony},
  title        = {{Continuous Software Engineering in the Wild}},
  booktitle    = {Software Quality: The Next Big Thing in Software Engineering and Quality},
  editor       = {Mendez, Daniel and Wimmer, Manuel and Winkler, Dietmar and Biffl, Stefan and Bergsmann, Judith},
  series       = {Lecture Notes in Business Information Processing},
  volume       = {439},
  year         = {2022},
  publisher    = {Springer, Cham},
  doi          = {10.1007/978-3-031-04115-0_1}
}

@article{parry2024EmpiricalOrgChange,
    author = {Warren Parry, Christina Kirsch, Paul Carey and Doug Shaw},
    title = {{Empirical Development of a Model of Performance Drivers in Organizational Change Projects}},
    journal = {Journal of Change Management},
    volume = {14},
    number = {1},
    pages = {99--125},
    year = {2014},
    publisher = {Routledge},
    doi = {10.1080/14697017.2012.745894},
}

@article{appelbaum2012back,
  author        = {Appelbaum, Steven H. and Habashy, Sally and Malo, Jean-Luc and Shafiq, Hisham},
  title         = {{Back to the future: revisiting Kotter's 1996 change model}},
  journal       = {Journal of Management Development},
  volume        = {31},
  number        = {8},
  pages         = {764--782},
  year          = {2012},
  doi           = {10.1108/02621711211253231}
}

@article{umarji2005predictingAcceptanceOfSPI,
    author = {Umarji, Medha and Seaman, Carolyn},
    title = {{Predicting acceptance of Software Process Improvement}},
    year = {2005},
    issue_date = {July 2005},
    publisher = {Association for Computing Machinery},
    address = {New York, NY, USA},
    volume = {30},
    number = {4},
    issn = {0163-5948},
    doi = {10.1145/1082983.1083121},
    journal = {SIGSOFT Softw. Eng. Notes},
    month = may,
    pages = {1--6},
    numpages = {6}
}

@inproceedings{davila2024AIbasedAssistants,
    author = {Davila, Nicole and Wiese, Igor and Steinmacher, Igor and Lucio da Silva, Lucas and Kawamoto, Andre and Favaro, Gilson Jose Peres and Nunes, Ingrid},
    title = {{An Industry Case Study on Adoption of AI-based Programming Assistants}},
    year = {2024},
    isbn = {9798400705014},
    publisher = {Association for Computing Machinery},
    address = {New York, NY, USA},
    doi = {10.1145/3639477.3643648},
    booktitle = {Proceedings of the 46th International Conference on Software Engineering: Software Engineering in Practice},
    pages = {92--102},
    numpages = {11},
    location = {Lisbon, Portugal},
    series = {ICSE-SEIP '24}
}

@inproceedings{bashir2024AI-centricRE,
    author = {Bashir, Sarmad},
    title = {{Towards AI-centric Requirements Engineering for Industrial Systems}},
    year = {2024},
    isbn = {9798400705021},
    publisher = {Association for Computing Machinery},
    address = {New York, NY, USA},
    doi = {10.1145/3639478.3639811},
    booktitle = {Proceedings of the 2024 IEEE/ACM 46th International Conference on Software Engineering: Companion Proceedings},
    pages = {242--246},
    numpages = {5},
    location = {Lisbon, Portugal},
    series = {ICSE-Companion '24}
}

@article{ajiga2024enhancingSEpractices,
    title={{Enhancing software development practices with AI insights in high-tech companies}},
    volume={5}, 
    DOI={10.51594/csitrj.v5i8.1450},
    number={8},
    journal={Computer Science \&; IT Research Journal},
    author={Daniel Ajiga and Patrick Azuka Okeleke and Samuel Olaoluwa Folorunsho and Chinedu Ezeigweneme},
    year={2024},
    month={Aug.},
    pages={1897-1919}
}

@article{nortje2020AFF,
  title={{A Framework for the Implementation of Artificial Intelligence in Business Enterprises: A Readiness Model}},
  author={M. A. Nortje and Sara S (Saartjie) Grobbelaar},
  journal={2020 IEEE International Conference on Engineering, Technology and Innovation (ICE/ITMC)},
  year={2020},
  pages={1-10},
  url={https://api.semanticscholar.org/CorpusID:221848830}
}

@inproceedings{bandara2025aiReasinessForPM,
  author={Bandara, R.B.M.T. and Wickramarachchi, Ruwan},
  booktitle={2025 5th International Conference on Advanced Research in Computing (ICARC)}, 
  title={{Developing an AI Readiness Model for Software Project Management: A Thematic Analysis}}, 
  year={2025},
  volume={},
  number={},
  pages={1-6},
  doi={10.1109/ICARC64760.2025.10962923}}

@article{uren2023AIReadiness,
  author  = {Uren, Victoria and Edwards, John S.},
  title   = {{Technology readiness and the organizational journey towards AI adoption: An empirical study}},
  journal = {International Journal of Information Management},
  volume  = {68},
  year    = {2023},
  pages   = {102588},
  issn    = {0268-4012},
  doi     = {10.1016/j.ijinfomgt.2022.102588}
}

@article{lavin2022TRLML,
  author  = {Lavin, Alexander and Gilligan-Lee, C. M. and Visnjic, Aleksandar and others},
  title   = {{Technology readiness levels for machine learning systems}},
  journal = {Nature Communications},
  volume  = {13},
  number  = {1},
  pages   = {6039},
  year    = {2022},
  doi     = {10.1038/s41467-022-33128-9},
}

@article{johnk2021AIcomes_AIorgReadiness,
  author = {J\"ohnk, Jan and Weißert, Martin and Wyrtki, Kristin},
  title = {{Ready or Not, AI Comes--- An Interview Study of Organizational AI Readiness Factors}},
  journal = {{Business \& Information Systems Engineering}},
  volume = {63},
  pages = {5--20},
  year = {2021},
  doi = {10.1007/s12599-020-00676-7}
}

@article{holmstrom2022AItoDigitalTransformation,
    title = {{From AI to digital transformation: The AI readiness framework}},
    journal = {Business Horizons},
    volume = {65},
    number = {3},
    pages = {329-339},
    year = {2022},
    issn = {0007-6813},
    doi = {https://doi.org/10.1016/j.bushor.2021.03.006},
    author = {Jonny Holmstr\"om},
}

@article{dwivedi_AI_2021,
    title = {{Artificial Intelligence (AI): Multidisciplinary perspectives on emerging challenges, opportunities, and agenda for research, practice and policy}},
    journal = {International Journal of Information Management},
    volume = {57},
    pages = {101994},
    year = {2021},
    issn = {0268-4012},
    doi = {https://doi.org/10.1016/j.ijinfomgt.2019.08.002},
    author = {Yogesh K. Dwivedi et al.}
}

@inproceedings{dolata2024freelancersGenerativeAI,
    author = {Dolata, Mateusz and Lange, Norbert and Schwabe, Gerhard},
    title = {{Development in times of hype: How freelancers explore Generative AI?}},
    year = {2024},
    isbn = {9798400702174},
    publisher = {Association for Computing Machinery},
    address = {New York, NY, USA},
    doi = {10.1145/3597503.3639111},
    booktitle = {Proceedings of the IEEE/ACM 46th International Conference on Software Engineering},
    articleno = {183},
    numpages = {13},
    location = {Lisbon, Portugal},
    series = {ICSE '24}
}

@article{polyportis2024navigating,
  author    = {Polyportis, A. and Pahos, N.},
  title     = {{Navigating the perils of Artificial Intelligence: a focused review on ChatGPT and responsible research and innovation}},
  journal   = {Humanities and Social Sciences Communications},
  volume    = {11},
  pages     = {107},
  year      = {2024},
  doi       = {10.1057/s41599-023-02464-6},
}

@book{creswell2017ResearchDesign,
  author    = {J. W. Creswell and J. D. Creswell},
  title     = {{Research Design: Qualitative, Quantitative, and Mixed Methods Approaches}},
  publisher = {Sage Publications},
  year      = {2017},
}

@inproceedings{ivanof2017theoryGapAndPractice,
    author = {Ivanov, Vladimir and Rogers, Alan and Succi, Giancarlo and Yi, Jooyong and Zorin, Vasilii},
    title = {What do software engineers care about? gaps between research and practice},
    year = {2017},
    isbn = {9781450351058},
    publisher = {Association for Computing Machinery},
    address = {New York, NY, USA},
    doi = {10.1145/3106237.3117778},
    booktitle = {Proceedings of the 2017 11th Joint Meeting on Foundations of Software Engineering},
    pages = {890--895},
    numpages = {6},
    location = {Paderborn, Germany},
    series = {ESEC/FSE 2017}
}

@article {moher2009PRISMA,
	author = {Moher, David and Liberati, Alessandro and Tetzlaff, Jennifer and Altman, Douglas G},
	title = {Preferred reporting items for systematic reviews and meta-analyses: the PRISMA statement},
	volume = {339},
	year = {2009},
	doi = {10.1136/bmj.b2535},
	publisher = {BMJ Publishing Group Ltd},
	journal = {BMJ}
}

@techreport{kasunic2005designingSurvey,
  author    = {M. Kasunic},
  title     = {{Designing an Effective Survey}},
  institution = {Software Engineering Institute, Carnegie Mellon University},
  address   = {Pittsburgh, PA, USA},
  year      = {2005},
  type      = {Technical Report}
}

@article{ghazi2019surveyResearchInSE,
  author={Ghazi, Ahmad Nauman and Petersen, Kai and Reddy, Sri Sai Vijay Raj and Nekkanti, Harini},
  journal={IEEE Access}, 
  title={{Survey Research in Software Engineering: Problems and Mitigation Strategies}}, 
  year={2019},
  volume={7},
  number={},
  pages={24703-24718},
  doi={10.1109/ACCESS.2018.2881041}
}

@inproceedings{molleri2016surveyGuidelines,
    author = {Molleri, Jefferson and Petersen, Kai and Mendes, Emilia},
    year = {2016},
    month = {09},
    pages = {},
    title = {{Survey Guidelines in Software Engineering: An Annotated Review}},
    doi = {10.1145/2961111.2962619}
}

@article{braun2006thematicAnalysis,
    author = {Virginia Braun and Victoria Clarke},
    title = {Using thematic analysis in psychology},
    journal = {Qualitative Research in Psychology},
    volume = {3},
    number = {2},
    pages = {77--101},
    year = {2006},
    publisher = {Routledge},
    doi = {10.1191/1478088706qp063oa}
}

@article{jelihovschi2014scottKnott,
    title={{ScottKnott: a package for performing the Scott-Knott clustering algorithm in R}},
    volume={15},
    ISSN={2179-8451},
    DOI={10.5540/tema.2014.015.01.0003},
    number={1},
    journal={TEMA (S\~ao Carlos)},
    publisher={Sociedade Brasileira de Matem\'atica Aplicada e Computacional},
    author={Jelihovschi, E.G. and Faria, J.C. and Allaman, I.B.},
    year={2014},
    month={Jan},
    pages={3--17}
}

@inproceedings{pulido2024changeManagementInDigitalTransformation,
    author="Pulido, Frida Lizbeth Ponce
    and Taherdoost, Hamed",
    editor="Moldovan, Liviu
    and Gligor, Adrian",
    title={{Employment of Change Management Models in the Digital Transformation Process}},
    booktitle="The 17th International Conference Interdisciplinarity in Engineering",
    year="2024",
    publisher="Springer Nature Switzerland",
    address="Cham",
    pages="392--405",
    isbn="978-3-031-54671-6"
}

@inproceedings{abdallah2016adkarModel,
  title={{Critical Thinking \& Lifelong Learning: An ADKAR Model-Based Framework for Managing a Change in Thinking \& English Language Learning Styles at the Secondary Stage.}},
  author={Mahmoud Mohammad Sayed Abdallah and Marwa Mohammad},
  year={2016},
  url={https://api.semanticscholar.org/CorpusID:147463015}
}

@article{kocaoglu2019mcKinsey7sModel,
    author = {Kocao\u{g}lu, Batuhan and Demir, Ezgi},
    year = {2019},
    month = {07},
    pages = {},
    title = {{The use of McKinsey's 7S framework as a strategic planning and economic assestment tool in the process of digital transformation}},
    volume = {9},
    journal = {Pressacademia},
    doi = {10.17261/Pressacademia.2019.1078}
}

@inproceedings{merino_barnacho2025understandingChangeManagement,
    author="Merino-Barbancho, Beatriz
    and Fico, Giuseppe",
    editor="Duffy, Vincent G.",
    title={{Understanding Change Management in Hospital Context with the Integration of Digital Technologies}},
    booktitle="HCI International 2024 -- Late Breaking Papers",
    year="2025",
    publisher="Springer Nature Switzerland",
    address="Cham",
    pages="144--153",
    isbn="978-3-031-76809-5"
}

@article{mukheli2023changeManagementInSouthAfrica,
    author = {Mukheli, Phatutshedzo and Naidoo, Vinessa},
    year = {2023},
    month = {11},
    pages = {294-304},
    title = {{Change Management in a South African Municipality: Can it Work?}},
    volume = {19},
    journal = {European Conference on Management Leadership and Governance},
    doi = {10.34190/ecmlg.19.1.1960}
}

@article{phillips2023changeManagementfromTheory,
  author    = {J. Phillips and J. D. Klein},
  title     = {{Change Management: From Theory to Practice}},
  journal   = {TechTrends},
  volume    = {67},
  pages     = {189--197},
  year      = {2023},
  doi       = {10.1007/s11528-022-00775-0}
}

@online{vonderlinn2009cap,
  author    = {Von Der Linn, Bob},
  title     = {Overview of GE's Change Acceleration Process (CAP)},
  year      = {2009},
  month     = {January 25},
  url       = {https://bvonderlinn.wordpress.com/2009/01/25/overview-of-ges-change-acceleration-process-cap/},
  note      = {Retrieved January 29, 2025},
  journal   = {Bob Von Der Linn's Change Management and Human Performance Technology Blog}
}

@article{owad2023leanSixSigmaAndKotterframework,
    author = {Al Owad, Ali and Yadav, Neeraj and Kumar, Vimal and Swarnakar, Vikas and Kandasamy, Jayakrishna and Haridy, Salah and Yadav},
    year = {2023},
    month = {12},
    pages = {},
    title = {Integrated Lean Six Sigma and Kotter change management framework for emergency healthcare services in Saudi Arabia},
    volume = {32},
    journal = {Benchmarking An International Journal},
    doi = {10.1108/BIJ-05-2023-0335}
}

@article{kirmizi2022ERP,
    author = {Kirmizi, Mehmet and Kocao\u{g}lu, Batuhan},
    year = {2022},
    month = {02},
    pages = {1089-1113},
    title = {The influencing factors of enterprise resource planning (ERP) readiness stage on enterprise resource planning project success: a project manager's perspective},
    volume = {51},
    journal = {Kybernetes},
    doi = {10.1108/K-11-2020-0812}
}

@article{spyropoulou2021businessExcellenceinLargeOrganizations,
    author = {Spyropoulou, Theodora and Panas, Antonios and Pantouvakis, John-Paris},
    year = {2021},
    month = {11},
    pages = {1452-1460},
    title = {Formulation of Change Management Model for Achieving Business Excellence in Large Organizations},
    volume = {18},
    journal = {WSEAS TRANSACTIONS ON BUSINESS AND ECONOMICS},
    doi = {10.37394/23207.2021.18.133}
}

@article{wan2020UNchangeManagement,
  author = {Wan, J. and Saade, R. and Wang, L.},
  title = {Deriving significant factors for managing change in UN},
  journal = {Journal of Organizational Change Management},
  volume = {33},
  number = {1},
  pages = {114-126},
  year = {2020},
  doi = {10.1108/JOCM-10-2018-0288}
}

@inproceedings{rohman2020changeManagementModelforIS,
  author={Rohmah, Manzilatul and Subriadi, Apol Pribadi},
  booktitle={2020 International Conference on Smart Technology and Applications (ICoSTA)}, 
  title={A Change Management Model for Information Systems Implementation}, 
  year={2020},
  volume={},
  number={},
  pages={1-6},
  doi={10.1109/ICoSTA48221.2020.1570613999}
}

@conference{li2019ChangeManagementAntibiotic,
	author = {Li, Mu and Yodmongkol, Pitipong},
	title = {Change management model to improve antibiotic prescribing in a small Chinese hospital},
	year = {2019},
	journal = {ECTI DAMT-NCON 2019 - 4th International Conference on Digital Arts, Media and Technology and 2nd ECTI Northern Section Conference on Electrical, Electronics, Computer and Telecommunications Engineering},
	pages = {255 -- 260},
	doi = {10.1109/ECTI-NCON.2019.8692284},
	type = {Conference paper}
}

@article{hourani2019proposed7Emodel,
    author = {Hourani, Hussam and Abdallah, Mohammad and Tamimi, Abdelfatah},
    year = {2019},
    month = {10},
    pages = {941-947},
    title = {7E: A PROPOSED CHANGE MANAGEMENT MODEL INTEGRATED WITH SOFTWARE DEVELOPMENT LIFECYCLE},
    volume = {13},
    journal = {ICIC Express Letters},
    doi = {10.24507/icicel.13.10.941}
}

@inproceedings{sulistiyani2018changeManagementForeGov,
  author={Sulistiyani, Endang and Susanto, Tony Dwi},
  booktitle={2018 Third International Conference on Informatics and Computing (ICIC)}, 
  title={Change Management Methodology for e-Government Project in Developing Countries: A Conceptual Model}, 
  year={2018},
  pages={1-5},
  doi={10.1109/IAC.2018.8780500}
}

@inproceedings{narciso2014overcomingStructuralResistanceinSPI,
  author={Narciso, Henrique and Allison, I.},
  booktitle={2014 9th International Conference on the Quality of Information and Communications Technology}, 
  title={Overcoming Structural Resistance in SPI with Change Management}, 
  year={2014},
  pages={8-17},
  doi={10.1109/QUATIC.2014.9}
}

@article{worley2014isChangeManagementObsolete,
    title = {Is change management obsolete?},
    journal = {Organizational Dynamics},
    volume = {43},
    number = {3},
    pages = {214-224},
    year = {2014},
    issn = {0090-2616},
    doi = {https://doi.org/10.1016/j.orgdyn.2014.08.008},
    author = {Christopher G. Worley and Susan A. Mohrman},
}

@article{bolbori2013businessExcellence,
    author = {Bolboli, Amir and Reiche, Markus},
    year = {2013},
    month = {06},
    pages = {},
    title = {A model for sustainable business excellence: Implementation and the roadmap},
    volume = {25},
    journal = {The TQM Journal},
    doi = {10.1108/17542731311314845}
}

@article{toherhi2012nineBlunders,
    author = {Toterhi, Tim and Recardo, Ronald J.},
    title = {Managing change: Nine common blunders---and how to avoid them},
    journal = {Global Business and Organizational Excellence},
    volume = {31},
    number = {5},
    pages = {54-69},
    doi = {https://doi.org/10.1002/joe.21438},
    year = {2012}
}

@article{calvert2006change_management_model_for_ERP,
    author = {Calvert, Cheryl},
    year = {2006},
    month = {01},
    pages = {},
    title = {A Change-Management Model for the Implementation and Upgrade of ERP Systems},
    journal = {ACIS 2006 Proceedings - 17th Australasian Conference on Information Systems}
}

@article{rajamanoharan2006sixSigmaImplementation,
    author = {Rajamanoharan, Indra Devi and Collier, Paul},
    year = {2006},
    month = {02},
    pages = {},
    title = {Six Sigma Implementation, Organisational Change and the impact on performance measurement systems: Case study evidence},
    volume = {2},
    journal = {International Journal of Six Sigma and Competitive Advantage}
}

@article{leppitt2006challengingTheCodeOfChange,
    author = {Nigel Leppitt},
    title = {{Challenging the code of change: Part 2. Crossing the rubicon: Extending the integration of change}},
    journal = {Journal of Change Management},
    volume = {6},
    number = {3},
    pages = {235--256},
    year = {2006},
    publisher = {Routledge},
    doi = {10.1080/14697010600683153}
}

@article{bamford2005caseStudyOfChangeManagement,
    author = {David Bamford and Stephen Daniel},
    title = {{A case study of change management effectiveness within the NHS}},
    journal = {Journal of Change Management},
    volume = {5},
    number = {4},
    pages = {391--406},
    year = {2005},
    publisher = {Routledge},
    doi = {10.1080/14697010500287360}
}

@article{victor2002theFiveDimensionsOfChange,
    author = {Victor, Paul and Franckeiss, Anton},
    title = {{The five dimensions of change: an integrated approach to strategic organizational change management}},
    journal = {Strategic Change},
    volume = {11},
    number = {1},
    pages = {35-42},
    doi = {https://doi.org/10.1002/jsc.567},
    year = {2002}
}

@article{tampoe1990drivingOrgChange,
    title = {Driving organisational change through the effective use of multi-disciplinary project teams},
    journal = {European Management Journal},
    volume = {8},
    number = {3},
    pages = {346-354},
    year = {1990},
    issn = {0263-2373},
    doi = {https://doi.org/10.1016/0263-2373(90)90012-U},
    author = {Mahen Tampoe}
}

@book{hayes2014theTheoryAndPracticeOfChangeManagement,
  author    = {John Hayes},
  title     = {The Theory and Practice of Change Management},
  edition   = {4th},
  year      = {2014},
  publisher = {Palgrave Macmillan},
  address   = {Basingstoke, UK},
  isbn      = {978-1-137-27534-9}
}

@book{cummings2013organizationDevelopmentAndChange,
  author    = {Cummings, Thomas G. and Worley, Christopher G.},
  title     = {Organization Development and Change},
  edition   = {10th},
  year      = {2013},
  publisher = {South-Western College Publishing},
  address   = {Cincinnati, OH}
}

@article{burke1992aCausalModelOfOrgChange,
  author  = {Burke, W. Warner and Litwin, George H.},
  title   = {A Causal Model of Organizational Performance and Change},
  journal = {Journal of Management},
  year    = {1992},
  volume  = {8},
  pages   = {523--546}
}

@article{mento2002changeManagementProcess,
  author  = {Mento, Anthony J. and Jones, Raymond M. and Dirndorfer, Walter},
  title   = {A Change Management Process: Grounded in Both Theory and Practice},
  journal = {Journal of Organizational Change Management},
  year    = {2002},
  volume  = {3},
  pages   = {45--59}
}

@misc{jick1993implementingChange,
  author = {Jick, Todd},
  title = {Implementing Change},
  note = {Note 9-191-114},
  year = {1993},
  publisher = {Harvard Business School Press},
  address = {Boston}
}

@article{whelan-berry2010linkingChangeDrivers,
  author = {Wheelan-Berry, Karen S. and Somerville, Karen A.},
  title = {Linking Change Drivers and the Organizational Change Process: A Review and Synthesis},
  journal = {Journal of Change Management},
  year = {2010},
  volume = {10},
  number = {2},
  pages = {175--193},
  doi = {10.1080/14697011003795651}
}

@article{bullock1985reviewAndSynthesisOfOrgChange,
  author = {Bullock, R. J. and Batten, D.},
  title = {It's just a phase we're going through: a review and synthesis of OD phase analysis},
  journal = {Group \& Organization Studies},
  year = {1985},
  volume = {10},
  pages = {383--412}
}

@book{judson1991changingBehaviorInOrganizations,
  author = {Judson, A. S.},
  title = {Changing Behavior in Organizations: Minimizing Resistance to Change},
  edition = {2nd},
  year = {1991},
  publisher = {Blackwell Publishing},
  address = {Hoboken, NJ}
}

@book{kanter1992challengesOfOrgChange,
  author = {Kanter, R. M. and Stein, B. A. and Jick, T. D.},
  title = {The Challenge of Organizational Change: How Companies Experience It and Leaders Guide It},
  year = {1992},
  publisher = {Free Press},
  address = {New York}
}

@article{armenakis2007orgChangeRecipientsBeliefScale,
  author = {Armenakis, A. A. and Bernerth, J. B. and Pitts, J. P.},
  title = {Organizational Change Recipients' Beliefs Scale: Development of an Assessment Instrument},
  journal = {Journal of Applied Behavioral Science},
  year = {2007},
  volume = {43},
  pages = {481--505}
}

@article{nadler1980congruenceModel,
  author = {Nadler, D. A. and Tushman, M. L.},
  title = {A Model for Diagnosing Organizational Behavior},
  journal = {Organizational Dynamics},
  year = {1980},
  volume = {9},
  pages = {35--51}
}

@book{carnall2007managingChangeInOrganizations,
  author = {Carnall, C.},
  title = {Managing Change in Organizations},
  edition = {5th},
  year = {2007},
  publisher = {Prentice Hall: Financial Times}
}

@book{beckhard1987organizationalTransitions_ManagingComplexChange,
  author = {Beckhard, R. and Harris, R. T.},
  title = {Organizational Transitions: Managing Complex Change},
  edition = {2nd},
  year = {1987},
  publisher = {Addison-Wesley},
  address = {Reading, MA}
}

@incollection{knoster2000frameworkForSystemsChange,
  author = {Knoster, T.},
  title = {A Framework for Thinking About Systems Change},
  booktitle = {Restructuring for Caring and Effective Education: Piecing the Puzzle Together},
  editor = {Villa, R. and Thousand, J.},
  publisher = {Paul H. Brookes Publishing Co},
  address = {Baltimore},
  year = {2000},
  pages = {93--128}
}

@book{bridges2003bridgesTransitionModel,
  author = {Bridges, W.},
  title = {Managing Transitions: Making the Most of Change},
  edition = {2nd},
  year = {2003},
  publisher = {Da Capo Press},
  address = {Cambridge, MA}
}

@book{schein2010organizational,
  author    = {Edgar H. Schein},
  title     = {Organizational Culture and Leadership},
  edition   = {4th},
  year      = {2010},
  publisher = {Wiley},
  address   = {San Francisco}
}

@inproceedings{modica2010portfolioPriotization,
    author = {Modica, Jose Eduardo and Rabechini, Roque, Jr. and Martins Braun, Edison},
    title = {{Prioritization of a Portfolio of Projects}},
    volume = {2010 8th International Pipeline Conference, Volume 3},
    series = {International Pipeline Conference},
    pages = {41-48},
    year = {2010},
    month = {09},
    doi = {10.1115/IPC2010-31509}
}

@article{venkatesh1996perceivedEaseOfUse,
    author = {Venkatesh, Viswanath and Davis, Fred D.},
    title = {{A Model of the Antecedents of Perceived Ease of Use: Development and Test}},
    journal = {Decision Sciences},
    volume = {27},
    number = {3},
    pages = {451-481},
    doi = {https://doi.org/10.1111/j.1540-5915.1996.tb00860.x},
    year = {1996}
}

@misc{harvard2022identifyBusinessNeeds,
  author       = {Harvard Business School Online},
  title        = {How to Identify Business \& Market Opportunities},
  year         = {2022},
  url          = {https://online.hbs.edu/blog/post/how-to-identify-business-opportunities},
  note         = {Accessed: 2025-02-19}
}

@article{remenyi1999continuousEvaluation,
  author = {Remenyi, D. and Sherwood-Smith, M.},
  title = {Maximise Information Systems Value by Continuous Participative Evaluation},
  journal = {Logistics Information Management},
  volume = {12},
  number = {1/2},
  pages = {14-31},
  year = {1999},
  doi = {10.1108/09576059910256222}
}

@article{abu-shanab2014knowledge_sharing_practices,
    author = {Abu-Shanab, Emad and Haddad, Maram and Knight, Michael},
    year = {2014},
    month = {01},
    pages = {},
    title = {Knowledge Sharing Practices and the Learning Organization: A Study},
    volume = {XII}
}

@article{calantone2002learning,
    title = {Learning orientation, firm innovation capability, and firm performance},
    journal = {Industrial Marketing Management},
    volume = {31},
    number = {6},
    pages = {515-524},
    year = {2002},
    issn = {0019-8501},
    doi = {https://doi.org/10.1016/S0019-8501(01)00203-6},
    author = {Roger J Calantone and S.Tamer Cavusgil and Yushan Zhao}
}

@article{kujala2022stakeholdersEngagement,
    author = {Kujala, Johanna and Sachs, Sybille and Leinonen, Heta and Heikkinen, Anna and Laude, Daniel},
    year = {2022},
    month = {05},
    pages = {1136-1196},
    title = {Stakeholder Engagement: Past, Present, and Future},
    volume = {61},
    journal = {Business \& Society},
    doi = {10.1177/00076503211066595}
}

@article{van_gemert1999designingCommunicationPlan,
    author = {Lisette Van Gemert and Egbert Woudstra},
    title = {Designing a strategic communication plan},
    journal = {Communicatio},
    volume = {25},
    number = {1-2},
    pages = {73--87},
    year = {1999},
    publisher = {NCA Website},
    doi = {10.1080/02500169908537883}
}

@misc{uoft2025AIevaluation,
    author       = {University of Toronto Library},
    title        = {Critically Evaluating AI Tools},
    year        = {2025},
  howpublished = {\url{https://guides.library.utoronto.ca/image-gen-ai/critical-evaluation}},
  note         = {Accessed: 2025-02-20}
}

@article{nazir2024collectiveOwnership,
    title = {{Understanding collective ownership in agile development: Turbo charging the process}},
    journal = {Information and Management},
    volume = {61},
    number = {6},
    pages = {104004},
    year = {2024},
    issn = {0378-7206},
    doi = {https://doi.org/10.1016/j.im.2024.104004},
    author = {Salman Nazir and Stephane Eric Collignon and Nanda Chingleput Surendra}
}

@book{kotter2008senseOfUrgency,
    author    = {Kotter, John P.},
    title     = {A Sense of Urgency},
    year      = {2008},
    publisher = {Harvard Business Press},
    address   = {Boston, MA}
}

@article{agarwal1998rewardSystems,
  author = {Agarwal, N. C.},
  title = {Reward systems: Emerging trends and issues},
  journal = {Canadian Psychology / Psychologie canadienne},
  year = {1998},
  volume = {39},
  number = {1-2},
  pages = {60--70},
  doi = {10.1037/h0086795}
}

@inproceedings{alami2024bestPracticesInRecruitment,
    author = {Alami, Adam and Zahedi, Mansooreh and Ernst, Neil},
    title = {{Are You a Real Software Engineer? Best Practices in Online Recruitment for Software Engineering Studies}},
    year = {2024},
    isbn = {9798400705670},
    publisher = {Association for Computing Machinery},
    address = {New York, NY, USA},
    doi = {10.1145/3643664.3648207},
    booktitle = {Proceedings of the 1st IEEE/ACM International Workshop on Methodological Issues with Empirical Studies in Software Engineering},
    pages = {52--57},
    numpages = {6},
    location = {Lisbon, Portugal},
    series = {WSESE '24}
}

@misc{dalrymple2024guaranteedsafeaiframework,
      title={{Towards Guaranteed Safe AI: A Framework for Ensuring Robust and Reliable AI Systems}}, 
      author={David Dalrymple and Joar Skalse and Yoshua Bengio and Stuart Russell and Max Tegmark and Sanjit Seshia and Steve Omohundro and Christian Szegedy and Ben Goldhaber and Nora Ammann and Alessandro Abate and Joe Halpern and Clark Barrett and Ding Zhao and Tan Zhi-Xuan and Jeannette Wing and Joshua Tenenbaum},
      year={2024},
      eprint={2405.06624},
      archivePrefix={arXiv},
      primaryClass={cs.AI}
}

@misc{laoyan2025importanceOfSettingShortTermGoals,
  author    = {Sarah Laoyan},
  title     = {{The Importance of Setting Short-Term Goals (with examples)}},
  journal   = {Asana},
  year      = {2025},
  url       = {https://asana.com/resources/short-term-goals},
  note         = {Accessed: 2025-02-21}
}

@article{adebayo2022AgileMindset,
    author = {Adebayo, Omowunmi},
    year = {2022},
    month = {12},
    pages = {672-681},
    title = {Agile and organizational culture: Fostering agile values and mindset},
    volume = {7},
    journal = {International Journal of Science and Research Archive},
    doi = {10.30574/ijsra.2022.7.2.0265}
}

@article{tallon2019search_for_organizational_agility,
    title = {Information technology and the search for organizational agility: A systematic review with future research possibilities},
    journal = {The Journal of Strategic Information Systems},
    volume = {28},
    number = {2},
    pages = {218-237},
    year = {2019},
    note = {SI: Review issue},
    issn = {0963-8687},
    doi = {https://doi.org/10.1016/j.jsis.2018.12.002},
    author = {Paul P. Tallon and Magno Queiroz and Tim Coltman and Rajeev Sharma}
}

@article{kumar2024fosteringInnovation,
    author = {Kumar, Parveen},
    year = {2024},
    month = {03},
    pages = {363-370},
    title = {The Influence of Fostering an Innovative Organisation Culture on Employee Engagement},
    volume = {16},
    journal = {NHRD Network Journal},
    doi = {10.1177/26314541231215398}
}

@misc{thomke2024buildingExperimentationCulture,
  title={The Critical Role of Leadership in Building a Culture of Experimentation},
  author={Thomke, Stefan H.},
  year={2024},
  month={May},
  url={https://www.exed.hbs.edu/blog/building-culture-experimentation},
  note={Accessed: 2025-02-22}
}

@article{mughairi2020coachingAndMentoring,
    author = {Hilali, Khalid and Al Mughairi, Badar and Kian, Mooi and Karim, Dr. Asif},
    year = {2020},
    month = {03},
    pages = {},
    title = {Coaching and Mentoring. Concepts and Practices in Development of Competencies: A Theoretical Perspective},
    volume = {10},
    journal = {International Journal of Academic Research in Accounting, Finance and Management Sciences},
    doi = {10.6007/IJARAFMS/v10-i1/6991}
}

@article{london1999empowerContinuousLearning,
    author = {London, Manuel and Smither, James W.},
    title = {Empowered self-development and continuous learning},
    journal = {Human Resource Management},
    volume = {38},
    number = {1},
    pages = {3-15},
    doi = {https://doi.org/10.1002/(SICI)1099-050X(199921)38:1<3::AID-HRM2>3.0.CO;2-M},
    year = {1999}
}

@article{floridi2024whyAIhypeIsTechBubble,
  author    = {Luciano Floridi},
  title     = {Why the {AI} Hype is Another Tech Bubble},
  journal   = {Philosophy \& Technology},
  year      = {2024},
  volume    = {37},
  number    = {4},
  pages     = {1--13},
  doi       = {10.1007/s13347-024-00817-w}
}

@inproceedings{mattmann2024AIhypeVsReality,
  author={Mattmann, Chris A. and Broderick, Daniel},
  booktitle={2024 IEEE Aerospace Conference}, 
  title={AI Hype versus Reality - Will It Work for You?}, 
  year={2024},
  volume={},
  number={},
  pages={1-10},
  doi={10.1109/AERO58975.2024.10521287}
}

@misc{nahar2023metasummarychallengesbuildingproducts,
      title={{A Meta-Summary of Challenges in Building Products with ML Components -- Collecting Experiences from 4758+ Practitioners}}, 
      author={Nadia Nahar and Haoran Zhang and Grace Lewis and Shurui Zhou and Christian K\"astner},
      year={2023},
      eprint={2304.00078},
      archivePrefix={arXiv},
      primaryClass={cs.SE}
}

@article{Eitel-Porter2021ethicalAI,
  author    = {Eitel-Porter, Rachel},
  title     = {Beyond the promise: implementing ethical AI},
  journal   = {AI Ethics},
  volume    = {1},
  pages     = {73--80},
  year      = {2021},
  doi       = {10.1007/s43681-020-00011-6}
}

@misc{google2025AIprinciples,
  author       = {{Google}},
  title        = {Our AI Principles},
  year         = {2025},
  url          = {https://ai.google/principles/},
  note         = {Google AI Responsibility},
  urldate      = {2026-04-21}
}

@misc{eu2025aiAct,
  author       = {{European Union}},
  title        = {AI Act - Artificial Intelligence Act},
  year         = {2025},
  url          = {https://artificialintelligenceact.eu},
  note         = {Accessed: 2025-05-07}
}

@article{parasuraman2000automationModel,
  author={Parasuraman, R. and Sheridan, T.B. and Wickens, C.D.},
  journal={IEEE Transactions on Systems, Man, and Cybernetics - Part A: Systems and Humans}, 
  title={A model for types and levels of human interaction with automation}, 
  year={2000},
  volume={30},
  number={3},
  pages={286-297},
  doi={10.1109/3468.844354}
}

@article{goodman2004mediumAndTheMessage,
    author = {Joanna Goodman and Catherine Truss \dag{} and},
    title = {The medium and the message: communicating effectively during a major change initiative},
    journal = {Journal of Change Management},
    volume = {4},
    number = {3},
    pages = {217--228},
    year = {2004},
    publisher = {Routledge},
    doi = {10.1080/1469701042000255392}
}

@book{wohlin2012experimentation,
  title={Experimentation in software engineering},
  author={Wohlin, Claes and Runeson, Per and H{\"o}st, Martin and Ohlsson, Magnus C and Regnell, Bj{\"o}rn and Wessl{\'e}n, Anders},
  year={2012},
  publisher={Springer Science \& Business Media}
}

\appendix{}
\section{}

Table \ref{tab:mapping_results} shows the outcome of the mapping process between existing change management frameworks and the identified BSE-related challenges of AI adoption.

Figure \ref{fig:proposed_framework} demonstrates the proposed early-stage BSE-informed framework, as this emerged from RQ\textsubscript{1} and RQ\textsubscript{2}.

\begin{figure*}[h!]
    \centerline{\includegraphics[scale=0.7]{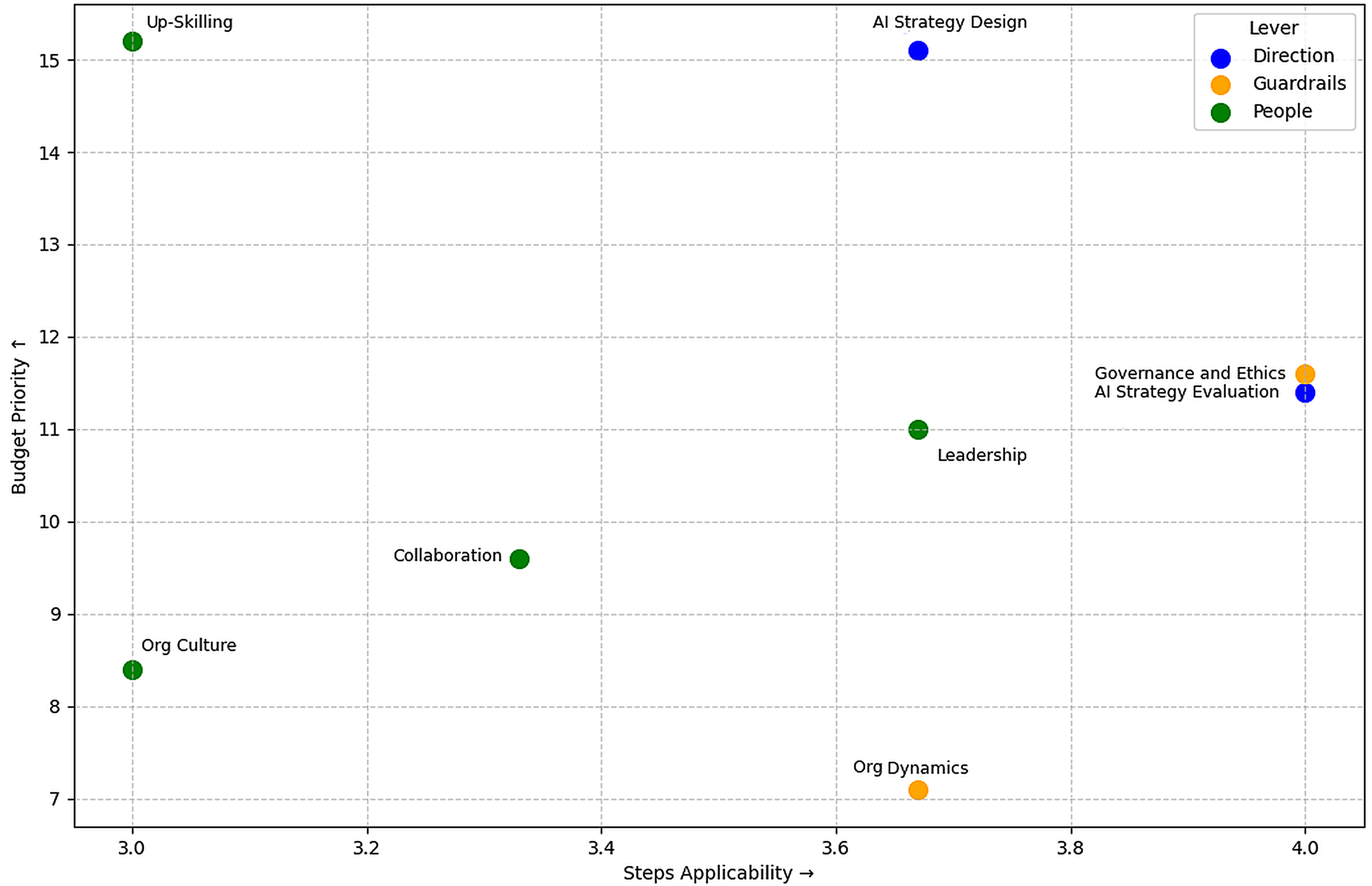}}
    \caption{Early-stage AI integration paradox}
    \label{fig:early-stage_paradox}
\end{figure*}

Figure \ref{fig:early-stage_paradox} presents the early-stage AI integration paradox, combining the perceived importance of each dimension (y-axis) and their perceived applicability (x-axis).

\onecolumn

\begin{small}
\begin{longtable}{@{}p{0.17\linewidth} p{0.20\linewidth} p{0.28\linewidth} p{0.27\linewidth}@{}}
    \caption{Mapping between BSE-related challenges and prior change management models.}
    \label{tab:mapping_results} \\
    \toprule
    \textbf{Challenge} & \textbf{Sub-challenge} & \textbf{Mapping to Change Management Models} & \textbf{Referred Model(s)} \\
    \midrule
    \endfirsthead

    \multicolumn{4}{l}{\textit{Table \ref{tab:mapping_results} continued from previous page}} \\
    \toprule
    \textbf{Challenge} & \textbf{Sub-challenge} & \textbf{Mapping to Change Management Models} & \textbf{Referred Model(s)} \\
    \midrule
    \endhead

    \midrule
    \multicolumn{4}{r}{\textit{Continued on next page}} \\
    \endfoot

    \bottomrule
    \endlastfoot

    \multirow{3}{=}{Change Management Strategy}
        & \multirow{2}{=}{External communication}
            & Constant communication to all stakeholders during change
            & S17, S19, S27 \\
        &   & Engage with all stakeholders
            & S16, S23, S26, S37, S43 \\
        \cmidrule(l){2-4}
        & \multirow{2}{=}{Internal communication}
            & Build the right communication plan
            & S35 \\
        &   & Communicate the change
            & S01, S03, S05, S09, S13, S17, S18, S19, S20, S22, S25, S27, S28, S29, S32, S33, S35, S36, S38, S40, S41, S43 \\
    \cmidrule(l){1-4}
    \multirow{2}{=}{Ethical Concerns}
        & Considerate use of external AI with company's data
            & \textit{No relevant steps or factors}
            & --- \\
        \cmidrule(l){2-4}
        & Risk aversion due to sensitive data
            & Encourage experimentation through pilot projects
            & S14, S33 \\
    \cmidrule(l){1-4}
    \multirow{6}{=}{Organizational Readiness for AI Adoption}
        & Adaptation to global and local data legislation
            & External environment (including regulations)
            & S08, S23 \\
        \cmidrule(l){2-4}
        & Cross-functional collaboration
            & Effective change team should be cross-functional
            & S09, S38 \\
        \cmidrule(l){2-4}
        & \multirow{2}{=}{Ensuring organizational alignment}
            & Create a sense of urgency
            & S05, S17, S19, S20, S22, S23, S31, S33, S39 \\
        &   & Create a shared need
            & S01, S02, S13, S14, S19, S43 \\
        \cmidrule(l){2-4}
        & \multirow{3}{=}{Handling differing views regarding AI transformation}
            & Line up political sponsorship
            & S12, S17, S19, S27 \\
        &   & Awareness of organizational timing (due to pressures)
            & S41 \\
        &   & Listen to employees' concerns
            & S32 \\
        \cmidrule(l){2-4}
        & Inter-generational collaboration
            & \textit{No relevant steps or factors}
            & --- \\
        \cmidrule(l){2-4}
        & Public sector's unique nature
            & Tailored solutions to the organization's context
            & S13, S23, S43 \\
    \cmidrule(l){1-4}
    \multirow{9}{=}{Resistance to Change}
        & \multirow{3}{=}{Job displacement}
            & Coaching
            & S13, S36 \\
        &   & Constant communication to all stakeholders during change
            & S17, S19, S27 \\
        &   & Training
            & S01, S09, S13, S19, S22, S26, S29, S30, S32, S33, S34, S41, S43 \\
        \cmidrule(l){2-4}
        & \multirow{3}{=}{Lack of deep AI knowledge and understanding}
            & Coaching
            & S13, S36 \\
        &   & Encourage experimentation
            & S14, S33 \\
        &   & Training
            & S01, S09, S13, S19, S22, S26, S29, S30, S32, S33, S34, S41, S43 \\
        \cmidrule(l){2-4}
        & \multirow{3}{=}{Lack of patience to work with AI}
            & Coaching
            & S13, S36 \\
        &   & Motivation using incentives and rewards
            & S01, S13, S19, S21 \\
        &   & Training
            & S01, S09, S13, S19, S22, S26, S29, S30, S32, S33, S34, S41, S43 \\
    \cmidrule(l){1-4}
    \multirow{4}{=}{Skills in the Era of AI}
        & Keeping up with AI advancements
            & Continuous learning initiatives
            & S14, S34, S41, S43 \\
        \cmidrule(l){2-4}
        & Lack of skills due to AI
            & \textit{No relevant steps or factors}
            & --- \\
        \cmidrule(l){2-4}
        & \multirow{2}{=}{Necessity to adjust skill-set}
            & Coaching
            & S13, S36 \\
        &   & Training
            & S01, S09, S13, S19, S22, S26, S29, S30, S32, S33, S34, S41, S43 \\
    \cmidrule(l){1-4}
    \multirow{2}{*}{\makecell[l]{\\\\Strategic Adoption of AI}}
        & \multirow{2}{=}{Balance automation with human supervision}
            & Continuously evaluate change
            & S05, S07, S13, S15, S19, S25, S27, S32, S33 \\
        &   & Use measurements, such as KPIs
            & S05, S28, S36 \\
        \cmidrule(l){2-4}
        & \multirow{3}{=}{Considering AI as a team member, instead of tool}
            & Coaching
            & S13, S36 \\
        &   & Encourage experimentation
            & S14, S33 \\
        &   & Training
            & S01, S09, S13, S19, S22, S26, S29, S30, S32, S33, S34, S41, S43 \\
        \cmidrule(l){2-4}
        & \multirow{2}{=}{Contextual use of AI}
            & Coaching
            & S13, S36 \\
        &   & Training
            & S01, S09, S13, S19, S22, S26, S29, S30, S32, S33, S34, S41, S43 \\
        \cmidrule(l){2-4}
        & Establish balance in using internal and external AI solutions
            & Use measurements, such as KPIs
            & S05, S28, S36 \\
        \cmidrule(l){2-4}
        & \multirow{2}{=}{Over-reliance on AI}
            & Coaching
            & S13, S36 \\
        &   & Training
            & S01, S09, S13, S19, S22, S26, S29, S30, S32, S33, S34, S41, S43 \\
        \cmidrule(l){2-4}
        & \multirow{2}{=}{Purpose-driven AI adoption}
            & Create short-term wins
            & S13, S22, S23, S27, S30, S32, S33 \\
        &   & Internal gap analysis
            & S30, S36, S37 \\

\end{longtable}
\end{small}

\twocolumn

\begin{landscape}
\thispagestyle{plain}

\vspace*{\fill}

\begin{center}
    \includegraphics[
        height=0.9\textheight,
        keepaspectratio
    ]{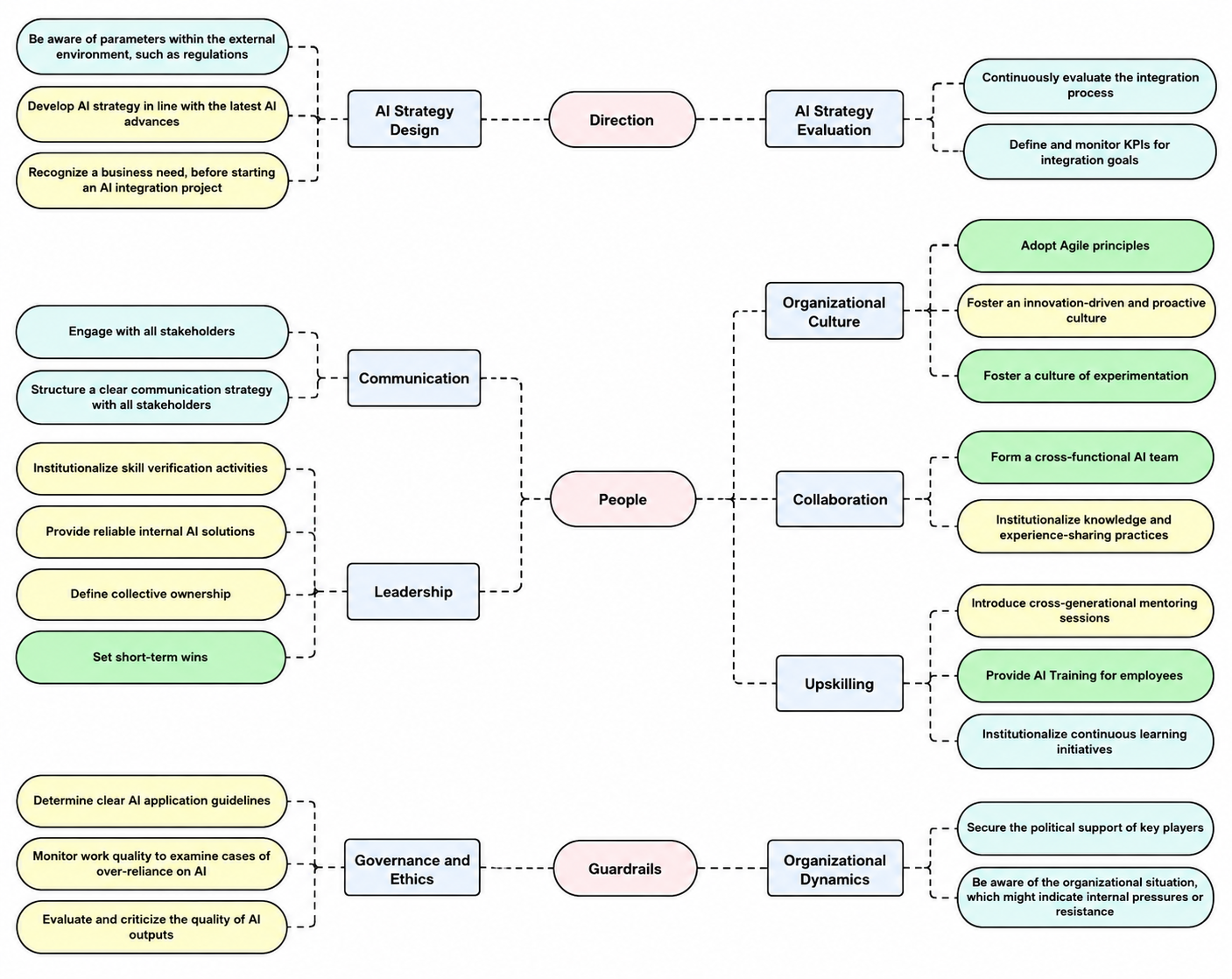}

    \captionof{figure}{Early-stage BSE-informed framework — RQ\textsubscript{2} outcome}
    \label{fig:proposed_framework}
\end{center}

\vspace*{\fill}

\end{landscape}

\end{document}